\documentclass[a4paper,12pt,leqno]{article}
\usepackage[T1]{fontenc}
\pdfoutput=1
\usepackage[utf8]{inputenc}
\usepackage[english]{babel}

\usepackage{titlesec}
\titleformat*{\section}{\Large\bfseries}
\titleformat*{\subsection}{\large\bfseries}

\usepackage{amssymb, amsmath, amsthm}
\usepackage{mathtools}
\usepackage{mathrsfs}
\usepackage{setspace}
\usepackage{longtable}
\usepackage{booktabs}
\usepackage{tabularx}
\usepackage{tabu}
\usepackage{upgreek}	
\usepackage{fixmath}
\usepackage{nccmath}
\usepackage{tensor}
\usepackage{accents}
\usepackage{stackengine}
\usepackage{esint}
\usepackage{chngcntr}
\usepackage{hyperref}
\hypersetup{
	colorlinks   = true,
	linkcolor = red,
	citecolor    = blue,
	urlcolor = black,	
	pdfstartview=
}
\usepackage[numbers,sort&compress]{natbib}
\usepackage{enumitem}
\usepackage{parcolumns}
\usepackage{lipsum,paracol}
\usepackage{geometry}
\usepackage{array}
\usepackage[toc, page]{appendix}
\def\g{\mbox{\normalfont g}}
\def\UOmega{\mathrm{\Omega}}
\DeclareRobustCommand{\lchi}{{\mathpalette\tempchi\relax}}
\newcommand{\tempchi}[2]{\raisebox{\depth}{$#1\chi$}}
\DeclareRobustCommand{\lupvarphi}{{\mathpalette\tempupvarphi\relax}}
\newcommand{\tempupvarphi}[2]{\raisebox{.5mm}{$#1\upvarphi$}}
\DeclareRobustCommand{\luppsi}{{\mathpalette\tempuppsi\relax}}
\newcommand{\tempuppsi}[2]{\raisebox{.3mm}{$#1\uppsi$}}
\DeclareRobustCommand{\lpsi}{{\mathpalette\temppsi\relax}}
\newcommand{\temppsi}[2]{\raisebox{.5mm}{$#1\psi$}}
\DeclareRobustCommand{\lvarphi}{{\mathpalette\tempvarphi\relax}}
\newcommand{\tempvarphi}[2]{\raisebox{.5mm}{$#1\varphi$}}
\makeatletter
\def\moverlay{\mathpalette\mov@rlay}
\def\mov@rlay#1#2{\leavevmode\vtop{%
		\baselineskip\z@skip \lineskiplimit-\maxdimen
		\ialign{\hfil$\m@th#1##$\hfil\cr#2\crcr}}}
\newcommand{\charfusion}[3][\mathord]{
	#1{\ifx#1\mathop\vphantom{#2}\fi
		\mathpalette\mov@rlay{#2\cr#3}
	}
	\ifx#1\mathop\expandafter\displaylimits\fi}
\makeatother

\newcommand{\bigcupdot}{\charfusion[\mathop]{\bigcup}{\cdot}}

\def\alp{\indices{_\alpha}}

\def\entprod{\tensor*[^s]{\hspace{-0.1em}\Sigma}{}}
\def\tipentflux{\tensor*[^s]{\Upphi}{^{\text{TIP}}}}
\def\tipentprod{\tensor*[^s]{\hspace{-0.1em}\Sigma}{^{\text{TIP}}}}
\newcommand{\total}[2]{\frac{\mathrm{d}#1}{\mathrm{d}#2}}
\newcommand{\diver}{\mathop{\mathrm{div}}\nolimits} 
\newcommand{\smatder}[1]{\frac{\delta#1}{\delta t}}
\newcommand{\indi}[1]{\indices{#1}}
\def\rhoAlpha{\rho\alp} 
\def\Norm#1{\lVert#1\rVert}
\def\zhet{{\mathbf{z}}}
\newcommand{\dcont}{\raisebox{.35mm}{$\,:\,$}}
\newcommand{\dx}{\mathrm{d}x}
\newcommand{\dt}{\mathrm{d}t}
\newcommand{\da}{\mathrm{d}a}
\newcommand{\dv}{\mathrm{d}v}
\newcommand{\dS}{\mathrm{d}A}
\newcommand{\dL}{\mathrm{dL}}

\newcommand{\weak}[2]{\Big\langle #1,\,#2 \Big\rangle}
\newcommand{\aweak}[2]{\left\langle #1,\,#2 \right\rangle}
\newcommand{\tweak}[2]{\langle #1,\,#2 \rangle}
\newcommand{\vardif}[2]{\frac{\delta #1}{\delta #2}}
\newcommand{\mvardif}[2]{\mfrac{\delta #1}{\delta #2}}
\newcommand{\tvardif}[2]{\tfrac{\delta #1}{\delta #2}}
\newcommand{\pd}[2]{\frac{\partial #1}{\partial #2}}
\newcommand{\mpd}[2]{\mfrac{\partial #1}{\partial #2}}
\newcommand{\tpd}[2]{\tfrac{\partial #1}{\partial #2}}
\newcommand{\mc}[1]{\mathcal{#1}}
\DeclareMathAlphabet{\mathpzc}{OT1}{pzc}{m}{it}
\AtBeginEnvironment{pmatrix}{\everymath{\displaystyle}}
\AtBeginEnvironment{bmatrix}{\everymath{\displaystyle}}

\makeatletter
\def\revddots{\mathinner{\mkern1mu\raise\p@
    \vbox{\kern7\p@\hbox{.}}\mkern2mu
    \raise4\p@\hbox{.}\mkern2mu\raise7\p@\hbox{.}\mkern1mu}}
\makeatother
\mathcode`@="8000 
{\catcode`\@=\active\gdef@{\mkern1mu}}
\newcommand{\beq}{\begin{equation}}
\newcommand{\eeq}{\end{equation}}
\newcommand {\en} {\end{equation}}
\newcommand{\be}{\begin{enumerate}}
\newcommand{\ee}{\end{enumerate}}
\newcommand{\bi}{\begin{itemize}}
\newcommand{\ei}{\end{itemize}}

\def\E{\mathbb{E}}

\def\R{\mathbb{R}}

\def\L{\mathbb{L}}

\def\N{\mathbb{N}}

\DeclareMathOperator{\trace}{tr}

\def\norm2#1{\|#1\|_2}

\numberwithin{equation}{section}
\numberwithin{table}{section}
\numberwithin{figure}{section}
\newtheorem{theorem}{Theorem}[section]

\newtheorem{lemma}[theorem]{Lemma}
\newtheorem{corollary}[theorem]{Corollary}

\newtheorem{remark}[theorem]{Remark}
\newcommand{\bmat}{\begin{bmatrix}}
\newcommand{\ebmat}{\end{bmatrix}}
\newcommand{\pmat}{\begin{pmatrix}}
\newcommand{\epmat}{\end{pmatrix}}
\newcommand{\smat}{\begin{smallmatrix}}
\newcommand{\esmat}{\end{smallmatrix}}

\mathchardef\Gamma="7100
\mathchardef\Delta="7101
\mathchardef\Theta="7102
\mathchardef\Lambda="7103
\mathchardef\Xi="7104
\mathchardef\Pi="7105
\mathchardef\Sigma="7106
\mathchardef\Upsilon="7107
\mathchardef\Phi="7108
\mathchardef\Psi="7109
\mathchardef\Omega="710A

\newcommand{\cD}{{\mathcal D}}

\newcommand {\eqn}  {\begin{eqnarray}}
\newcommand {\enn}  {\end{eqnarray}}
\newcommand {\bstar}    {\begin{eqnarray*}}
\newcommand {\estar}    {\end{eqnarray*}}

\newcommand {\mat}  [1] {\left[\begin{array}{#1}}
\newcommand {\rix}      {\end{array}\right]}
\catcode`@=11     
\font\tenex=cmex10 
\newdimen\p@renwd
\setbox0=\hbox{\tenex B} \p@renwd=\wd0 
\def\bmat#1{\begingroup \m@th
  \setbox\z@\vbox{\def\cr{\crcr\noalign{\kern2\p@\global\let\cr\endline}}%
    \ialign{$##$\hfil\kern2\p@\kern\p@renwd&\thinspace\hfil$##$\hfil
      &&\quad\hfil$##$\hfil\crcr
      \omit\strut\hfil\crcr\noalign{\kern-\baselineskip}%
      #1\crcr\omit\strut\cr}}%
  \setbox\tw@\vbox{\unvcopy\z@\global\setbox\@ne\lastbox}%
  \setbox\tw@\hbox{\unhbox\@ne\unskip\global\setbox\@ne\lastbox}%
  \setbox\tw@\hbox{$\kern\wd\@ne\kern-\p@renwd\Bigl[\kern-\wd\@ne
    \global\setbox\@ne\vbox{\box\@ne\kern2\p@}%
    \vcenter{\kern-\ht\@ne\unvbox\z@\kern-\baselineskip}\,\Bigr]$}%
  \null\;\vbox{\kern\ht\@ne\box\tw@}\endgroup}
\catcode`@=12    

\begin {document}
	
	
	\title{Operator-GENERIC Formulation of\\ Thermodynamics of Irreversible Processes}
	\author{Arbi Moses Badlyan\footnotemark[1]\and Christoph Zimmer\footnotemark[1]}
	\date{November 10, 2018}

	\maketitle\thispagestyle{empty}

	\begin{abstract}\noindent 
		Metriplectic systems are state space formulations that have become well-known under the acronym GENERIC.
		In this work we present a GENERIC based state space formulation in an operator setting that encodes a weak-formulation of the field equations describing the dynamics of a homogeneous mixture of compressible heat-conducting Newtonian fluids consisting of reactive constituents.
		We discuss the mathematical model of the fluid mixture formulated in the framework of continuum thermodynamics.
		The fluid mixture is considered an open thermodynamic system that moves free of external body forces.
		As closure relations we use the linear constitutive equations of the phenomenological theory known as Thermodynamics of Irreversible Processes (TIP).
		The phenomenological coefficients of these linear constitutive equations satisfy the Onsager-Casimir reciprocal relations.
		We present the state space representation of the fluid mixture, formulated in the extended GENERIC framework for open systems, specified by a symmetric, mixture related dissipation bracket and a mixture related Poisson-bracket for which we prove the Jacobi-identity. 
		\noindent
	\end{abstract}
	
	\vskip .3truecm
	\noindent
	{\bf AMS(MOS) subject classification:} 35Q35, 37K05, 37L99

	\footnotetext[1]{
		Institut f\"ur Mathematik, TU Berlin, Germany, \texttt{$\{$badlyan,zimmer$\}$@math.tu-berlin.de}.
	}


\begingroup
\let\clearpage\relax
\section{Introduction}\label{sec:introduction}
Various physical phenomena are mathematically described as finite or infinite dimensional dynamical systems.~The corresponding mathematical models often contain terms which can be identified as being part of some Hamiltonian system, whereas other terms can be related to dissipation phenomena, see e.g. \cite{Kau84,KauMor82}. In the Hamiltonian formulation of classical mechanics the time evolution of dynamical quantities is typically expressed by means of Poisson brackets \cite{Abr87,MarR99}.
The main objective of the bracket formulation is to obtain a representation in which the dynamical system is endowed with certain structural properties which in general are geometric by nature \cite{Mor09}.
 
A double-bracket formalism for the description of dissipative dynamical systems containing aspects of generalized Hamiltonian and gradient flows was presented in \cite{Mor86}. The dynamical systems described in this framework were named metriplectic systems (see also \cite{Mor82,Mor84}).
For further details regarding gradient flows see e.g. \cite{AGS08}.
The acronym GENERIC \cite{GrmOet97I,Oet05,OetGrm97II} stands for a state space formalism built around a rate equation, that in its original form is formulated for isolated systems, and represents the additive combination of a Hamiltonian flow and a gradient flow.
The so-called 'General Equation for the Non-Equilibrium Reversible-Irreversible Coupling' (GENERIC) is based on the same geometric structures used to define metriplectic systems, cf.~\cite[p.~325]{Bl13}.
This formalism is able to cover the dynamics of a large variety of physical systems~\cite{Mie11,Mie15,MiM17,Oet11}. The double bracket formulation induced by GENERIC via duality pairing is a so called two-generator double-bracket formulation, where the dynamics of the system under consideration is described by means of two generating functionals, supplemented by two complimentary degeneracy conditions that are also called non-interacting conditions. 
In~\cite{Oet06} the GENERIC framework has been extended to a formalism that can cover the dynamics of open non-equilibrium thermodynamical systems. In this extended framework for open systems the boundary contributions to the generalized
brackets are incorporated by means of boundary variables in the spirit of Stokes-Dirac structures~\cite{SchM02}.

In this work we present an operator based formalism that is closely related to the extended GENERIC framework for open systems. We prove that this operator formalism is able to encode a weak-formulation of the field equations describing the dynamics of a homogeneous mixture of heat-conducting compressible Newtonian fluids consisting of reactive constituents.
We present the functionals and operators for two distinct compositions of an abstract state variable that is related to a number of fields, which in our consideration are given by functions of time and implicit of space. These fields determine the macroscopic state of the fluid mixture. The two compositions of the state variable conform with two representations of hydrodynamics in form of dynamical systems modeled in the GENERIC framework. In the first case energy constitutes the thermodynamic potential, and in the second case entropy constitutes the thermodynamic potential. Note that the fields associated with the macroscopic state of the latter case correspond to the fields of classical hydrodynamics, see e.g.~\cite[Sec.~II]{OetGrm97II} for the one-component system, and for mixtures~\cite[p.~417~f.]{MeiR59}. The corresponding field equations, given in form of differential balance laws which are complemented by material specific closure relations, are modeled in the framework of the phenomenological theory known as Thermodynamics of Irreversible Processes (TIP)~\cite{Mazur,Eck1,Eck2,Eck3,Mei41,MeiR59}.
We formulate matrices by means of the phenomenological coefficients appearing in the linear constitutive relations of TIP, which are equipped with the Onsager-Casimir reciprocal relations \cite{Cas45,Ons1,Ons2}. According to the Onsager-Casimir reciprocal relations, the phenomenological coefficients related to volume variation and chemical reaction are coupled by an anti-symmetric relation \cite[p.~1790]{BotDre15}.
In case the differential mixture related entropy balance is used to model the dissipation operator of the GENERIC formulation, this typically results in a skew-symmetric block contained in the dissipation operator. Since the dissipation bracket induced by this dissipation operator is required to be symmetric, the Onsager-Casimir reciprocal relations prevent the formulation of a proper GENERIC model of the mixture. The general solution to this problem such that all Onsager-Casimir reciprocal relations can be transformed into pure Onsager reciprocal relations \cite{Mei73} requires the reinterpretation of the entropy production and the introduction of the concept of parities for the thermodynamical quantities (for an application see e.g.~\cite{BotDre15}). In this context the classical notion of thermodynamic fluxes and thermodynamic forces is replaced by the notion of cofactors of positive and negative parity. Since the classical notion of thermodynamic forces and fluxed conforms with the system theoretic description of thermodynamic systems, where fluxes and forces are related to efforts and flows (see e.g.~\cite{EMV05,EMV07,SchM01}), we develop an alternative way to deal with the Onsager-Casimir reciprocal relations. This results in a dynamically equivalent model of the mixture in the GENERIC framework with a mixture related self-adjoint dissipation and a skew-adjoint operator. The anti-symmetric bracket induced via duality paring by this mixture related skew-adjoint operator is proven to constitute a Poisson bracket.

The paper is organized as follows. In Section~\ref{sec:GENERIC} we introduce GENERIC in more detail and show how it can be rewritten as a collection of operator equations.
For isolated systems this collection corresponds with the time evolution equation of the GENERIC formalism, cf.~\cite[Eq.~(1)]{OetGrm97II}, and in case of open systems reflects a dynamical system that interacts with its environment in a system theoretic sense through boundary ports, see e.g.~\cite{BAW91,Dui09,Mas96,Wil72I,Wil72II}.
Section~\ref{sec:MixtureTheory} is devoted to the continuum theoretical motivation of our results. Here the field equations and the mixture related constitutive relations of TIP are presented. The first and second law of thermodynamics for continuous media are motivated and related to the balance law representing the conservation of energy and the entropy balance for open thermodynamical systems. 
For the sake of overview, the most important formulas and statements are summarized in Section~\ref{sec:formulas}. 
The Operator-GENERIC formulations, describing the dynamics of the reactive fluid mixture for isolated and open systems, are presented in Section~\ref{sec:mixture}.
Therein, we consider two cases, i.e. whether energy or entropy constitutes the thermodynamic potential. In Section~\ref{sec:mixture_GENERIC} the dynamics of the fluid mixture is formulated in the classical GENERIC framework. Concluding remarks and an outlook are presented in Section~\ref{sec:conclusion}.

\subsection*{Notation and conventions}\label{sec:NotCon}
\subsubsection*{List of frequently used symbols}
\begin{longtabu}{cl}
			Symbol & Name\\
			\cmidrule(lr){1-1}\cmidrule(lr){2-2}
			\addlinespace
			$\begin{aligned}\rho_\alpha\end{aligned}$ & Field of the mass density of constituent $\alpha$ in the mixture\\ 
			$\begin{aligned}\rho\end{aligned}$ & Field of the mixture mass density, $\begin{aligned}\textstyle \rho =\sum_{\alpha=1}^\nu \rho_\alpha\end{aligned}$ \\
			$\begin{aligned}\mathbf{M}\end{aligned}$ & Field of the mixture linear momentum density \\
			$\begin{aligned}\mathbf{v}\end{aligned}$ & Barycentric velocity field,  $\mathbf{v}=\mathbf{M}/(\sum_{\alpha=1}^\nu \rho_\alpha)$\\
			$\begin{aligned}T\end{aligned}$ & Absolute temperature field of the mixture \\
			$\begin{aligned}s\end{aligned}$ & Field of the mixture entropy density \\
			$\begin{aligned}\mu_\alpha\end{aligned}$ & Chemical potential of constituent $\alpha$ in mixture \\
			$\begin{aligned}\lambda\end{aligned}$ & Bulk (or volume) viscosity coefficient \\
			$\begin{aligned}\zeta\end{aligned}$ & Shear (or dynamic) viscosity coefficient \\	
			$\begin{aligned}u\end{aligned}$ & Field of the mixture internal energy density \\
			$\begin{aligned}p\end{aligned}$ & Thermodynamic equilibrium pressure \\	
			$\begin{aligned}\mathbf{T}\end{aligned}$ & Mixture Cauchy stress tensor	\\
			$\begin{aligned}~\mathbf{T}^\mathrm{d}\end{aligned}$ & Deviator of mixture Cauchy stress tensor \\
			$\begin{aligned}\pi\end{aligned}$ & Dynamic pressure \\ 
			$\begin{aligned}\kappa\end{aligned}$ & 	Heat conductivity coefficient \\
			$\begin{aligned}\mathbf{q}\end{aligned}$ & Mixture heat flux vector (non-convective flux of internal energy) \\
			$\begin{aligned}\mathbf{S}\end{aligned}$ & Viscosity (part of the) mixture stress tensor $\mathbf{T}$ \\
			$\begin{aligned}\tau_\alpha\end{aligned}$ & Mass production density of constituent $\alpha$\\
			$\begin{aligned}\mathbf{J}_\alpha\end{aligned}$ & Diffusion flux vector of constituent $\alpha$ in the mixture\\
			$\begin{aligned}m_\alpha\end{aligned}$ & 	Molecular mass of constituent $\alpha$ \\
			$\begin{aligned}\gamma_\alpha^k\end{aligned}$ & Stoichiometric coefficient of constituent $\alpha$ w.r.t. the $k$-th reaction \\
			$\begin{aligned}\Lambda\indices{^k}\end{aligned}$ &	Reaction rate density of $k$-th reaction \\
			$\begin{aligned}\mathbf{u}\end{aligned}$ &	Combined input - control \\
			$\begin{aligned}\mathbf{y}_H\end{aligned}$ & Outputs with respect to the change of the Hamiltonian \\
			$\begin{aligned}\mathbf{y}_S\end{aligned}$ & Outputs with respect to the change of the total entropy
\end{longtabu}

By the term \textbf{system} we mean a separable part of the physical universe which is defined by a set of macroscopic boundary conditions.
By an \textbf{isolated system} we mean a system that is \emph{materially}, \emph{mechanically}, and \emph{adiabatically closed}. As \emph{materially closed} systems we define systems without exchange of matter with the~\emph{environment}. As \emph{mechanically closed} systems we define systems without exchange of work with the environment, and as \emph{adiabatically closed} systems we define systems enclosed in thermally isolated walls.~If none of these conditions is fulfilled then we call the system under consideration~\textbf{open}.
\section{GENERIC and its Operator Formulation}\label{sec:GENERIC}

In this section we introduce the GENERIC formulation in more detail. As mentioned in the introduction, GENERIC induces a double bracket formulation in which the dynamics is split in two parts, a reversible and an irreversible one. The reversible part of the dynamics is described by the total energy functional (or Hamiltonian) $H$ of the system, mathematically described by means of a Poisson bracket $\{\cdot,\cdot\}$. The irreversible part is described by the \emph{entropy} $S$ of the system with the help of a so-called dissipation bracket $[\cdot,\cdot]$.

To introduce the brackets, let $\mc{Z}$ represent the ensemble of all macroscopic state variables of the thermodynamic system under consideration in their local representation such that at each fixed time parameter the state variable corresponds with an element (of a subset) of the Cartesian product of some normed spaces.
Let the total energy $H$ and the entropy $S$ of the system under consideration be known and given as smooth real valued functions defined on $\mc{Z}$, which we call~\emph{state space}.
The Poisson bracket $\{\cdot,\cdot\}$ is a bilinear map of the form
\begin{alignat*}{4}
	  &\{\cdot,\cdot\}\colon C^\infty(\mc{Z})\times C^\infty(\mc{Z})\rightarrow C^\infty(\mc{Z}),
	  &\qquad &
	  (A,B)\mapsto \{A,B\}.
\end{alignat*}
The dissipation bracket $[\cdot,\cdot]$ is a bilinear mapping defined analogously.
The Poisson bracket $\{\cdot,\cdot\}$ by definition has the following properties \\[0.5em]
\begin{tabu} to \textwidth {lX[c]}
	\textbf{Anti-symmetry:} &  $\{A,B\}=-\{B,A\}$,\\
	\textbf{Leibniz rule:} & $\{AB,C\}=A\{B,C\}+B\{A,C\}$,\\
	\textbf{Jacobi identity:} & $\{A,\{B,C\}\} + \{B,\{C,A\}\} + \{C,\{A,B\}\} =0$,
\end{tabu}\\[0.5em]
with $A,B,C\in C^ \infty(\mc{Z})$.
The Jacobi identity reflects the time-structure invariance of the Poisson bracket, see e.g. \cite{BriOet97} or \cite[Sec.~1.2]{Sur03}. The Leibniz rule states that the Poisson bracket $\{\cdot,\cdot\}$ is a derivation in each argument~\cite{Sur03}.

The dissipation bracket $[\cdot,\cdot]$ should also fulfill the Leibniz rule, cf.~\cite[p.~14~ff.]{Oet05}. Moreover, the dissipation bracket is required to be symmetric and non-negative. This results in the three properties \\[0.5em]
\begin{tabu} to \textwidth {lX[c]}
	\textbf{Symmetry:} & $[A,B]=[B,A]$,\\
	\textbf{Leibniz rule:} & $[AB,C]=A[B,C]+B[A,C]$,\\
	\textbf{Non-Negativity:} & $[A,A] \geq 0$,
\end{tabu}\\[0.5em]
which the dissipation bracket has to satisfy. 

The bracket-formalism of GENERIC for isolated systems requires that the Poisson and the dissipation bracket satisfy the following two non-interacting (or degeneracy) conditions
\begin{alignat}{6}	\label{eq:noninteracting}
\{F,S\}= 0 &\quad& \text{and} &\quad &[F,H]= 0 &\quad & \text{for all $F\in C^\infty(\mc{Z})$}.
\end{alignat}
These conditions separate the reversible and the irreversible evolution,~\cite[p.~2]{PavKG14}.
The time evolution of an arbitrary smooth observable $A\in C^\infty(\mc{Z})$ is given by
\begin{equation}\label{eq:GENERIC_isolated}
\frac{\mathrm{d}A}{\mathrm{d}t} = \{A,H\} + [A,S].
\end{equation}

The bracket formalism of GENERIC, in case of an isolated system, leads to the following two consequences,\\[1em]
\begin{subequations}\textit{Energy conservation} 
	\begin{alignat}{4}
	\label{eq:EnergyCons}
	\frac{\mathrm{d}H}{\mathrm{d}t} &= \{H,H\} + [H,S]= 0,	\intertext{\textit{Entropy production}} 
	\label{eq:EntropyProd}
	\frac{\mathrm{d}S}{\mathrm{d}t} &= \{S,H\} + [S,S]\geq 0,
	\end{alignat}
\end{subequations}
which are direct implications of the properties of the brackets, and the non-inter\-acting conditions~\eqref{eq:noninteracting} and~\eqref{eq:GENERIC_isolated}.
For further details see e.g.~\cite{Bl13}.

For systems confined to a time-independent \emph{domain} $\mathrm{\Omega} \subset \R^d$, i.e. an open -- in a topological sense -- and connected set, with boundary $\partial \mathrm{\Omega}$, the GENERIC formalism was extended in~\cite{Oet06} to open non-equilibrium thermodynamic systems that interact with their environment. To this end abstract Poisson and dissipation brackets were introduced. These abstract brackets, which we call \emph{full brackets}, are split into bulk- and boundary contributions
	\begin{align*}
		\{A,B\} &= \{A,B\}_{\text{bulk}}+\{A,B\}_{\text{boundary}},\\
		[A,B] &= [A,B]_{\text{bulk}}+[A,B]_{\text{boundary}}.
	\end{align*}
It is shown that the proper time evolution equation for a sufficiently smooth observable~$A$ in case of an open system is of the form 
\begin{equation}\label{eq:GENERIC_open}
\frac{\mathrm{d}A}{\mathrm{d}t} = \{A,B\} - \{A,B\}_{\text{boundary}} + [A,B] -[A,B]_{\text{boundary}}.
\end{equation}
Note that in \cite{Oet06} the time evolution equation \eqref{eq:GENERIC_open} is formulated by means of the bulk contributions of the corresponding brackets.

Considering open thermodynamic systems, results from continuum physics indicate that the total entropy of the system cannot be affected by any reversible dynamics, and that the total energy of the system is conserved under irreversible dynamics. It follows that for general open thermodynamic systems where the systems energy $H$ and entropy $S$ are known, the degeneracy conditions are given~-- in contrast to isolated systems and equation~\eqref{eq:noninteracting}~-- by
\begin{equation}
\label{eq:noninteracting_bulk}
\{A,S\}_{\text{bulk}}=0, \quad \qquad  [A,H]_{\text{bulk}}=0,
\end{equation}
for every observable $A$. Note that in general there can be reversible flux of entropy and irreversible flux of energy at the boundary \cite{Oet06}. So, the non-interaction conditions~\eqref{eq:noninteracting_bulk} are in general not requested for the boundary contributions of the brackets and therefore also not for the full brackets. However, in~\cite[Sec.~III]{Oet06} it is shown that for hydrodynamics the degeneracy is required for bulk and boundary contributions of both brackets.

The brackets that are constructed in this paper are written as integral expressions where we use the functional derivative $\vardif{A}{\zhet}$ for observables~$A$. The functional derivative is defined by  $\total{}{\varepsilon} A(\zhet + \varepsilon \boldsymbol{h})|_{\varepsilon=0}= \tweak{\vardif{A}{\zhet}}{\boldsymbol{h}}$, where $\zhet + \varepsilon \boldsymbol{h}$ is an element of the same space as $\zhet$ for small~$\varepsilon$.  The functional derivative is unique if it exists and it can be determined by the \emph{partial functional derivatives} of $A$ with respect to functions associated with~$\zhet$. For further details on the functional derivative, which is also called \emph{Volterra variational derivative} or \emph{Fr\'{e}chet-Volterra derivative}, we refer to~\cite[Sec.~2.4~f.]{AbrMarRat88} and \cite{Ham82,Pfe85}. 

With the functional derivative, the brackets can be written as
\begin{equation}\label{eq:mathfrak_J_and_R}
\{A,B\}(\zhet)=\! \int_\Omega \vardif{A}{\zhet} \mathfrak{J}(\zhet) \vardif{B}{\zhet}\, \dx \quad \text{and} \quad [A,B](\zhet)=\! \int_\Omega \vardif{A}{\zhet} \mathfrak{R}(\zhet) \vardif{B}{\zhet}\, \dx,
\end{equation}
where $\mathfrak{J}(\zhet)$ and $\mathfrak{R}(\zhet)$ are linear spatial-differential operators. Since it holds that, cf.~\cite[p.~224]{AveSmo67}, 
\begin{align}\label{eq:tmp1}
	\vardif{AB}{\zhet(x)}=B(\zhet)\vardif{A}{\zhet(x)} + A(\zhet)\vardif{B}{\zhet(x)},
\end{align}
the expressions of the brackets given by~\eqref{eq:mathfrak_J_and_R} in combination with~\eqref{eq:tmp1} allow us to prove easily the Leibniz rule. For the Poisson bracket we obtain
\begin{align*}
\int_\Omega \vardif{AB}{\zhet} \mathfrak{J}(\zhet) \vardif{C}{\zhet}\, \dx &= \int_\Omega \Big(B\vardif{A}{\zhet}+A\vardif{B}{\zhet} \Big) \mathfrak{J}(\zhet) \vardif{C}{\zhet}\, \dx\\
&= B\int_\Omega \vardif{A}{\zhet} \mathfrak{J}(\zhet) \vardif{C}{\zhet}\, \dx + A\int_\Omega \vardif{B}{\zhet} \mathfrak{J}(\zhet) \vardif{C}{\zhet}\, \dx,
\end{align*}
and an analogous formula holds for the dissipation bracket. In the following we will identify a special type of brackets which fulfill the Jacobi identity. All Poisson brackets appearing in this paper have this form. 

\begin{theorem}\label{th:jacobi_identity}
Let the state space $\mc{Z}$ consist of sufficiently smooth enough functions which not necessarily vanish at the boundary. Suppose that $\mc{Z}$ is (re-)arranged such that $\mc{Z} \ni \mathbf{z}=[\mathbf{\tilde{z}}^\top,\mathbf{M}^\top]^\top =[\tilde{z}_1,\ldots,\tilde{z}_{\mu+1},\mathbf{M}^\top]^\top$, where~$z_{\alpha}$, $\alpha=1,\ldots,{\mu+1}$, is a scalar field and~$\mathbf{M}$ is a vector field with a $d$-dimensional coordinate representation. Suppose that there exist continuously differentiable functions $f_\alpha\colon \R^{\mu+1} \to \R$, $\alpha=1,\ldots,{\mu+1}$, and define the bracket~$\{\cdot,\cdot\}$ by
\begin{align}\label{eq:jacobi_identity_bracket}
\begin{split}
\{A,B\}(\zhet) :=
\int_\Omega & - \sum_{\alpha=1}^{\mu+1} \tilde{z}_\alpha \Big[\Big(\vardif{A}{\mathbf{M}}\cdot \nabla\Big) \vardif{B}{\tilde{z}_\alpha} - \Big(\vardif{B}{\mathbf{M}}\cdot\nabla \Big) \vardif{A}{\tilde{z}_\alpha}\Big]\\
 & - \mathbf{M} \cdot \Big[\Big(\vardif{A}{\mathbf{M}}\cdot \nabla\Big) \vardif{B}{\mathbf{M}} - \Big(\vardif{B}{\mathbf{M}}\cdot\nabla\Big) \vardif{A}{\mathbf{M}}\Big]\\
& + \sum_{\alpha=1}^{\mu+1} \Big[\Big(\vardif{A}{\mathbf{M}}\cdot \nabla\Big)\Big(f_\alpha(\mathbf{\tilde{z}}) \vardif{B}{\tilde{z}
_\alpha}\Big) - \Big(\vardif{B}{\mathbf{M}}\cdot \nabla\Big)\Big(f_\alpha(\mathbf{\tilde{z}}) \vardif{A}{\tilde{z}_\alpha}\Big)\Big] \,\dx.
\end{split}
\end{align}
Then the bracket fulfills the Jacobi identity
$$\{A,\{B,C\}\} + \{B,\{C,A\}\} + \{C,\{A,B\}\} =0$$
for all observables $A,B,C\in C^{\infty}(\mc{Z})$.
\end{theorem}

\begin{proof}
To streamline the readability of the paper the proof is presented in the \hyperref[sec:proof_Jacobi_identity]{Appendix}.
\end{proof}

The expression~\eqref{eq:mathfrak_J_and_R} can be used to write the GENERIC formulations~\eqref{eq:GENERIC_isolated} and~\eqref{eq:GENERIC_open} as  operator equations. For this, let the state $\zhet$ evolve in an open subset $\mc{Z}$ of the space $\mc{D}_\zhet$, i.e., $\zhet\colon \mathbb{I} \to \mc{Z}$ for a bounded time interval $\mathbb{I}$. By the definition of the brackets~\eqref{eq:mathfrak_J_and_R} the space $\mc{D}_\zhet$ is a set of spatially dependent functions mapping from the domain $\Omega \subset \R^d$ into $\R^n$ with weak derivatives. We will denote by~$C^{\infty}(\Omega)$ the space of all infinitely differentiable functions and by~$W^{1,p}(\Omega)$, $p\geq 1$, the Sobolev space of all functions with a weak derivative where the function itself and its derivative are measurable and integrable up to the power of $p$, i.\,e., elements of~$L^p(\Omega)$. For further details we refer to~\cite{Ada75}. 

We will write $W^{1,p}(\Omega)^n$ for the $n$-fold Cartesian product of $W^{1,p}(\Omega)$ and $W^{1,p}(\Omega)^{n\ast}$ for its dual space, i.\,e., the space of all continuous linear functionals of $W^{1,p}(\Omega)^n$, $n\in \mathbb{N}$.
The duality pairing of $W^{1,p}(\Omega)^{n\ast}$ and $W^{1,p}(\Omega)^n$ is denoted by $\tweak{\cdot}{\cdot}$. For $\mathcal{F} \in L^2(\Omega)^n$ and $p \geq \frac{2d}{2+d}$ we define $\tweak{v}{\mathcal{F}} = \int_{\Omega} \mathcal{F} \cdot v\,\dx $ for all $v \in W^{1,p}(\Omega)^n$. Note that we have used the Sobolev embedding into the space $L^2(\Omega)$,~\cite[Lem.~5.12]{Ada75}.
We will assume that $\mc{D}_\zhet$ is a closed subspace of $W^{1,p}(\Omega)^n$ and $\zhet = [z_1, \ldots, z_n]^\top$. Let 
$\boldsymbol{\varphi}_k = [\varphi_{k,1}, \ldots, \varphi_{k,n}]^\top \in C^{\infty}(\Omega)^n \cap \mc{D}_\zhet$
 be arbitrary and define $A_{\boldsymbol{\varphi}_k}(\zhet)=\int_\Omega \varphi_{k,i} z_i\,\dx$, $k=1,2$. Then  $\tvardif{A_{\boldsymbol{\varphi}_k}}{\zhet}=\boldsymbol{\varphi}_k$ 
and therefore $\{A_{\boldsymbol{\varphi}_1},A_{\boldsymbol{\varphi}_2}\} = \int_{\Omega} \boldsymbol{\varphi}_1 \mathfrak{J}(\zhet)\boldsymbol{\varphi}_2 \,\dx$. Since $C^\infty(\Omega)\cap W^{1,p}(\Omega)$ is dense in $W^{1,p}(\Omega)$,~\cite{MeyS64}, so is $C^{\infty}(\Omega)^n \cap \mc{D}_\zhet$ in $\mc{D}_\zhet$. This allows us to define (under some regularity assumptions) the linear, bounded operator $\mc{J}(\zhet)\colon \mc{D}_\zhet \to \mc{D}_\zhet^\ast$ by 
$\tweak{\boldsymbol{\varphi}}{\mc{J}(\zhet)\boldsymbol{\psi}}= (\mc{J}(\zhet)\boldsymbol{\psi})[\boldsymbol{\varphi}]= \int_{\Omega} \boldsymbol{\varphi} \mathfrak{J}(\zhet)\boldsymbol{\psi} \,\dx$ with $\boldsymbol{\varphi},\boldsymbol{\psi} \in \mc{D}_\zhet$. Analogously we define the linear, bounded operator $\mc{R}(\zhet)\colon \mc{D}_\zhet \to \mc{D}_\zhet^\ast$ with $\mathfrak{R}$.

Let $\mc{V}_1$ and $\mc{V}_2$ be two real, reflexive Banach spaces. The adjoint $\mc{A}^\ast$ of a linear continuous operator~$\mc{A}\colon \mc{V}_1 \to \mc{V}_2^\ast$ is defined as the unique linear continuous operator mapping from $\mc{V}_2$ into $\mc{V}_1^\ast$ which satisfies $\tweak{v_1}{\mc{A}^\ast v_2}_{\mc{V}_1, \mc{V}_1^\ast} = \tweak{v_2}{\mc{A} v_1}_{\mc{V}_2, \mc{V}_2^\ast}$ for all $v_i\in\mc{V}_i$, $i=1,2$.
Note that, the symmetry of the dissipation bracket requires self-adjointness of the operator $\mc{R}$, i.e. $\mc{R}^\ast=\mc{R}$ or equivalently $\tweak{\boldsymbol{\varphi}}{\mc{R}\boldsymbol{\psi}} = \tweak{\boldsymbol{\psi}}{\mc{R}\boldsymbol{\varphi}}$. The claim that $\mc{R}$ is semi-elliptic, i.e. $\tweak{\boldsymbol{\varphi}}{\mc{R}\boldsymbol{\varphi}} \geq 0$, follows from the non-negativeness of the dissipation bracket.
The anti-symmetry of the Poisson bracket instead translates to a skew-adjointness of $\mc{J}$, i.e. $\mc{J}^\ast= - \mc{J}$ or $-\tweak{\boldsymbol{\varphi}}{\mc{J}\boldsymbol{\psi}} = \tweak{\boldsymbol{\psi}}{\mc{J}\boldsymbol{\varphi}}$. An equivalence property for a skew-adjoint operator is given in the following lemma.

\begin{lemma}\label{lem:equi_skew}
	Let $\mathcal{V}$ be a real, reflexive Banach space and $\mathcal{A}\colon \mathcal{V} \to \mathcal{V}^\ast$ be linear and continuous. Then, $\mathcal{A}$ is skew-adjoint
	if and only if $\langle  v, \mathcal{A}v \rangle_{\mathcal{V},\mathcal{V}^\ast} = 0$ for all $v\in \mathcal{V}$. 
\end{lemma}
\begin{proof}
	Let $\mathcal{A}$ be skew-adjoint and $v \in \mathcal{V}$ arbitrary. Then we choose $v_1=v_2=v$ and get $\langle  v, \mathcal{A}v \rangle_{\mathcal{V},\mathcal{V}^\ast} = -\langle  v, \mathcal{A}v \rangle_{\mathcal{V},\mathcal{V}^\ast}$. The other direction follows by
	\begin{align*}
		0 &= \langle (v_1+v_2),\mathcal{A} (v_1+v_2) \rangle_{\mathcal{V},\mathcal{V}^\ast} - \langle v_1,\mathcal{A}v_1 \rangle_{\mathcal{V},\mathcal{V}^\ast}  - \langle v_2,\mathcal{A}v_2 \rangle_{\mathcal{V},\mathcal{V}^\ast}\\
		 &= \langle v_1,\mathcal{A}v_2 \rangle_{\mathcal{V},\mathcal{V}^\ast} + \langle  v_2,\mathcal{A} v_1 \rangle_{\mathcal{V},\mathcal{V}^\ast}. \qedhere
	\end{align*}
\end{proof}%

To derive operator expressions for the GENERIC formulations~\eqref{eq:GENERIC_isolated} and \eqref{eq:GENERIC_open}, we use again a density argument and get
\begin{equation}
\label{eq:Lemma2_2}
	\tweak{\boldsymbol{\varphi}}{\dot{\zhet}}=\frac{\mathrm{d}A_{\boldsymbol{\varphi}}}{\mathrm{d}t} \overset{\eqref{eq:GENERIC_open}}{=}
	   \weak{\boldsymbol{\varphi}}{\mc{J}\vardif{H}{\zhet} + \mc{R}\vardif{S}{\zhet} + \mc{B} \mathbf{u}}
\end{equation}
for every $\boldsymbol{\varphi}\in \mc{D}_\zhet$, where  $\mc{B}\mathbf{u} \in \mc{D}_\zhet^\ast$ describe the boundary contributions and~$\mathbf{u}$ is a so-called port variable. Since the dual pairing is non-degenerate \cite[p.~774]{Zei86}, \eqref{eq:Lemma2_2} can be written as
\begin{equation}\label{eq:operator_equation_open_dynamics}
\dot{\zhet} = \mc{J}(\zhet) \vardif{H}{\zhet}+\mc{R}(\zhet) \vardif{S}{\zhet}+\mc{B}(\zhet)\mathbf{u} \qquad \text{in } \mc{D}_\zhet^\ast.
\end{equation}
In addition to $\mathbf{u}$ we introduce the two port variables $\mathbf{y}_H$ and $\mathbf{y}_S$. The port variable $\mathbf{y}_H$ together with $\mathbf{u}$ will then describe the change of the total energy $H$ by the interaction with the environment. Under the assumption that $\mathbf{u}$ is a function from the same bounded time interval $\mathbb{I}$ as $\zhet$ into a reflexive space $\mc{D}_\mathbf{u}$, the port variable $\mathbf{y}_H$ is then given as $\mc{B}^\ast \tvardif{H}{\zhet}$. The meaning and definition of $\mathbf{y}_S$ is analogous for the entropy $S$. Together with the evolution equation~\eqref{eq:operator_equation_open_dynamics} we get the system
\begin{subequations}\label{eq:operator_equation_open}
\begin{alignat}{3}
\dot{\mathbf{z}} &= \,\mc{J}(\mathbf{z})\vardif{H}{\mathbf{z}}\,+\, &&\,\mc{R}(\mathbf{z})\vardif{S}{\mathbf{z}} + \mc{B}(\zhet)\mathbf{u} \quad &&\text{ in } \mc{D}_\zhet^\ast,
\label{eq:operator_equation_open_I}\\[5pt]
\mathbf{y}_{H} &= \mc{B}^\ast(\mathbf{z}) \vardif{H}{\mathbf{z}} && &&\text{ in } \mc{D}_{\mathbf{u}}^\ast,\label{eq:operator_equation_open_II}\\[5pt]
\mathbf{y}_{S} &= &&\mc{B}^\ast(\mathbf{z}) \vardif{S}{\mathbf{z}} &&\text{ in } \mc{D}_{\mathbf{u}}^\ast.\label{eq:operator_equation_open_III}
\end{alignat}
\end{subequations}
This is the general form of the operator equations which will describe the dynamics for open systems. Note that for an isolated system we have to restrict ourselves to a subspace of $\mc{D}_\zhet$ which covers the properties of such systems. These restrictions lead to vanishing boundary contributions, i.e. $\mc{B}\mathbf{u}=0$ in $\mc{D}_\zhet^\ast$, and finally to an operator equation
\begin{equation}\label{eq:operator_equation_isolated}
\dot{\zhet} = \mc{J}(\zhet) \vardif{H}{\zhet}+\mc{R}(\zhet) \vardif{S}{\zhet} \qquad \text{in } \mc{D}_\zhet^\ast.
\end{equation}

\section{Continuum Thermodynamics}\label{sec:MixtureTheory}
In this section we motivate the field equations and complementary closure relations that represent the mathematical model of the fluid mixture in the framework of classical continuum physics. Furthermore, we provide relations which can be used to verify the results obtained from the GENERIC based structured weak formulation of the field equations that we present in Section \ref{sec:mixture}. 
We start this section by briefly recalling the notion of \emph{integral curves} of vector fields defined on (possibly infinite-dimensional) manifolds~\cite[Ch.~3.1]{AbrMarRat88}.

Let $\mc{N}$ be a differentiable manifold and let $\mathscr{U}\subset\mathcal{N}$ be a local manifold. Assume that the \emph{tangent space} $T_p(\mc{N})$ at each $p\in \mathscr{U}$ is defined, and denote the set $T(\mathscr{U})= \left.T(\mc{N})\right|\!\mathscr{U}:=\bigcupdot_{p\in \mathscr{U}} T_p(\mc{N})$ (disjoint union) as the \emph{tangent bundle} restricted to $\mathscr{U}$. Let $\left.\uppi\right|_{\mathscr U}\!\colon T(\mathscr{U})\to \mathscr{U}$ be the corresponding \emph{tangent bundle projection map}, i.e., a surjective map with the property such that $ \left.\uppi\right|_{\mathscr U}\!(X(p))=p$ for all \emph{tangent vectors} $X(p)\in T_p(\mc{N})$ and all $p\in\mathscr{U}$. A \emph{section} of the tangent bundle $T({\mathscr U})$ is a  mapping $\psi\colon \mathscr U \to T(\mathscr U)$ such that $\left.\uppi\right|_{\mathscr U}\circ\psi=\text{id}_\mathscr{U}$. 
We denote by $\mathrm{\Gamma}(T\mathscr{U}):=\left\{\psi\colon\mathscr U\to T(\mathscr U) \,\vert\, \left.\uppi\right|_{\mathscr U}\circ\psi=\text{id}_\mathscr{U}\right\}$ the set of all sections of the tangent bundle $T(\mathscr U)$. The elements of $\mathrm{\Gamma}(T\mathscr{U})$ are called \emph{vector field} on $\mathscr{U}$. For details we refer to \cite{AbrMarRat88}. 

Let $\mathscr I\subset\R$ be an open interval considered as one-dimensional differentiable manifold. Then $(r,s)\in \mathscr I\times\R=T(\mathscr I)$ is a tangent vector to $\mathscr I$. Let $c\colon \mathscr{I}\to \mathscr{U}$ be a smooth curve. Its tangent $Tc\colon T(\mathscr I)\to T(\mathscr U)$ is given by the vector $Tc(r,s)=(c(r),(dc/dt)(r)s)$, which evaluated at $(r,1)$ becomes $Tc(r,1)=(c(r),(dc/dt)(r))\in T_{c(r)}\!(\mathscr{U})$. The tangent vector to the curve on a manifold, understood as generalized directional derivative, is therefore given by $Tc(\cdot,1)\colon \mathscr{I}\to T(\mathscr U)$.  Under abuse of notation one often writes $(dc/dt)$ instead of $Tc(\cdot,1)$.
A curve $c\colon \mathscr I\to \mathscr U$ is called \emph{integral curve} of the vector field $X\in\mathrm{\Gamma}(T\mathscr U)$ at $p\in\mathscr U$ with $c(t_0)=p$ and $t_0\in\mathscr I$, if it is a solution of the equation \cite[Ch.~1.6]{MarHug94} 
\begin{gather}\label{eq:integralcurves}
	  Tc(t,1)=X(c(t)) \quad\text{for all $t\in\mathscr I$.}
\end{gather}

Recall that a \emph{reference frame} from a mathematical point of view is characterized as \emph{time-like future-pointing} vector field defined on the spacetime manifold, see e.g. \cite[Axiom~2.1]{Rod95}. Each integral curve of any given reference frame (vector field) is called an \emph{observer} in the terminology used in space-time theories \cite[Def.~3.1]{Rod95}. For this consider Equation \eqref{eq:integralcurves} and note that, in case the underlying manifold $\mc{N}$ is finite-dimensional, the coordinate representation of Equation \eqref{eq:integralcurves} will correspond to a system of ordinary differential equations. For different initial conditions this ODE system will in general have different solutions, under the assumption that existence and uniqueness conditions for solutions are satisfied. These solutions will typically not be defined on the whole of the space-time domain but only on sufficiently small spatial regions and sufficiently small time intervals.
Therefore a reference frame in \emph{Newtonian spacetime} is also introduced as the infinite collection of observer considered as \emph{sparsed} over the spacetime manifold. Briefly speaking, a reference frame should be thought of as part of some mathematical apparatus that allows the (local) \emph{trivialization} of the spacetime manifold. This is required since the spacetime manifold, due to the structure of the underlying mathematical theory of Newtonian spacetime, consists of abstract point called \emph{events}, that cannot be characterized by the instant and the location of their occurrence, cf. \cite[p.~884]{Rod95} and \cite[Ch.~9.2.1]{ZeiQFT3}.  
For further details and an introduction of the mathematical structure of Newtonian spacetime see e.g. \cite[Ch.~12]{Mis17} and \cite{Rod95}. 

We restrict our considerations to Newtonian spacetime and assume that a (globally defined, Euclidean rigid) \emph{reference frame} is always chosen.
We call the mathematical model of the physical space of our experience \emph{spatial manifold} and assume that the spatial manifold at each instance of time has the structure of a finite-dimensional \emph{Hilbert manifold}, i.e., a manifold modeled over a Hilbert space \cite[Sec.~3.1]{AbrMarRat88}.
We express this manifold through the pair $(\E^3,E)$, where $\E^3$ is the three-dimensional \emph{Euclidean manifold} and $E$ is the three-dimensional linear space called \emph{Euclidean space}, a Hilbert space that has the additional structure of a Lie algebra \cite[Ch.~1]{ZeiQFT3},  see \cite[Sec.~I.2]{Noll1973} and \cite[Sec.~6]{Tra70}. 
Recall that the $d$-dimensional Euclidean manifold $\E^d$ is a $d$-dimensional \emph{Riemannian manifold} whose elements are called \emph{points} and whose tangent spaces are isomorphic to the $d$-dimensional Euclidean space $E$. 
Furthermore, we assume that the spatial manifold at each instance of time is endowed with a torsion- and curvature free metric compatible affine connection, for details we refer to \cite[Ch.~74.18]{ZeidIV} and \cite[Ch.~8.9~f.]{ZeiQFT3}.
The Euclidean manifold is a flat Riemannian manifold that allows global parallel transport, see \cite[Ch.~11.5]{Mis17} and \cite[p.~71]{ZeiQFT3}. In the literature related to continuum mechanics, the Euclidean manifold is sometimes called \emph{Euclidean point space} and the Euclidean space $E$ is called \emph{translation space}, see e.g. \cite[App.~A.2.1]{Liu02} and \cite[App.~II.B]{TruesRM}.
The spatial manifold $(\E^d,E)$ can be identified with $\R^d$ \cite[Sec.~2.2]{Sil97}, in this case the $d$-dimensional real Euclidean space $\R^d$ is considered an affine space over itself whose elements represent both, points and (coordinate representations of) vectors, and whose additive identity is considered as the origin of coordinates \cite[Def.~1.14]{KosM97}.

Let $\mathcal{U}\subset\mathbb{R}^{N}$ be an open set and $(F,\Norm{\cdot}_F)$ be a Banach space over $\R$. We write $\text{clos}(\mc{U})$ or $\bar{\mc{U}}$ for the closure of a set $\mc{U}$. Recall that the \emph{support} of an $F$-valued function on $\mc{U}$ is given by the set $\text{supp}(f):=\text{clos}(\left\{u\in\mc{U}\,\vert\, f(u)\neq 0\right\})\subset\text{clos}(\mc{U})$.
We denote by $C_c^\infty(\mc{U},F):=\left\{f \in C^\infty(\mc{U},F)\,\vert\, \text{supp}(f)\subset\subset\mc{U}\right\}$
the set of all smooth $F$-valued functions on $\mc{U}$, whose support is compactly contained in $\mc{U}$.
We denote $\mathscr{D}(\mc{U},F):=C_c^\infty(\mc{U},F)$ and call its elements $F$-valued \emph{test functions} on $\mc{U}$, and for the case $F=\R$ write $\mathscr{D}(\mc{U}):=\mathscr{D}(\mc{U};\R)$.
Following \cite[Def.~2.2]{AltD14} we call any linear map $T\colon \mathscr{D}(\mc{U},F)\to \R$ a \emph{distribution} on $\mc{U}$ if for every subset $U\subset\subset\mc{U}$ there exists a constant $A_U\geq 0$ and an order $b_U\in \N$ such that $\big|T(f)\big| \leq A_U\Norm{f}_{C^{b_U}\!(\bar{U},F)}$ for all $f\in C_c^\infty(\mc{U},F)$ with $\text{supp}(f)\subset U$.
We denote by $\mathscr{D}^\prime(\mc{U};F)$ the set of all these distributions on $\mc{U}$, see also \cite[Ch.~5.17]{Alt16}. An alternative but equivalent definition of distributions is to define a topology on $\mathscr{D}(\mc{U},F)$ by means of a Fr{\'e}chet metric, for details see \cite[Ch.~12]{AltD14}. 

Let $\mathrm{L}^{N}$ denote the $N$-dimensional Lebesgue measure on $\R^N$, and let $N=d+1>1$. We consider a space-time domain $\mc{U}\subset\R\times\R^d$ ($d=3$ for the physical application) with coordinates $(t,x)\in\mc{U}$. Let $\UOmega_t=\left\{x\in \R^d\,\vert\,(t,x)\in\mc{U}\right\}$ be an open set in the $d$-dimensional real Euclidean space for each $t\in\R$ with $\{t\}\times\UOmega_t\subset\mc{U}$. For integrable functions $f\in L_{\text{loc}}^{1}(\mathcal{U})$ and test functions $\phi\in\mathscr{D}(\mc{U})$ we introduce the integral identity
\begin{gather*}
	\int_\mathcal{U} f\,\phi\, \mathrm{dL}^{d+1} = \int_{\R}\int_{\mathrm{\Omega}_t}f(t,x)\phi(t,x)\, \dx\, \dt,
\end{gather*}
which we use to simplify notation. For further details see \cite{AltCM17}.
Recall that locally integrable functions can be identified with (regular) distributions $T_f\in \mathscr{D}^\prime(\mathcal{U})$ defined through the assignment $T_f:=(\phi\mapsto\int_{\mathcal{U}} f \phi\,\mathrm{dL}^{d+1})\in \mathscr{D}^\prime(\mathcal{U})$ \cite{AltD14}. For each fixed $t\in\R$ with $\{t\}\times\UOmega_t\subset\mc{U}$ let $[\g\indi{_i_j}]\in L_{\text{loc}}^1(\UOmega_t;\R^{d\times d}_\text{sym})$ be the coordinate representation of the Euclidean metric tensor, cf. \cite[Ch.~7.10]{AmaEschII08} and \cite[Ch.~11.5]{AmaEschIII09}, which in a global Euclidean rigid reference frame will be independent of the time parameter \cite[Prop.~3.60]{Rod95}. Note that in Euclidean geometry, for any non-empty open and arcwise connected subset of $\R^d$ the matrix $[\g\indi{_i_j}(x)]\in\R^{d\times d}_\text{sym}$ exists, and in case of regular points can be inverted with $[\g\indi{^i^j}(x)]:=[\g\indi{_i_j}(x)]^{-1}$, such that $[\g\indi{^i^k}(x)][\g\indi{_k_j}(x)]=[\delta\indi{^i_j}]$. For singular points, where the matrix $[\g\indi{_i_j}(x)]$ cannot be inverted, the consideration is complemented by a limiting process, for details see e.g. \cite[Ch.~9.2~f.]{ZeiQFT3}.

For vector valued locally integrable functions $\mathbf{f}\in L_{\text{loc}}^{1}(\mathcal{U};\R^{d})$ and vector valued test functions $\lupvarphi\in\mathscr{D}(\mc{U};\R^{d})$ we require their contraction $\mathbf{f}\cdot\lupvarphi=\aweak{\mathbf{f}}{\lupvarphi}$ \cite[Ch.~2.2]{Tal02} to hold
\begin{gather*}
	\int_\mathcal{U}\mathbf{f}\cdot\lupvarphi\,\mathrm{dL}^{d+1}=\int_{\R}\int_{\mathrm{\Omega}_t}\sum_{k,l=1}^{d}\g\indi{_k_l}(x) f\indi{^k}(t,x)\varphi\indi{^l}(t,x)\,\dx\,\dt=\int_\mathcal{U}\aweak{\mathbf{f}}{\lupvarphi}\mathrm{dL}^{d+1},
\end{gather*}
 almost everywhere in a space-time domain $\mc{U}\subset\R\times\R^d$. The identification of a vector valued integrable function with a (regular) distribution is given by
\begin{gather*}
	(\lupvarphi\mapsto\int_{\mathcal{U}} \mathbf{f}\cdot\lupvarphi\,\mathrm{dL}^{d+1})\in \mathscr{D}^\prime(\mathcal{U};\R^d).
\end{gather*}
For matrix valued locally integrable functions $\mathbf{F}\in L_{\text{loc}}^{1}(\mathcal{U};\R^{d\times d})$ and matrix valued test functions $\luppsi\in\mathscr{D}(\mc{U};\R^{d\times d})$ their double contraction $\mathbf{F}\dcont\luppsi=\aweak{\mathbf{F}}{\luppsi}$, cf. \cite[Def.~4.12]{MarHug94}
\begin{align*}
	\int_{\mathcal{U}}\mathbf{F}\dcont\luppsi\,\mathrm{dL}^{d+1}&=\int_{\R}\int_{\mathrm{\Omega}_t}\sum_{i,k,l,m=1}^{d} \g\indi{_m_k}(x)\g\indi{_l_i}(x)F\indi{^i^k}(t,x)\psi\indi{^l^m}(t,x)\,\dx\,\dt\\
	&=\int_\mathcal{U}\aweak{\mathbf{F}}{\luppsi}\mathrm{dL}^{d+1}\,,
\end{align*}%
is required to hold almost everywhere in $\mc{U}$. The identification with a (regular) distribution is through the assignment
 \begin{gather*}
 	(\luppsi\mapsto\int_{\mathcal{U}} \mathbf{F}\dcont\luppsi\,\mathrm{dL}^{d+1})\in \mathscr{D}^\prime(\mathcal{U};\R^{d\times d}).
 \end{gather*}

Note that $\mathbf{A}\dcont\mathbf{B}=\sum_{i,j=1}^{d} A\indi{^i^j}B\indi{_i_j}=\operatorname{trace}(\sum_{j=1}^{d}A\indi{^k^j}(\tensor*[]{\mathbf{B}}{^\top})\indi{_j_l})=\trace(\mathbf{A}\cdot\mathbf{B}^\top)$ 
for arbitrary $\mathbf{A},\mathbf{B}\in L_{\text{loc}}^{1}(\mathcal{U};\R^{d\times d})$. The divergence of a velocity vector field is related to the trace of the velocity gradient $(\nabla\mathbf{ v})\in L_{\text{loc}}^{1}(\mathcal{U};\R^{d\times d})$ via the family of identity relations
\begin{equation}\label{eq:div_trace_relation}
\begin{split}
\diver(\mathbf{v}(t,\cdot))&=\sum_{i,j=1}^d(\nabla\mathbf{ v}(t,\cdot))\indi{^i^j}\g\indi{_j_i}(\,\cdot\,)\\
&= \operatorname{trace}(\sum_{j=1}^{d}(\nabla\mathbf{ v}(t,\cdot))\indi{^i^j}\g\indi{_j_k}(\,\cdot\,))=\trace(\nabla\mathbf{v}(t,\cdot)).
\end{split}
\end{equation}
For the remainder of this section we will make no distinction in notation between the function $f$ and the distribution $T_f:=(\phi\mapsto\int f\phi dx)$ and denote both by the same symbol.

\subsection{Differential Balance Laws}\label{sec:differential_balance_laws}
In continuum mechanics and thermodynamics one postulates integral balance laws for a number of \emph{extensive} quantities, which then are monitored through the \emph{fields} of their \emph{densities} \cite{Liu02,Mue85}. These fields typically correspond to macroscopic state variables and in this work are formulated in spatial (also called~\emph{Eulerian}) representation \cite{Daf93}. Their time evolution and spatial distribution is described by \emph{field equations}, which in this work are given as partial differential equations representing differential balance laws that are complemented by closure relations, cf. \cite{Daf16}. 

Let $\mc{U}\subset\R\times\R^d$ be a space-time domain and  $\mathbf{v}\in C^\infty(\mc{U};\R^d)$ be the \emph{spatial velocity} vector field. In order to formulate a differential balance law for some real valued scalar field $\lpsi\in L_{\text{loc}}^{1}(\mc{U})$, suppose that an associated scalar $\tensor*[^\lpsi]{\!\sigma}{}\in L_{\text{loc}}^{1}(\mc{U})$ and a vector field $\tensor*[^\lpsi]{\!\Upphi}{}\in L_{\text{loc}}^{1}(\mc{U};\R^d)$ are given. Then the evolution of $\lpsi$ is described by a partial differential equation of the form
   \begin{align}\label{eq:scalar_differential_balance}
   	\partial_t \lpsi +  \diver(\lpsi\mathbf{v} + \tensor*[^\lpsi]{\!\Upphi}{})= \tensor*[^\lpsi]{\!\!\sigma}{}
   \end{align}
satisfied in $\mathscr{D}^\prime (\mathcal{U})$,
   (in the sense of distributions), i.e., we require
   	   \begin{gather*}
   	   \int_{\mathcal{U}}
   	   \big(\lpsi\,\partial_t \lvarphi
   	   + (\lpsi\mathbf{v} + \tensor*[^\lpsi]{\!\Upphi}{})\cdot\nabla\lvarphi
   	   + \lvarphi\tensor*[^\lpsi]{\!\sigma}{}
   	   \big)\,\dL^{d+1} = 0 
   	   \end{gather*}
to hold for all test functions $\lvarphi\in\mathscr{D}(\mc{U})$.  
   
All scalar differential balance laws considered in this work have the form of \eqref{eq:scalar_differential_balance} where $\lpsi$ is the \emph{field of some density}, $(\lpsi\mathbf{ v})$ is the \emph{convective flux}, $\tensor*[^\lpsi]{\!\Upphi}{}$ is the \emph{non-convective flux} and $\tensor*[^\lpsi]{\!\!\sigma}{}$ is the \emph{source} (or \emph{total production}).
The source term $\tensor*[^\lpsi]{\!\sigma}{}$ is split in two type of contributions, internal sources denoted $\tensor*[^\lpsi]{\!\Sigma}{}$ and called \emph{production density}, and external sources $\tensor*[^\lpsi]{\mc{P}}{}$, which we call \emph{influx density}. External sources (e.g. body forces due to gravity or electromagnetic fields) are typically considered as known. In the absence of internal sources the balanced extensive quantity (e.g. total energy, mass or linear momentum) are conserved quantities. Therefore, we define $\tensor*[^\lpsi]{\!\sigma}{} = \tensor*[^\lpsi]{\!\Sigma}{} + \tensor*[^\lpsi]{\mc{P}}{}$ to distinguish between internal and external sources, i.e., between production densities and influx densities. 

Note that the spatial velocity field $\mathbf{v}\in C^\infty(\mc{U};\R^d)$ appearing in \eqref{eq:scalar_differential_balance} is an independent quantity that needs to be prescribed. Therefore, for the description of a compressible fluid flow we need at least the balance of mass and the balance of linear momentum.
Assume that the field of the \emph{mass density} $\rho\in L_{\text{loc}}^{1}(\mathcal{U})$ is a non-negative scalar field governed by a differential balance law of the form \eqref{eq:scalar_differential_balance} and the field of the \emph{linear momentum density} $\mathbf{M}\in L_{\text{loc}}^{1}(\mathcal{U},\R^d)$ a vector field with $\rho\mathbf{v}=\mathbf{M}$. Let $\tensor*[]{\mathbf{T}}{}\in  L_{\text{loc}}^{1}(\mathcal{U};\R^{d\times d}_\text{sym})$ be the corresponding non-convective flux and $\tensor*[]{\mathbf{b}}{}\in L_{\text{loc}}^{1}(\mc{U};\R^d)$ represent the external source.
Then the evolution of $\mathbf{M}$ is described by
\begin{alignat}{4}\label{eq:vector_balance_law}
&\partial_t \mathbf{M}  +  \diver(\mathbf{M}\otimes\mathbf{ v}-\mathbf{T})  = \mathbf{b}
\end{alignat}
satisfied in $\mathscr{D}^\prime(\mc{U},\R^d)$ (in the sense of distributions), i.e., we require
  \begin{gather*}
	\int_{\mc{U}}\left(
	\mathbf{M}\cdot\partial_t \lupvarphi
	+(\mathbf{M}\otimes\mathbf{ v}-\mathbf{T})\dcont\nabla\lupvarphi+\lupvarphi\cdot\mathbf{b}\right)\dL^{d+1} = 0
\end{gather*}
to hold for every test function $\upvarphi\in\mathscr{D}(\mc{U},\R^d)$, cf. \cite{FeiN17}. 

\subsection{Diffusion Model}\label{subsec:diffusion_model}
We are interested in a homogeneous mixture of heat-conducting compressible Newtonian fluids consisting of a finite number $\nu \geq 2$ of reactive constituents,~$\nu\in \N$. It is assumed that the mixture moves free of external body forces. 
For the considered fluid mixture the collection of field equations contains the differential balance equations for the fields $\rho_\alpha$, $\alpha=1,\dots,\nu$, of the~\emph{constituent mass densities}, the differential balance equations for the field $\mathbf{M}$ of the~\emph{mixture linear momentum density}, the field $u$ of the~\emph{mixture internal energy density}, and the field $s$ of the~\emph{mixture entropy density}. From now on and for the remainder of this work we will call these fields by the name of their densities.

Although chemical reactions may change the total amount of mass of certain constituents in the mixture, the related processes do not result in real~\emph{production} or~\emph{destruction} of matter, but rather have to be seen as exchange and interaction processes~\cite{Trues}. For this, we assume that the total mass of the whole mixture is conserved. Based on this assumption we define the mixture mass density $\rho$ as the sum of the constituent mass densities $\rho=\sum_{\alpha=1}^{\nu}\rho_\alpha$ and require that $0<\rho$ holds almost everywhere in a space-time domain associated with the fluid flow.
With this we define the \emph{barycentric velocity} field by $\mathbf{v}=\mathbf{M}/(\sum_{\alpha=1}^\nu \rho_\alpha)$, cf. \cite[p.~49]{Hut09}.

We call the following collection of differential balance laws the \emph{diffusion model}:
	\begin{subequations}\label{diffusionModel}
	\begin{align}
		\label{bal:partmass}
		\hspace*{1cm}&\partial_t\rho\indi{_\alpha} +\diver\!\left(\rho\indi{_\alpha}\mathbf{v} + \mathbf{J}\indi{_\alpha}\right) = \tau\indi{_\alpha}, \hspace*{.1cm}(\alpha=1,\dots,\nu)\\
		\label{bal:mixmom}
		&\partial_t\mathbf{M} + \diver\!\left(\mathbf{M}\otimes\mathbf{v} - \mathbf{T}\right)= 0, \\
		\label{bal:mixint}
		&\partial_t u + \diver\!\left(u\mathbf{v} + \mathbf{q} \right) = \mathbf{T}\dcont\nabla\mathbf{v},\\
		\label{bal:entrop}
		&\partial_t s + \diver\!\left(s\mathbf{v}+\tensor*[^s]{\Upphi}{}\right) =
		\tensor*[^s]{\hspace{-0.1em}\Sigma}{},
		\end{align}
	\end{subequations}
	where $\mathbf{T}$ is the~\emph{mixture stress tensor}, $\mathbf{q}$ is the~\emph{mixture heat flux vector}, $\mathbf{J}_\alpha$ is the~\emph{diffusion flux} and $\tau_\alpha$ the density of \emph{mass production} with respect to constituent $\alpha$ in the mixture. The~\emph{non-convective entropy flux} $\tensor*[^s]{\Upphi}{}$ and the \emph{entropy production density} $\tensor*[^s]{\hspace{-0.1em}\Sigma}{}$ are discussed and specified in Section \ref{sec:TIP}.
\begin{remark}\normalfont
	Note that the diffusion model~\eqref{diffusionModel} does not contain any~\emph{influx} terms, since the fluid mixture is assumed to move free of external body forces. Also, the influx of internal energy due to thermal radiation is neglected.
\end{remark}

The density of mass production $\tau_\alpha$ appearing at the right-hand side of the differential mass balance~\eqref{bal:partmass} can be expressed with respect to $n$ independent chemical reactions~\cite{IngoM} as there are only as many independent mass productions as there are independent chemical reactions~\cite[p.~267]{Trues}, $n\in \N$. 
This is taken into account via 
\begin{align}
\label{eq:massprod}
\tau\indi{_\alpha}=\sum\limits_{k=1}^{n}\tensor*{\gamma}{*^k_\alpha}m\indi{_\alpha}\Lambda^k,
\end{align}
where each of the $n$ independent chemical reactions has distinct stoichiometric coefficients $\gamma_\alpha^k$ and reaction rate densities $\Lambda^k,~k=1,\dots,n$.

\subsection{Thermodynamics of Irreversible Processes (TIP)}\label{sec:TIP}
The diffusion model~\eqref{diffusionModel} as a collection of partial differential equations in its present form is not a closed system. Closure relations are required, which in continuum physics are typically given in form of \emph{constitutive relations} and relate the \emph{constitutive quantities} to the fields \cite{Liu02,IngoM,TruesdNoll}. Constitutive quantities can, roughly speaking, be identified as those quantities that appear in the field equation but do not belong to the set of fields \cite{BotDre15}. For the diffusion model, constitutive equations for the mixture stress tensor $\mathbf{T}$, the mixture heat flux vector $\mathbf{q}$, the diffusion fluxes $\mathbf{J}_\alpha$, and the reaction rate densities $\Lambda^k$ are required. Complemented by proper constitutive relations, the system of partial differential equations~\eqref{diffusionModel} becomes formally well-posed, that is, there are as many equations as unknowns \cite[Ch.~3]{MarHug94}.

In this section we specify the non-convective entropy flux
 $	\tensor*[^s]{\Upphi}{}$ and the entropy production density $\tensor*[^s]{\!\Sigma}{}$ that appear on the right-hand side of the field equation~\eqref{bal:entrop}. Also, the required closure relations given in form of linear constitutive equations are motivated and their relation to the second law of thermodynamics is explained. This is done by means of the phenomenological theory called~\emph{Thermodynamics of Irreversible Processes} (TIP), also known as \emph{Classical Irreversible Thermodynamics} (CIT). The fundamental assumption made in TIP is the \emph{local equilibrium hypothesis}, see e.g. \cite[Ch.~2]{Leb08}. Consider a \emph{fluid particle} as being a region of physical space, occupied by the fluid or parts of it at some fixed time instant, that from a microscopic perspective contains a large number of molecules but from a macroscopic perspective is point-like.
The local equilibrium hypothesis may be interpreted as some local statistical homogenization processes such that
at any time instant the fluid can locally, i.e. restricted to fluid particles, be considered as in thermodynamic equilibrium. The term \emph{local} emphasizes that the equilibrium states of two distinct fluid particles will in general be related to different local equilibrium state variables~\cite{MeiR59,Oet05}, resulting in interaction and exchange processes between the fluid particles. From the perspective of kinetic theory, the local equilibrium hypotheses may be considered as a situation where the probability distribution function takes the form of a local expression, where each local probability distribution function approximately is given by a local Maxwell-Boltzmann distribution, for details see e.g.~\cite[Ch.~9~f.]{Sch06}.
In each region of local equilibrium, the local equations of state are assumed to have the same form as in global equilibrium. Consequently, one assumes that the Gibbs equation stays locally valid \cite[Ch.~2.2]{Leb08}. The locally formulated Gibbs equation yields a way to derive a differential entropy balance without the requirement of postulating an entropy balance as done in other continuum thermodynamic theories, see e.g. the~\emph{Clausius-Duhem Inequality}~\cite[p.~76]{Sil97}. 

For the considered fluid mixture,
the locally expressed Gibbs equation written with the differentials of the thermodynamic state variables takes the form~\cite{MeiR59}
\begin{align}
	\label{eq:Gibbs}ds&=\frac{1}{T}du-\sum\limits_{\alpha=1}^{\nu}\frac{\mu_\alpha}{T}d\rho_\alpha\,,
\end{align}
where the absolute temperature $T$ and the chemical potentials $\mu_\alpha$ are accessible through the following identities
\begin{alignat}{6}
		\label{rel:entpot}
		&\frac{1}{T}= \frac{\partial s}{\partial u},&\qquad&-\frac{\mu_\alpha}{T}= \frac{\partial s}{\partial \rho_\alpha},
\end{alignat}
given as functions of the constituent mass densities and the mixture internal energy density 
\begin{gather*}
	 T=T(\rho_1,\dots,\rho_\nu,u),\qquad \mu_\alpha=\mu_\alpha(\rho_1,\dots,\rho_\nu,u),
\end{gather*} 
respectively. The differentials appearing in the Gibbs equation~\eqref{eq:Gibbs} imply that the macroscopic state of the fluid mixture is specified by the thermodynamic state variables $(\rho_1,\dots,\rho_\nu,u)$ plus the complementary state variable $\mathbf{M}$ \cite[Ch.~2.4]{Leb08}. Thus we define the macroscopic state $\mathbf{z}\in\mc{Z}$ of the fluid mixture to be given by the following block-vector of unknown fields
\begin{equation}\label{entpot:stat}
	\mathbf{z} = \begin{bmatrix}
	\rho\indi{_1}  & \dots &\rho\indi{_\nu} & \mathbf{M}^\top & u
	\end{bmatrix}^\top\,.
\end{equation}
Note that due to the local equilibrium hypothesis of TIP, which justifies the use of the Gibbs equation in its local form~\eqref{eq:Gibbs}, the macroscopic state of the system will be determined in case the unknowns $(\rho_1,\dots,\rho_\nu, \mathbf{M}^\top)$ plus either the thermodynamic state variable $u$ (internal energy density) or $s$ (entropy density) are known. The case where the internal energy density $u$ is one of the independent state variables associated with the macroscopic state $\zhet$, as reflected in relation~\eqref{entpot:stat}, conforms with~\emph{classical hydrodynamics} in the sense that in classical hydrodynamics the internal energy density is one of the independent state variables and the entropy density is the thermodynamic potential field, see e.g.~\cite[Sec.~II]{OetGrm97II} for the one-component system and e.g.~\cite[p.~419~f.]{MeiR59} for mixtures. 
\begin{remark}\normalfont
If the mixture entropy density is chosen to be one of the independent state variables, the macroscopic state $\zhet\in\mc{Z}$ associated with the unknown fields will be given by the following block vector
\begin{equation}\label{energypot:stat}
	\mathbf{z} = \begin{bmatrix}
	\rho\indi{_1}  & \dots &\rho\indi{_\nu} & \mathbf{M}^\top & s
\end{bmatrix}^\top.
\end{equation}
Here the internal energy density $u=u(\rho_1,\dots,\rho_\nu,s)$ will constitute the thermodynamic potential field.
In Section~\ref{sec:formulas} the formulas for the latter case are presented and in Section~\ref{sec:mixture} its operator formulation in the framework of GENERIC is discussed.
\end{remark}
We proceed to discuss the case where the state is of the form~\eqref{entpot:stat}.
A further important relation that holds under the assumption of local equilibrium is the determination of the thermodynamic equilibrium pressure~$p$ through the thermodynamic constitutive relation~\cite[Ch.~4.3~f.]{BerE94},
\begin{equation}\label{cr:thermoeqpressure}
p = -u + Ts  + \sum_{\alpha=1}^{\nu}\rho\indi{_\alpha}\mu\indi{_\alpha}.
\end{equation}	
 The chemical potentials $\mu_\alpha$ and the absolute temperature $T$ appearing in~\eqref{cr:thermoeqpressure} are specified through~\eqref{rel:entpot}.
 

Next, we motivate the differential balance law for the entropy density by specifying the form of the non-convective entropy flux $\tensor*[^s]{\Upphi}{}$ as well as the form of the entropy production density $\tensor*[^s]{\hspace{-0.1em}\Sigma}{}$ appearing in~\eqref{bal:entrop} with respect to the fluid mixture modeled in the theoretical framework of TIP. For this, we recall the notion of the \emph{spatial material time derivative} for scalar fields. Let the scalar field $f\in L_{\text{loc}}^{1}(\mathcal{U})$ and the barycentric velocity $\mathbf{v}\in C^\infty(\mc{U};\R^d)$ be given. Following \cite{IVM16} and restricted to a fixed observation point, the spatial material time derivative $\tfrac{\delta f}{\delta t}$ becomes
\begin{gather}\label{eq:spatialMaterialDerivative}
\smatder{f}=\total{f}{t}+\,\mathbf{v}\cdot\nabla f,
\end{gather}
where $\total{}{t}=\left.\pd{}{t}\right|_{\mathrm{x}=\text{const.}}$ in case of a fixed observation point.
With the notion of the spatial material time derivative Equations \eqref{bal:partmass} and \eqref{bal:mixint} are rewritten and we obtain:\\[5pt]
\begin{subequations}
	\textit{partial mass balance}
	\begin{align} 					
	&\smatder{\rho\indi{_\alpha}} = -\rho\indi{_\alpha}\diver\!\left(\mathbf{v}\right)-\diver\!\left(\mathbf{J}\indi{_\alpha}\right)+\tau\indi{_\alpha},\hspace*{0.1cm}(\alpha=1,\dots,\nu)\label{bal:dermas}
	\intertext{\itshape mixture internal energy balance}	
	&\smatder{u}=
	-u\diver\!\left(\mathbf{v}\right)
	-\diver\!\left(\mathbf{q}\right)+\mathbf{T}\dcont\nabla\mathbf{v}\label{bal:derinten}
	\end{align}
\end{subequations}
Following~\cite[p.~41]{Leb08} we assume that the Gibbs relation stays valid if expressed with material time derivatives such that
\begin{align}\label{eq:matdirGibbs}
\smatder{s}=\frac{1}{T}\smatder{u}-\sum_{\alpha=1}^{\nu}\frac{\mu\indi{_\alpha}}{T}\smatder{\rho\indi{_\alpha}}.
\end{align}
Inserting~\eqref{bal:dermas} and~\eqref{bal:derinten} into~\eqref{eq:matdirGibbs} results in
\begin{equation}
\label{eq:entrop1}
\begin{split}
\smatder{s}=&\frac{1}{T}(-u+\sum_{\alpha=1}^{\nu}\mu\indi{_\alpha}\rho\indi{_\alpha})\diver\mathbf{v}-\frac{1}{T}\diver\mathbf{q}
+\frac{1}{T}\mathbf{T}\dcont\nabla\mathbf{v}
+\sum_{\alpha=1}^{\nu}\frac{\mu\indi{_\alpha}}{T} \left(\diver\mathbf{J}\indi{_\alpha}-\tau\indi{_\alpha}\right).
\end{split}
\end{equation}
For the next step of reformulation, the following two identities are required
\begin{align}
\label{id:1}
&\diver\!\left(\frac{1}{T}\mathbf{q}\right)=\nabla\left(\frac{1}{T}\right)\cdot\mathbf{q}+\frac{1}{T}\diver\left(\mathbf{q}\right)\,,\\
\label{id:2}
&\sum_{\alpha=1}^{\nu}\diver\!\left(\frac{1}{T}\mathbf{J}\indi{_\alpha}\mu\indi{_\alpha}\right)=\frac{1}{T}\sum_{\alpha=1}^{\nu}\diver\!\left(\mathbf{J}\indi{_\alpha}\right)\mu\indi{_\alpha}+
\sum_{\alpha=1}^{\nu}\mathbf{J}\indi{_\alpha}\cdot\nabla\left(\frac{\mu\indi{_\alpha}}{T}\right)\,.
\end{align}
Now we are able to formulate the differential entropy balance law out of \eqref{eq:entrop1}. For this we use relation \eqref{eq:spatialMaterialDerivative} for the spatial material time-derivative, the identity relations \eqref{id:1} and \eqref{id:2}, and the thermodynamic constitutive relation \eqref{cr:thermoeqpressure}.
By means of these relations Equation~\eqref{eq:entrop1} is transformed into the following form
\begin{align}\label{eq:entbal2}
	&\partial_t s +\diver\!\left(s\mathbf{v}\right)+\diver\Biggr\{\frac{1}{T}\left[\mathbf{q}-\sum\limits_{\alpha=1}^{\nu}\mathbf{J}\alp\mu\alp\right]\Biggl\}\\
	&\quad = \frac{1}{T}\mathbf{T}\dcont\nabla\mathbf{v}
	+\frac{p}{T}\diver\!\left(\mathbf{v}\right)
	+\mathbf{q}\cdot\nabla\left(\frac{1}{T}\right)
	-\sum\limits_{\alpha=1}^{\nu}\mathbf{J}\alp\cdot\nabla\left(\frac{\mu\alp }{T}\right)
	-\frac{1}{T}\sum\limits_{\alpha=1}^{\nu}\tau\alp\mu\alp.\notag
\end{align}
Based on results from kinetic theory of gases~\cite{IngoM} the expression in the argument of the divergence operator on the left-hand side of Equation~\eqref{eq:entbal2} is identified as the spatial representation of the non-convective entropy flux in the framework of the phenomenological theory of TIP, $\tensor*[^s]{\Upphi}{}=\tensor*[^s]{\Upphi}{*^{\text{TIP}}}$, given by
\begin{gather}
\label{eq:entropyflux}
 \tensor*[^s]{\Upphi}{*^{\text{TIP}}}=\frac{1}{T}\left[\mathbf{q}-\sum\limits_{\alpha=1}^{\nu}\mathbf{J}\alp\mu\alp\right].
\end{gather}
Since the density of supply rate of internal energy through thermal radiation (influx of internal energy) has been neglected in \eqref{diffusionModel}, the right-hand side of Equation~\eqref{eq:entbal2} has to be the spatial representation of the entropy production density, $\tensor*[^s]{\hspace{-0.1em}\Sigma}{}=\tensor*[^s]{\hspace{-0.1em}\Sigma}{*^{\text{TIP}}}$, given by
\begin{gather}\label{eq:entropyprod}
\tensor*[^s]{\hspace{-0.1em}\Sigma}{*^{\text{TIP}}}=
	\frac{1}{T}\mathbf{T}\dcont\nabla\mathbf{v}
	+\frac{p}{T}\diver\!\left(\mathbf{v}\right)+\mathbf{q}\cdot\nabla\left(\frac{1}{T}\right)
	-\!\sum\limits_{\alpha=1}^{\nu}\mathbf{J}\alp\cdot\nabla\left(\frac{\mu\alp }{T}\right) -\frac{1}{T}\sum\limits_{\alpha=1}^{\nu}\tau\alp\mu\alp.
\end{gather}

\subsection{Closure Relations}\label{sec:consti}
In the following, the right-hand side of the entropy production density~\eqref{eq:entropyprod} is manipulated via a tensor decomposition rule resulting in an expression of the entropy production density which then is given as a sum with summands containing products of so called \emph{thermodynamic fluxes} and \emph{thermodynamic forces}.
For this we consider the velocity gradient $\nabla\mathbf{v}\in L_{\text{loc}}^{1}(\mc{U};\R^{3\times 3})$ and assume that it can be decomposed into a symmetric and a skew-symmetric part, $\nabla\mathbf{v}=\mathbf{D}+\mathbf{W}$ with
\begin{subequations}
\begin{alignat}{4}
	 \label{eq:symVelGrad}
	 \mathbf{D}&:=\frac{1}{2}(\nabla\mathbf{v}+\nabla\mathbf{v}^\top), &\qquad&(\text{symmetric part})\\
	 \label{eq:skewSymVelGrad}
	 \mathbf{W}&:=\frac{1}{2}(\nabla\mathbf{v}-\nabla\mathbf{v}^\top). &\qquad&(\text{skew-symmetric part})
\end{alignat}
\end{subequations}
Furthermore, we require the notion of the \emph{deviator} of the coordinate representation of a second order tensor field, which in our consideration is a locally integrable matrix valued function defined on the space-time domain $\mc{U}\subset\R\times\R^d$. We denote the deviator of an arbitrary $\mathbf{A}\in L_{\text{loc}}^{1}(\mc{U};\R^{d\times d})$ by $\mathbf{A}^\mathrm{d}$ and define it through the relation
\begin{gather}\label{eq:decomposition_rule_deviator}
	 \mathbf{A}^{\mathrm{d}}=\mathbf{A}-\frac{1}{\trace({\mathbf{I}})}\trace(\mathbf{A})\mathbf{I},
\end{gather}
	where $\mathbf{I}\in L_{\text{loc}}^{1}(\mc{U};\R^{d\times d}_\text{sym})$ is the \emph{identity}, cf. \cite[p.~342]{AbrMarRat88}.
	Note that the deviator is trace-free, i.e. $\trace(\mathbf{A}^\mathrm{d})=0$. 
	Using decomposition rule \eqref{eq:decomposition_rule_deviator} we express the mixture stress tensor, which in our consideration equals $\mathbf{T}\in L_{\text{loc}}^{1}(\mc{U};\R^{3\times 3}_{\text{sym}})$, and the velocity gradient $\nabla\mathbf{v}$ through their deviatoric part and obtain
\begin{align}
	\label{eq:decomposed_stresstensor}
			\mathbf{T}&= \mathbf{T}^{\mathrm{d}}+\frac{1}{3}\trace(\mathbf{T})\mathbf{I},\\
	\label{eq:decomposed_velocitygradient}		
	\nabla\mathbf{v}&= (\nabla\mathbf{v})^\mathrm{d} + \frac{1}{3}\diver(\mathbf{ v})\mathbf{I},
\end{align}
	where in \eqref{eq:decomposed_velocitygradient} we have used relation \eqref{eq:div_trace_relation} for the divergence, i.e. $\diver(\mathbf{ v})=\trace(\nabla\mathbf{ v})$. 
	
One can show that the operation of double contraction for two arbitrary matrix valued locally integrable mappings $\mathbf{A},\mathbf{B}\in L_{\text{loc}}^{1}(\mc{U};\R^{d\times d})$ satisfies the relation, cf. \cite[p.~31]{Tal02}
\begin{gather}\label{eq:tensor_contraction_identity_relation}
	\mathbf{A}\dcont\mathbf{B}= \frac{1}{2}(\mathbf{A}+\mathbf{A}^\top)\dcont\frac{1}{2}(\mathbf{B}+\mathbf{B}^\top) +  \frac{1}{2}(\mathbf{A}-\mathbf{A}^\top)\dcont\frac{1}{2}(\mathbf{B}-\mathbf{B}^\top).
\end{gather}
	
Next we formulate the deviator of the symmetric velocity gradient $\mathbf{D}$ \eqref{eq:symVelGrad} and in order to conform with the notation used in classical continuum mechanics denote it by 
\begin{gather}\label{eq:deviator_symmetric_velocity_gradient}
	\mathbf{L}^\mathrm{d} = \frac{1}{2}(\nabla\mathbf{v}+\nabla\mathbf{v}^\top)-\frac{1}{3}\diver(\mathbf{v})\mathbf{I}.
\end{gather}
Since the mixture stress tensor is a symmetric tensor field, its double contraction with the velocity gradient gives 
\begin{gather}\label{eq:contraction_stress_vgradient}
	\mathbf{T}\dcont\nabla\mathbf{v}=\mathbf{T}^\mathrm{d}\dcont 	\mathbf{L}^\mathrm{d} + \frac{1}{3}\trace(\mathbf{T})\diver(\mathbf{v}).
\end{gather}
Relation \eqref{eq:contraction_stress_vgradient} can be proven with the help of relation \eqref{eq:tensor_contraction_identity_relation} in combination with \eqref{eq:deviator_symmetric_velocity_gradient}.\par
For continuum thermodynamical systems the second law of thermodynamics has been used to deduce that the entropy production density has to be nonnegative for all thermodynamic processes, $\tensor*[^s]{\hspace{-0.1em}\Sigma}{}\geq 0$ \cite[Ch.~III]{Mazur}. This statement is assumed to hold also for thermodynamic systems that do not satisfy the local equilibrium hypothesis of TIP, as is discussed in more detail in Section \ref{sec:secondlaw}. 
Using~\eqref{eq:contraction_stress_vgradient} the entropy production density $\tensor*[^s]{\hspace{-0.1em}\Sigma}{*^{\text{TIP}}}$~\eqref{eq:entropyprod} is rewritten and results in a form which is given as a sum. Its summands contain products of~\emph{thermodynamic fluxes} and~\emph{thermodynamic forces}, cf.~\cite[p.~174]{Mueller2012},
\begin{align}
\label{bal:entropyproduction}
\tensor*[^s]{\hspace{-0.1em}\Sigma}{*^{\text{TIP}}}&=\begin{aligned}[t]
&\frac{1}{T}\mathbf{T}^\mathrm{d}\dcont\mathbf{L}^\mathrm{d}
-\frac{1}{T}\pi\diver\left(\mathbf{v}\right)
+ \mathbf{q}\cdot\nabla\!\left(\frac{1}{T}\right)\\
&-\sum\limits_{\alpha=1}^{\nu}\mathbf{J}\alp\cdot\nabla\!\left(\frac{\mu\alp}{T}\right) -\frac{1}{T}\sum\limits_{k=1}^{n}\!\Big(\sum\limits_{\alpha=1}^{\nu}\mu\alp\gamma\alp{\!\!\!}^k m\alp\Big) \Lambda^k \geq 0.
\end{aligned} 
\end{align}
The term $\pi$ appearing in \eqref{bal:entropyproduction} is called \emph{dynamic pressure}, defined by the relation
\begin{align}
\label{eq:dynp}
	-\pi=\mfrac{1}{3}\trace(\mathbf{T})+p,
\end{align}
where $p$ is the thermodynamic equilibrium pressure \eqref{cr:thermoeqpressure}. The products appearing on the right-hand side of \eqref{bal:entropyproduction} are interpreted as follows \cite[p.~81]{IngoM}:
\begin{parcolumns}[colwidths={2=0.53\linewidth}]{2}	
	\linespread{2.2}\selectfont
	\colchunk{\textit{Thermodynamic Fluxes}}
	\colchunk{\textit{Thermodynamic Forces}}
	\colplacechunks
	\colchunk{Mixture heat flux vector, $\mathbf{q}$\,;}
	\colchunk{Reciprocal temperature gradient, $\nabla\!\left(\displaystyle\frac{1}{T}\right)$\,;}
	\colplacechunks
	\colchunk{Mixture stress tensor deviator, $\mathbf{T}^\mathrm{d}$\,;}
	\colchunk{Barycentric velocity gradient deviator, $\mathbf{L}^\mathrm{d}$\,; }
	\colplacechunks
	\colchunk{Dynamic pressure, $-\pi$\,;}
	\colchunk{Divergence of barycentric velocity, $ \diver\!\left(\mathbf{v}\right)$\,;}
	\colplacechunks
	\colchunk{Diffusion flux vector, $ \mathbf{J}\alp$\,;}
	\colchunk{Gradient of chemical potential, $\nabla\!\left(\displaystyle\frac{\mu\alp}{T}\right)$\,; }
	\colplacechunks
	\colchunk{Reaction rate density, $ \Lambda^k$\,;}
	\colchunk{Chemical affinity, $\Big(\sum\limits_{\alpha=1}^{\nu}\mu\alp \gamma\alp{\!\!\!}^k m\alp\Big)$\,.}
	\colplacechunks
\end{parcolumns}
\vspace{.5cm}
In case of isotropic fluid mixtures the following constitutive relations, defining the thermodynamic fluxes to be homogeneous linear functions of the thermodynamic forces~\cite[p. 1789 f.]{BotDre15}, guarantee the non-negativity of the entropy production density $\tensor*[^s]{\hspace{-0.1em}\Sigma}{*^{\text{TIP}}}$~\cite[p.~175]{Mueller2012}:
\begin{subequations}\label{eq:const}
	\begin{align}
	\intertext{\textit{Mixture stress tensor deviator}}
	\label{const:stressdev}
	\mathbf{T}^\mathrm{d}&=2\zeta\left[\frac{1}{2}(\nabla\mathbf{v}+\nabla\mathbf{v}^\top)-\frac{1}{3}\diver(\mathbf{v})\mathbf{I}\right];
	\intertext{\textit{Mixture heat flux vector}}
	\label{const:heatflux} \mathbf{q}&=\kappa T^2 \nabla\!\left(\frac{1}{T}\right) 
	- \sum\limits_{\beta=1}^{\nu}B\indices{_\beta}\,\nabla\!\left(\frac{\mu\indices{_\beta}}{T}\right);
	\intertext{\textit{Diffusion fluxes}}
	\label{const:diffflux}  \mathbf{J}\alp &= B\alp\nabla\!\left(\frac{1}{T}\right) 
	- \sum\limits_{\beta=1}^{\nu}B\indices{_\alpha_\beta}\nabla\!\left(\frac{\mu\indices{_\beta}}{T}\right);
	\intertext{\textit{Reaction rate densities}}
	\label{const:rRate}
	\Lambda^k&=-\sum\limits_{b=1}^n L^{kb} \Big(\sum\limits_{\alpha=1}^{\nu}\mu\alp\gamma\alp{\!\!\!}^b m\alp\Big)+L^k\diver\!\left(\mathbf{v}\right);
	\intertext{\textit{Dynamic pressure}} 
	\label{const:dynpr} -\pi&=\sum\limits_{b=1}^n  L^b\Big(\sum\limits_{\alpha=1}^{\nu}\mu\alp\gamma\alp{\!\!\!}^b m\alp\Big)+\lambda \diver\!\left(\mathbf{v}\right).
	\end{align}
\end{subequations}
Constitutive relations~\eqref{const:heatflux} and \eqref{const:diffflux} are generalizations of the laws of Fourier and Fick. The non-convective transport of internal energy caused by the gradients of the chemical potentials in~\eqref{const:heatflux} is known as \emph{Dufour-effect}. The influence of the temperature gradient on the diffusion flux is called \emph{thermal diffusion}.~The phenomenological coefficients $B\indices{_\beta}, B\indices{_\alpha_\beta}$ and $\kappa T^2$ are transport coefficients with respect to heat conduction and diffusion, while~$L\indices{^b}$ is interpreted as chemical viscosity~\cite{IngoM}. 
The phenomenological coefficients $L\indices{^k^b}$ and $L\indices{^k}$ appearing at the right-hand sides of constitutive relations~\eqref{const:rRate} and~\eqref{const:dynpr} are used to define the coefficients
\begin{alignat}{4}\label{eq:newcoff}
&\mathbb{L}\alp := \sum_{k=1}^{n}\gamma\alp{\!\!\!}^k m\alp L\indices{^k},  &\qquad& \mathbb{L}\indices{_\alpha_\beta}:=\sum_{k=1}^{n}\sum_{b=1}^{n}\gamma\alp{\!\!\!}^k m\alp L\indices{^k^b}\gamma\indices{_\beta}{\!\!\!}^b m\indices{_\beta}\,,
\end{alignat}
where $n$ is the number of independent chemical reactions and $\alpha,\beta=1,\ldots,\nu$.
Then the two matrices 
\begin{equation}\label{mat:pheno}
\begin{bmatrix}
\kappa T^2 & B_1 & \cdots & B_{\nu}\\ 
B_1 & B_{1,1} & \cdots & B_{1,\nu}\\
\vdots & \vdots & \ddots  & \vdots\\
B_\nu & B_{\nu,1}& \cdots  & B_{\nu,\nu}
\end{bmatrix} 
\quad \text{and} \quad 
\left[\begin{array}{c|ccc}
\lambda & -\mathbb{L}_1\phantom{-}\ & \cdots & -\mathbb{L}\indi{_\nu}\\ \hline
\mathbb{L}_1 & \mathbb{L}_{1,1} & \cdots & \phantom{-}\mathbb{L}_{1,\nu}\\
\vdots & \vdots & \ddots & \phantom{-}\vdots\\
\mathbb{L}_{\nu} & \mathbb{L}_{\nu,1} & \cdots &  \phantom{-}\mathbb{L}_{\nu,\nu}
\end{array}\right].
\end{equation} 
contain the phenomenological coefficients, which describe a linear relation between the thermodynamic fluxes and forces, see the right-hand sides of~\eqref{const:heatflux} to~\eqref{const:dynpr}. 
In order for the constitutive relations~\eqref{eq:const} to satisfy the inequality given by the entropy production density~\eqref{bal:entropyproduction}, i.e., to guarantee a non-negative entropy production density such that the resulting mathematical model does not violate the second law of thermodynamics, both matrices~\eqref{mat:pheno} have to be positive semi-definite and furthermore $\zeta\geq 0$, cf.~\cite{IngoM,Mueller2012}.
Note that the first matrix of~\eqref{mat:pheno} and the submatrix of the second matrix specified by $\mathbb{L}_{\alpha \beta}$ are symmetric, cf. \cite[p.~1790]{BotDre15}. Also note that the symmetry properties of these matrices reflect the Onsager-Casimir reciprocal relations which the phenomenological coefficients of the constitutive relations satisfy, see e.g. \cite{BotDre15,Mei73}.
\subsubsection*{Mixture Stress Tensor}
In order to derive the closure relation for the mixture stress tensor $\mathbf{T}$ we combine Equation \eqref{eq:decomposed_stresstensor} with \eqref{eq:dynp} and obtain the following relation for the mixture stress tensor
\begin{align}\label{eq:stressig}
\mathbf{T} = \mathbf{T}^{\mathrm{d}}-(\pi + p)\mathbf{I}.
\end{align}
The constitutive relation for the dynamic pressure $\pi$ specified by~\eqref{const:dynpr} can be written in the following alternative form $-\pi=\sum_{\alpha=1}^{\nu}\mathbb{L}_\alpha\mu_\alpha+\lambda\diver(\mathbf{ v})$. In combination with the constitutive equation for the mixture stress tensor deviator $\mathbf{T}^{\mathrm{d}}$, which in \eqref{const:stressdev} is specified with respect to the three dimensional physical space, relation \eqref{eq:stressig} results in the following constitutive equation for the stress tensor of the mixture
\begin{align}\label{const:MixStrTensor}
\mathbf{T}= (-p+\sum_{\alpha=1}^{\nu}\mathbb{L}\indi{_\alpha}\mu\indi{_\alpha})\mathbf{I}+\lambda\diver(\mathbf{v})\mathbf{I}+2\zeta\left[\frac{1}{2}(\nabla\mathbf{v}+\nabla\mathbf{v}^\top)-\frac{1}{3}\diver(\mathbf{v})\mathbf{I}\right],
\end{align}
where we once again have used the relation $\mathbf{v}=\mathbf{M}/\sum_{\alpha=1}^{\nu}\rho_\alpha$. We denote the viscosity part of the stress tensor of the mixture~\eqref{const:MixStrTensor} by $\mathbf{S}$ such that 
\\[0.5em]
\textit{Mixture stress tensor}
	\begin{align*}
	\mathbf{T} &= (-p+\sum_{\alpha=1}^{\nu}\mathbb{L}\indi{_\alpha}\mu\indi{_\alpha})\mathbf{I} + \mathbf{S},\\
	\mathbf{S}&=(\lambda-\frac{2\zeta}{3})\diver(\mathbf{v})\mathbf{I}+\zeta(\nabla\mathbf{v}+\nabla\mathbf{v}^\top).
	\end{align*}

\subsection{First Law of Thermodynamics and Energy Balance}\label{sec:FirstSecond}
In the following we reinterpret the energy balance and the entropy balance of classical continuum mechanics in order to show their relation to the~\emph{first}- and the~\emph{second law of thermodynamics}, respectively. For the sake of simplicity we follow \cite{BotDre15} and assume that the differential balance laws for the mixture are of the same form as the corresponding differential balances of the one-component system, cf.~\cite[p.~119~f.]{Trues}. We start this section by briefly recalling the notion of an integral balance law in spatial representation. For a rigorous development of the theory of balance laws based on first principles see e.g. \cite{Gur87}

Let $\UOmega\subset\R^d$ be a domain and let $\mc{U}\subset\R\times\UOmega$ be an open set containing $\left\{\lambda\right\}\times\UOmega$, with $\lambda\in\R$.
Let $\mathbf{ v}\in C^k(\mc{U}; \R^d)$ be a given time-dependent spatial velocity vector field, $k\geq 1$. The collection of all maps $\lchi\indi{_t_,_\lambda}$ defined by the requirement that for each $\lambda$ and $x\in\UOmega$ the map $t\mapsto \lchi\indi{_t_,_\lambda}(x):=\lchi(t;\lambda,x)$ is an integral curve of $\mathbf{ v}$ starting at $x$ at time $t=\lambda$, i.e.
\begin{gather*}
\total{\lchi\indi{_t_,_\lambda}(x)}{t}=\mathbf{ v}(t,\lchi\indi{_t_,_\lambda}(x)) \quad \text{with}\quad \lchi\indi{_\lambda_,_\lambda}(x)=x,
\end{gather*}
is called \emph{time-dependent flow} or \emph{evolution operator} of $\mathbf{ v}$ \cite[Ch.~4~f.]{AbrMarRat88}. Usually, one writes $\lchi\indi{_t}=\lchi\indi{_t_,_0}$ and considers $x\in\UOmega$ as the position of some fluid particle at initial time $\lambda=0$.

Let $\mathbb{I}$ be a time interval containing the initial time (here $\lambda=0$) for which the flow of $\mathbf{v}$ starting at $\lambda$ is defined. By an \emph{admissible} subset we mean a bounded measurable set for which the notion of a boundary is defined, see e.g. \cite{CZT09}. Let $V\subset\UOmega$ be an admissible subset. Denote by $\lchi_t(V):=\left\{\lchi(t;x)\in\R^d\,\vert\, x\in V \right\}$ its image under the flow of $\mathbf{ v}$ at time parameter $t\in\mathbb{I}$, and let $\partial\lchi_t(V)$ denote its boundary.
In classical continuum physics, a general integral balance law in spatial description is a family of integral identities, cf. \cite{Daf93} 
\begin{gather}\label{eq:general_spatial_integral_balance}
\int\limits_{\lchi\indi{_t}(V)}\lpsi\, \dv\Biggr]_{t_1}^{t_2} =-  \int\limits_{t_1}^{t_2}\oint\limits_{\partial\lchi\indi{_t}(V)}\tensor*[]{\hat{\mathbf{F}}}{}\cdot\mathbf{n}\,\da \,\dt + \, \int\limits_{t_1}^{t_2}\int\limits_{\lchi\indi{_t}(V)} \tensor*[]{\!\sigma}{}\, \dv\,\dt,
\end{gather} 
required to hold for any admissible subset $V$ of $\UOmega$ and any subinterval $(t_1,t_2)$ of $\mathbb{I}$ with $t_1\leq t_2$. The scalar function $\lpsi(t,\cdot)$ is the volume-specific (i.e. per unit volume) density of some extensive observable, $\tensor*[]{\sigma(t,\cdot)}{}$ represents the rate of total production per unit volume, the vector field $\hat{\mathbf{F}}(t,\cdot)$ is the total flux and $\mathbf{n}$ is the outer normal vector to $\partial\lchi_t(V)$.

Note that in the literature related to classical continuum mechanics the general integral balance is typically presented in a slightly different form, namely as the following family of time-dependent integral identities \cite{Liu02,MarHug94,Mue85} 
\begin{gather}\label{eq:general_spatial_traditional_balance}
\total{}{t}\int\limits_{\lchi\indi{_t}(V)}\lpsi\, \dv =-  \oint\limits_{\partial\lchi\indi{_t}(V)}\tensor*[]{\mathbf{F}}{}\cdot\mathbf{n}\,\da  + \, \int\limits_{\lchi\indi{_t}(V)} \tensor*[]{\!\sigma}{}\, \dv,
\end{gather} 
where the left-hand side is required to be a $t$-differentiable mapping, $t\mapsto \int_{\lchi_t(V)}\lpsi(t,\cdot)\, \dv$. Also note that the vector field $\tensor*[]{\mathbf{F}}{}(t,\cdot)$ appearing in \eqref{eq:general_spatial_traditional_balance} is the \emph{non-convective flux}, and hence differs from the total flux vector field $\hat{\mathbf{F}}(t,\cdot)$ of \eqref{eq:general_spatial_integral_balance}. 

Suppose that the conditions for the validity of the \emph{transport theorem} are satisfied, cf. \cite[Thm.~2.11~\&~Rem.~2.15]{AmaEschIII09} and \cite{AbrMarRat88,Mue85}. Then the left hand-side of~\eqref{eq:general_spatial_traditional_balance} holds
\begin{align}\label{eq:transportheorem_general_balance}
\total{}{t}\int\limits_{\lchi\indi{_t}(V)}\lpsi(t,\cdot)\,\dv &=
\int\limits_{\lchi\indi{_t}(V)}\Big[\partial_t\lpsi(t,\cdot) + \diver\left(\lpsi(t,\cdot)\mathbf{v}(t,\cdot)\right)\!\Big]\dv \quad\text{for } t\in \R.
\end{align}
Let the boundary of $\lchi_t(V)$ move with the velocity $\mathbf{ v}$. Neglecting jump discontinuities on singular surfaces and under the assumption that the fields are regular enough such that the divergence theorem \cite[p.~124~f.]{MarHug94} (or the Gauss-Green theorem \cite{CZT09}) can be applied, the right-hand side of \eqref{eq:transportheorem_general_balance} is rewritten and we obtain
\begin{align}\label{eq:transportheorem_general_balance2}
\total{}{t}\int\limits_{\lchi\indi{_t}(V)}\lpsi\,\dv &= \int\limits_{\lchi\indi{_t}(V)}\partial_t\lpsi\, \dv 
	+\oint\limits_{\partial\lchi\indi{_t}(V)}\left(\lpsi\mathbf{v}\right)\cdot\mathbf{n}\, \da, 
\end{align}
where the term $(\lpsi\mathbf{v})(t,\cdot)$ is the \emph{convective flux}, see Section~\ref{sec:differential_balance_laws}. By means of \eqref{eq:transportheorem_general_balance2} integral balance \eqref{eq:general_spatial_traditional_balance} becomes
\begin{gather*}
\int\limits_{\lchi\indi{_t}(V)}\partial_t\lpsi\, \dv =-  \oint\limits_{\partial\lchi\indi{_t}(V)}\big(\lpsi\mathbf{v} +\tensor*[]{\mathbf{F}}{}\big)\cdot\mathbf{n}\,\da  + \, \int\limits_{\lchi\indi{_t}(V)} \tensor*[]{\!\sigma}{}\, \dv,
\end{gather*}
where the relation to the total flux in \eqref{eq:general_spatial_integral_balance} is given by $\hat{\mathbf{F}}=\big(\lpsi\mathbf{v} +\tensor*[]{\mathbf{F}}{}\big)$.
On the other hand, by means of the transport theorem \eqref{eq:transportheorem_general_balance} and the divergence (or Gauss-Green) theorem integral balance \eqref{eq:general_spatial_traditional_balance} can be reformulated into
\begin{gather*}
	\int\limits_{\lchi\indi{_t}(V)}\!\!\big(\partial_t\lpsi + \diver\big(\lpsi\mathbf{v} + \mathbf{F}\big) -\sigma \big)\, \dv = 0,
\end{gather*}
required to hold for all admissible subsets $V$ of $\UOmega$ and all $t\in\mathbb{I}$. From this follows the partial differential equation 
\begin{gather*}
	\partial_t\lpsi + \diver\big(\lpsi\mathbf{v} + \mathbf{F}\big) -\sigma = 0,
\end{gather*}
if singular surfaces are not taken into account.
Now we are ready to connect the integral total energy balance in spatial representation with the first law of thermodynamics. For this we introduce the quantities:\\[0.5em]
\begin{tabular}{cl}
	$\begin{aligned}e\end{aligned}$ & internal energy per unit mass\\
	$\begin{aligned}r\end{aligned}$ & influx of internal energy per unit mass\\
	$\begin{aligned}\mathbf{f}\end{aligned}$ & influx of linear momentum per unit mass (mass-specific body forces)
\end{tabular}\\[0.5em]
\noindent
The mass-specific internal energy $e(t,\cdot)$ is related to the volume-specific internal energy density $u(t,\cdot)$ via the mass density, $\rho e= u$. Similarly, the mass-specific body forces $\mathbf{f}(t,\cdot)$ are related to the influx density of linear momentum $\mathbf{b}(t,\cdot)$ \eqref{eq:vector_balance_law} via $\rho\mathbf{f}=\mathbf{b}$. 

Formulated for the total energy, integral balance \eqref{eq:general_spatial_traditional_balance} takes the form
	\begin{gather}\label{eq:integral_balance_total_energy_spatial}
	\total{}{t}\int\limits_{\lchi\indi{_t}(V)}\rho\bigr(e+\frac{1}{2}\mathbf{v}\cdot\mathbf{v}\bigr)\, \dv =-  \oint\limits_{\partial\lchi\indi{_t}(V)}\bigr(\mathbf{q}-\mathbf{T}^\top\!\!\cdot\mathbf{v}\bigr)\cdot\mathbf{n}\,\da  + \, \int\limits_{\lchi\indi{_t}(V)} \rho\bigr(\mathbf{f}\cdot\mathbf{v}+r\bigr) \dv,
	\end{gather} 
where $(\rho\mathbf{f}\cdot\mathbf{v})(t,\cdot)$ is called the~\emph{influx of kinetic energy} and the term $(\rho r)(t,\cdot)$ is called~\emph{influx of internal energy}. These terms describe the rate of change of kinetic and internal energy through the influx of mechanical and non-mechanical power, respectively.

We introduce the following terminologies:
\begin{alignat*}{4}
\mathfrak{K}(t)&=\int\limits_{\lchi\indi{_t}(V)} \frac{1}{2}\rho\mathbf{v}\cdot\mathbf{v}\,\dv ,
&\qquad&\text{(kinetic energy)}\\
\mathfrak{E}(t)&=\int\limits_{\lchi\indi{_t}(V)} \rho e\,\dv ,
&\qquad&\text{(internal energy)}\\
\mathfrak{Q}(t)&=\oint\limits_{\partial\lchi\indi{_t}(V)}\!\!\!-\,\mathbf{q}\cdot\mathbf{n}\,\da\,+ \int\limits_{\lchi\indi{_t}(V)}\rho r\,\dv ,
&\qquad&\text{(non-mechanical power)}\\
\mathfrak{W}(t)&=\oint\limits_{\partial\lchi\indi{_t}(V)} (\mathbf{T}^\top\!\!\cdot\mathbf{v})\cdot\mathbf{n}\,\da\,+\int\limits_{\lchi\indi{_t}(V)}\rho\mathbf{f}\cdot\mathbf{v}\,\dv,
&\qquad&\text{(mechanical power)}
\end{alignat*}
which we use to write the integral total energy balance~\eqref{eq:integral_balance_total_energy_spatial} in form of the~\emph{first law of thermodynamics}, see e.g. \cite[Ch.~2.3]{MarHug94} and~\cite{Trues}
\begin{align*}
	\total{}{t}(\mathfrak{E}+\mathfrak{K})=\mathfrak{Q} +\mathfrak{W}.
\end{align*}
With the help of the transport-theorem \eqref{eq:transportheorem_general_balance} the left hand-side of~\eqref{eq:integral_balance_total_energy_spatial} is transformed into the following family of integral identities
\begin{equation}\label{eq:lhs}
\begin{split}
	&\total{}{t}\int\limits_{\lchi\indi{_t}(V)}\rho\bigr(e+\frac{1}{2}\mathbf{v}\cdot\mathbf{v}\bigr)\,\dv\\
	 & \qquad =
	\int\limits_{\lchi\indi{_t}(V)}\partial_t\bigr(\rho\bigr(e+\frac{1}{2}\mathbf{v}\cdot\mathbf{v}\bigr)\bigr)\dv
	+\oint\limits_{\partial\lchi\indi{_t}(V)} \bigr(\rho\bigr(e+\frac{1}{2}\mathbf{v}\cdot\mathbf{v}\bigr)\mathbf{v}\bigr)\cdot\mathbf{n}\,\da.
\end{split}
\end{equation}
Due to notational convenience we denote the total energy density by $\mathcal{E}=  \rho\bigr(e + \frac{1}{2}\mathbf{v}\cdot\mathbf{v}\bigr)$. Then with~\eqref{eq:lhs} total energy balance~\eqref{eq:integral_balance_total_energy_spatial} is rewritten into
\begin{align*}
\int\limits_{\lchi\indi{_t}(V)}\!\!\partial_t\mathcal{E}\,\dv =
-\!\!\!\oint\limits_{\partial\lchi\indi{_t}(V)}\bigr(\mathcal{E}\mathbf{v}
+\bigr(\mathbf{q}-\mathbf{T}^\top\!\cdot\mathbf{v}\bigr)\bigr)\cdot\mathbf{n}\,\da
+\!\!\!\int\limits_{\lchi\indi{_t}(V)}\!\!\rho\bigr(\mathbf{f}\cdot\mathbf{v}+r\bigr)\dv.
\end{align*}   
On the other hand we have the family of integral identities
\begin{align*}
\int\limits_{\lchi\indi{_t}(V)}\big(\partial_t\mathcal{E}
+\diver\bigr(\mathcal{E}\mathbf{v}
+\bigr(\mathbf{q}-\mathbf{T}^\top\!\cdot\mathbf{v}\bigr)\bigr)- \rho\bigr(\mathbf{f}\cdot\mathbf{v}+r\bigr)\big)\dv = 0,
\end{align*}
required to hold for all admissible subsets $V$ of $\UOmega$ and all $t\in\mathbb{I}$. Hence it follows that
\begin{gather}\label{eq:differetial_total_energy}
\partial_t\mathcal{E}
+\diver\bigr(\mathcal{E}\mathbf{v}
+\bigr(\mathbf{q}-\mathbf{T}^\top\!\cdot\mathbf{v}\bigr)\bigr)-\rho\bigr(\mathbf{f}\cdot\mathbf{v}+r\bigr)= 0.
\end{gather}
We integrate \eqref{eq:differetial_total_energy} with respect to an admissible domain $V$ that is fixed in time and by means of the divergence theorem (or Gauss-Green theorem) obtain
\begin{gather}\label{eq:total_energy_bal_fixed_domain1}
\int_V \partial_t\mathcal{E} \dx = - \int_{\partial V}\,\bigr(\mathcal{E}\mathbf{v}
+\bigr(\mathbf{q}-\mathbf{T}^\top\!\cdot\mathbf{v}\bigr)\bigr)\cdot\mathbf{n}\,\dS + \int_V \rho\bigr(\mathbf{f}\cdot\mathbf{v}+r\bigr) \dx.
\end{gather}
Next we use the spatial material time derivative \cite{IVM16} that relates the partial time derivative with the total time derivative. Note that this relation, that we mentioned in Section \ref{sec:TIP}, reflects a special case of the \emph{Lie derivative} of a time-dependent tensor field with respect to a time-dependent vector field, see \cite[p.~95~f.]{MarHug94}. In case of the total energy density and restricted to a fixed observation point (think of an observer sitting on $V$), the spatial material time derivative allows us to express the integrand of the integral at the left hand side of \eqref{eq:total_energy_bal_fixed_domain1} by means of the total time derivative such that
\begin{gather}
\int_V \total{\mathcal{E}}{t} \dx = - \int_{\partial V}\bigr(\mathcal{E}\mathbf{v}
+\bigr(\mathbf{q}-\mathbf{T}^\top\!\cdot\mathbf{v}\bigr)\bigr)\cdot\mathbf{n}\,\dS + \int_V \rho\bigr(\mathbf{f}\cdot\mathbf{v}+r\bigr) \dx.\notag
\intertext{The total time derivative can be \emph{pulled out} of the integral in accordance to the integral transformation rule \cite[p.~22]{IVM16} such that}
\label{eq:total_energy_bal_fixed_domain3}
\total{}{t}\int_V \mathcal{E} \dx = - \int_{\partial V}\bigr(\mathcal{E}\mathbf{v}
+\bigr(\mathbf{q}-\mathbf{T}^\top\!\cdot\mathbf{v}\bigr)\bigr)\cdot\mathbf{n}\,\dS + \int_V \rho\bigr(\mathbf{f}\cdot\mathbf{v}+r\bigr) \dx.
\end{gather}
With $\mathcal{E}=  \rho\bigr(e + \frac{1}{2}\mathbf{v}\cdot\mathbf{v}\bigr)$ and neglected influx terms 
\eqref{eq:total_energy_bal_fixed_domain3} becomes
\begin{equation}
\label{eq:total_energy_bal_fixed_domain5}
\total{}{t}\int_V \rho\bigr(e + \frac{1}{2}\mathbf{v}\cdot\mathbf{v}\bigr) \dx = - \int_{\partial V}\Big[\rho\bigr(e+\frac{1}{2}\mathbf{v}\cdot\mathbf{v}\bigr)\mathbf{v} + \mathbf{q}-\mathbf{T}^\top\!\cdot\mathbf{v}\Big]\cdot\mathbf{n}\,\dS. 
\end{equation}
Using the relation between the spatial linear momentum density and the spatial velocity, $\mathbf{M}=\rho\mathbf{v}$, as well as the relation between mass-specific and volumetric internal energy, $\rho e= u$, we rewrite integral total energy balance \eqref{eq:total_energy_bal_fixed_domain5} and obtain the following equivalent total energy balance
\begin{equation}\label{eq:energyintbal2}
\total{}{t}\int_{V}\Big(\frac{\mathbf{M}\cdot\mathbf{v}}{2}+ u\Big)\,\dx = -\int_{\partial V}\,\Big[\Big(\frac{\mathbf{M}\cdot\mathbf{v}}{2}+ u\Big)\mathbf{v}
+\Big(\mathbf{q}-\mathbf{T}^\top\!\cdot\mathbf{v}\Big)\Big]\cdot\mathbf{n}\,\dS,
\end{equation}
which is in accordance to Equation (38) in~\cite{Oet06}.
 \subsection{Second Law of Thermodynamics and Entropy Balance}\label{sec:secondlaw}
As mentioned in Section \ref{sec:consti}, the linear closure relations of TIP guarantee a non-negative entropy production density $\entprod$.~In~\cite[Ch.~III]{Mazur} it has been shown that the local mathematical representation of the second law of thermodynamics has the form of the differential balance law~\eqref{bal:entrop}, with a non-negative entropy production density $\entprod\geq 0$. Hence, the constitutive relations of TIP are in accordance with the second law of thermodynamics. However, this statement is restricted to the local equilibrium hypothesis of TIP. A generalization can be achieved through the~\emph{entropy principle} used in different continuum thermodynamic theories~\cite{BotDre15,Liu02,Liu83,Mue85,IngoM}. The entropy principle is also supported by the kinetic theory of gases~\cite{Rug08}. These continuum theories in contrast to the phenomenological theory of TIP are not based on the local equilibrium hypothesis. In these theories the differential entropy balance law is postulated and the so called~\emph{entropy principle} is exploited. According to this principle, the entropy production density $\entprod$ has to become non-negative for every \emph{thermodynamic process}~\cite{Liu02,Liu83,Mue85}. For the measure theoretical foundations of the second law of thermodynamics and a generalization of the Clausius-Duhem inequality see e.g.~\cite{Mar02}. It has been shown that the linear closure relations of TIP can be obtained in the framework of the constitutive theory of these thermodynamic continuum theories, where they represent a special case \cite[Sec.~13~f.]{BotDre15}. 

Motivated by the entropy principle and under the assumption that the balance laws of the mixture are of the same form as the corresponding balance laws of the single-body system, we consider the differential entropy balance equation~\eqref{bal:entrop} and formulate the integral with respect to an admissible bounded domain that does not change in time
\begin{align}\label{eq:entbal}
\total{}{t}\int_{V}s\,\dx =
-\int_{\partial V}\bigr(s\mathbf{v}+\tensor*[^s]{\Upphi}{}\bigr)\cdot\mathbf{n}\,\dS
+\int_V \tensor*[^s]{\Sigma}{}\,\dx.
\end{align}
In accordance with the entropy principle we require the entropy production density at the right-hand side of the integral entropy balance~\eqref{eq:entbal} to becomes non-negative for all thermodynamic processes, i.e. $\tensor*[^s]{\Sigma}{}\geq 0$, from which we obtain the inequality
\begin{align}\label{ineq:entba}
\total{}{t}\int_{V}s\,\dx \geq
-\int_{\partial V}\bigr(s\mathbf{v}+\tensor*[^s]{\Upphi}{}\bigr)\cdot\mathbf{n}\,\dS.
\end{align}
In case of the fluid mixture considered in this work, the non-convective flux and the entropy production density derived in the framework of the phenomenological theory of TIP are of the form  $\tensor*[^s]{\Upphi}{}=\tensor*[^s]{\Upphi}{^{\text{TIP}}}$ and $\tensor*[^s]{\Sigma}{}=\tensor*[^s]{\Sigma}{^{\text{TIP}}}$ given by~\eqref{eq:entropyflux} and~\eqref{eq:entropyprod}, respectively. With these relations, the non-convective flux term and the entropy production density of the general entropy balance inequality~\eqref{ineq:entba} are specified, resulting in
\begin{align}\label{ineq:entbal2}
\total{}{t}\int_{V}s\, \dx \geq
-\int_{\partial V}\Big[s\mathbf{v}+\frac{1}{T}\Big(\mathbf{q}-\sum\limits_{\alpha=1}^{\nu}\mathbf{J}\indi{_\alpha}\mu\indi{_\alpha}\Big)\Big]\cdot\mathbf{n}\,\dS\,.
\end{align}

In view of the preceding considerations we conclude that a proper mathematical model of the fluid mixture in the Operator-GENERIC framework has to reflect the first law of thermodynamics in form of the energy balance~\eqref{eq:energyintbal2} and the second law of thermodynamics in form of the balance inequality~\eqref{ineq:entba}, which for isolated systems reflect energy conservation~\eqref{eq:EnergyCons} and entropy production~\eqref{eq:EntropyProd}, respectively.

\subsection{Summary of Important Formulas}\label{sec:formulas}
In the following we summarize the relations for the two different compositions of the state variable considered in this work, distinguished in form of two cases. For these cases, the state evolution equations may be different but the closure relations for the mixture stress tensor $\mathbf{T}$, the heat flux vector $\mathbf{q}$, the diffusion fluxes $\mathbf{J}_\alpha$, and the reaction rate density $\Lambda^k$ with respect to the $k$-th reaction are always given by the constitutive equations~\eqref{const:MixStrTensor}, ~\eqref{const:heatflux},~\eqref{const:diffflux}, and~\eqref{const:rRate}, respectively. 
We start with the first case where the (internal) energy of the system constitutes the thermodynamic potential, followed by the second case where the entropy of the system constitutes the thermodynamic potential. 
\subsection*{Energy as Thermodynamic Potential}
Consider the entropy density as one of the independent state variables.~The evolution of the macroscopic state $\zhet\in\mc{Z}$ is described through the time evolution and spatial distribution of the state variables associated with $\zhet$ through relation~\eqref{energypot:stat}.~The corresponding governing equations, which we call \emph{state evolution equations}, are given by the following field equations:
{\allowdisplaybreaks\\[+1em]
	\textit{State Evolution Equations}
	\begin{subequations}\label{goveq:EnergyPot}~
		\vspace*{-2pt}\begin{align}
		\label{ba:partmass}
		\partial_t\rho\alp&= -\diver\!\left(\rhoAlpha\mathbf{v}+\mathbf{J\!}\alp\right)+\tau\indi{_\alpha}, \hspace*{.2cm}(\alpha=1,\dots,\nu)\\[5pt]
		\label{ba:mixmom}
		\partial_t\mathbf{M}&=-\diver\!\left(\mathbf{M}\otimes\mathbf{v}-\mathbf{T}\right),\\
		\label{ba:entdens}
		\partial_t s&=-\diver(s\mathbf{v}+\tipentflux) + \tipentprod,
				\end{align}
				\begin{align}
		\begin{split}
		\tipentflux&=\frac{1}{T}\Big[\mathbf{q}-\sum_{\alpha=1}^{\nu}\mathbf{J}\alp\mu\alp\Big],\\
		\tipentprod&=\mathbf{q}\cdot\nabla\!\left(\frac{1}{T}\right)
		\!+\!\frac{1}{T}\Big(\mathbf{T}\dcont\nabla\mathbf{v} 
		+p\diver(\mathbf{v})\Big)
		\!-\!\sum_{\alpha=1}^{\nu}\!\left(\mathbf{J}\indi{_\alpha}\cdot\nabla\!\left(\frac{\mu\indi{_\alpha}}{T}\right) \!+\! \tau\indi{_\alpha}\frac{\mu\indi{_\alpha}}{T}\right)\!\geq 0.
		\end{split}
		\end{align}
	\end{subequations}}%
Note that the barycentric velocity is given by the relation $\mathbf{v}=\mathbf{M}/\sum_{\alpha=1}^{\nu}\rhoAlpha$, and the mass~\emph{production} densities $\tau_\alpha$ by~\eqref{eq:massprod}.
The internal energy density $u=u(\rho_1,\dots,\rho_\nu,s)$ is a~\emph{thermodynamic constitutive relation} and has to be specified with respect to the concrete problem under consideration. Then the absolute temperature field $T$ and the chemical potentials $\mu_\alpha$ become accessible through the following identity relations
\begin{alignat}{4}
	\label{eq:partInt}
	&{T}= \frac{\partial u}{\partial s},
	&\qquad&
	\mu\indi{_\alpha}=\frac{\partial u}{\partial \rho_\alpha},
\end{alignat}
with $\alpha=1,\dots,\nu$. The thermodynamic constitutive relation for the thermodynamic equilibrium pressure $p$ \eqref{cr:thermoeqpressure} then takes the form
\begin{align}
	p=- u + \left(\pd{u}{s}\right) s +\sum_{\alpha=1}^{\nu}\left(\pd{u}{\rho_\alpha}\right)\rho\indi{_\alpha}\,.
\end{align} 

\subsection*{Entropy as Thermodynamic Potential}
Consider the internal energy density as one of the independent state variables. The evolution of the macroscopic state $\zhet\in\mc{Z}$ is described through the evolution of the state variables associated with $\zhet$ through relation~\eqref{entpot:stat}.~The corresponding state evolution equations are:
{\allowdisplaybreaks\\[+1em]
	\textit{State Evolution Equations}
	\begin{subequations}\label{eq:stateEvoEntPot}~
		\vspace*{-2pt}\begin{align}
		\label{entopt:partmass}
		\partial_t\rho\alp &=- \diver\left(\rhoAlpha\mathbf{v}+\mathbf{J\!}\alp\right)+\tau\indi{_\alpha}, \hspace*{.2cm}(\alpha=1,\dots,\nu)\\[1pt]
		\label{entpot:mixmom}
		\partial_t\mathbf{M} &=-\diver\left(\mathbf{M}\otimes\mathbf{v} - \mathbf{T}\right),\\[1pt]
		\label{entopot:mixint}
		\partial_t u &= - \diver\left(u\mathbf{v} + \mathbf{q}\right) + \mathbf{T}\dcont\nabla\mathbf{v}.
		\end{align}
\end{subequations}}%
The entropy density $s=s(\rho_1,\dots,\rho_\nu,u)$ is a~\emph{thermodynamic constitutive relation} and has to be specified with respect to the concrete problem under consideration. Then the absolute temperature field $T$ and the chemical potentials $\mu_\alpha$ become accessibly through relations~\eqref{rel:entpot}.
The constitutive relation for the thermodynamic equilibrium pressure $p$~\eqref{cr:thermoeqpressure} takes the form, cf.~\cite[Eq.~29]{Ed98},
\begin{align*}
p=- u + \left(\pd{s}{u}\right)^{-1}\left[s  - \sum_{\alpha=1}^{\nu}\left(\pd{s}{\rho_\alpha}\right)\rho\indi{_\alpha}\right]\,.
\end{align*}

\section{TIP as Operator-GENERIC Formulation}\label{sec:mixture}
In Section~\ref{sec:MixtureTheory} we presented the field equations and closure relations for the fluid mixture consisting of $\nu\in \N$, $\nu \geq 2$ constituents. In this section we introduce weak formulations of the corresponding differential balance equations and show that in the operator setting they are encoded in the Operator-GENERIC formulation of the mixture mentioned in Section~\ref{sec:GENERIC}. We will consider the two cases summarized in Section~\ref{sec:formulas}, viz. the case where energy constitutes the thermodynamic potential, and the case where entropy represents the thermodynamic potential. 
The evolution equations that we are looking for are given by~\eqref{eq:operator_equation_isolated} and~\eqref{eq:operator_equation_open_dynamics} depending on whether we consider isolated or open systems, respectively. The associated~$\mc{J}$ and $\mc{R}$ should satisfy the properties stated in Section~\ref{sec:GENERIC}. 

We start with the case where energy constitutes the thermodynamic potential. To derive a weak formulation of the corresponding partial differential equations~\eqref{goveq:EnergyPot} we have to treat time and space separately~\cite[Ch.~23.1]{Zei90}. Let $\mathbb{I}$ be the considered time interval and $\Omega$ be a time-independent, bounded domain with \emph{Lipschitz boundary}~\cite[p.~232]{Zei86}. Then the state variable%
\begin{equation}%
\mathbf{z} = \begin{bmatrix}
\rho_1 & \ldots & \rho_\nu & \mathbf{M}^\top & s 
\end{bmatrix}^\top
\end{equation}%
maps from~$\mathbb{I}$ into the open subset
\begin{equation}\label{eq:set_Z}
\mc{Z}:=\Big\{ \zhet \in \mc{D}_{\mathbf{z}} \,\Big|\, \sum_{\alpha=1}^\nu \rho_\alpha \geq \varrho \text{ allmost everywhere for a } \varrho >0 \Big\}
\end{equation}
of the space~$\mc{D}_{\mathbf{z}}$ which will be a Cartesian product of~$W^{1,3}(\Omega)$ and its subspaces which will be defined later.  The space~$\mc{D}_{\mathbf{z}}$ contains implicitly the dependence on the spatial coordinates. Note that $\mc{Z}$ is equal to the set of all $\mc{D}_{\zhet}$-function with $\rho=\sum_{\alpha=1}^\nu \rho_\alpha$ be almost everywhere greater or equal zero and its reciprocal be an element of $L^\infty(\Omega)$. Furthermore, the subspaces of the space~$W^{1,3}(\Omega)$ which define $\mc{D}_{\mathbf{z}}$ are chosen such that $\mathbf{v}=\mathbf{M} / \rho $ and its derivative given by $\nabla\mathbf{v}=(\rho \nabla \mathbf{M} - \mathbf{M} \otimes \nabla \rho)/ \rho^2 $ are component-wise elements of $L^{p}(\Omega)$ for every $p<3$ by the continuous embedding of $W^{1,3}(\Omega)$ into $L^q(\Omega)$, $1\leq q<\infty$, \cite[Cor.~5.13]{Ada75}. 
Therefore, $\mathbf{v} \cdot \mathbf{v}$, $\mathbf{M} \cdot \mathbf{v}$, and the component functions of $\mathbf{v}$ are also $W^{1,3}$-functions, if one assumes the densities~$\rho_\alpha$ and the momentum~$\mathbf{M}$ to be slightly more regular, such that they are elements of $W^{1,3+\varepsilon}(\Omega)$ with arbitrary $\varepsilon >0$. 
This will be of importance, since the functional derivatives of the energy and entropy have to map into $\mc{D}_\zhet$. Out of the same reason, we choose $s \in W^{1,3}(\Omega)$, since the gradient of $\mu_\alpha=\mu_\alpha(\rho_1,\ldots,\rho_\nu,s)$ and therefore~$\nabla s$ should be in the same space as the gradient of $\rho_\alpha$.
To derive the weak formulation as well as to get rid of the second derivatives, which are hidden in the derivatives of $\mathbf{q}$, $\mathbf{J}_\alpha$, and $\mathbf{T}$, cf.~\eqref{eq:const}, we multiply the
equations of~\eqref{goveq:EnergyPot} with an arbitrary test function 
$$\boldsymbol{\varphi} = \begin{bmatrix}
\varphi_{\rho_1} & \ldots & \varphi_{\rho_\nu} & \varphi_{\mathbf{M}}^\top & \varphi_s 
\end{bmatrix}^\top \in [C^\infty(\Omega) \cap W^{1,3}(\Omega)]^{\nu + 4}.$$
Using then integration by parts we obtain
\begin{subequations}
	\label{eq:weak_mixture_energy}
	\begin{alignat}{2}
	\label{eq:weak_mixture_energy_I}
	\tweak{\lvarphi_{\rho_\alpha}}{\partial_t \rho_\alpha} &= \int_\Omega \left(\rho_\alpha\mathbf{v} + \mathbf{J}\indi{_\alpha}\right)\cdot \nabla \lvarphi_{\rho_\alpha}  + \tau\indi{_\alpha}\lvarphi_{\rho_\alpha}\, \dx
	-\! \int_{\partial \Omega} \mathbf{n}\cdot\left(\rho_\alpha\mathbf{v} 
	+ \mathbf{J}\indi{_\alpha}\right) \lvarphi_{\rho_\alpha}  \, \dS,\\
	\label{eq:weak_mixture_energy_II}	
	\tweak{\lvarphi\indi{_{\mathbf{M}}}}{\!\partial_t\mathbf{M}} &= \int_\Omega(\mathbf{M}\otimes\mathbf{v})\dcont\nabla\lvarphi\indi{_{\mathbf{M}}}
	+\lvarphi\indi{_{\mathbf{M}}}\cdot\!\nabla  \Big(\!\!-p+\sum_{\alpha=1}^{\nu}\mathbb{L}\indi{_\alpha}\mu\indi{_\alpha}\Big)
	+\mathbf{S}\dcont\nabla\lvarphi\indi{_{\mathbf{M}}} \, \dx\\
	& - \int_{\partial \Omega}\mathbf{n}\cdot\left(\mathbf{M}\otimes\mathbf{v}-\mathbf{S}\right)\cdot\lvarphi\indi{_{\mathbf{M}}}\, \dS,\notag\\
	\label{eq:weak_mixture_energy_III}
	\tweak{\lvarphi\indi{_s}}{\partial_t s}&=\begin{aligned}[t]\int_\Omega &s\mathbf{v} \cdot \nabla \lvarphi\indi{_s} + \mathbf{q}\cdot \nabla \left(\frac{\varphi\indi{_s}}{T}\right)+ \mathbf{S}\dcont \nabla\mathbf{v} \left(\frac{\varphi\indi{_s}}{T}\right)\\ 
	& - \sum_{\alpha=1}^{\nu}\left(\mathbf{J}\indi{_\alpha}\cdot\nabla\left(\frac{\mu_\alpha\varphi\indi{_s}}{T}\right) + \Big(\tau_\alpha - \mathbb{L}\indi{_\alpha}\diver(\mathbf{v})\Big)\left(\frac{\mu_\alpha\varphi\indi{_s}}{T}\right)\right)\dx\end{aligned}\\
	& - \int_{\partial \Omega}\mathbf{n}\cdot\Big(s\mathbf{v}\varphi\indi{_s} + \left(\frac{\varphi\indi{_s}}{T}\right)\Big[\mathbf{q} - \sum_{\alpha=1}^{\nu}\mathbf{J}\indi{_\alpha}\mu\indi{_\alpha}\Big]\Big)\, \dS,\notag 
	\end{alignat}
\end{subequations}
with $\alpha=1,\ldots,\nu$. Equation~\eqref{eq:weak_mixture_energy_I} corresponds to~\eqref{ba:partmass} as well as~\eqref{eq:weak_mixture_energy_II} to~\eqref{ba:mixmom} and~\eqref{eq:weak_mixture_energy_III} to~\eqref{ba:entdens}. Note that in~\eqref{eq:weak_mixture_energy_III} the identity relations \eqref{id:1} and \eqref{id:2} have been used. We require that~\eqref{eq:weak_mixture_energy} is satisfied almost everywhere on $\mathbb{I}$, where the time derivative of $\mathbf{z}$ is understood in the weak sense, i.e., $\dot{\mathbf{z}}(t) \in \mc{D}_\zhet^\ast$ for almost every $t\in \mathbb{I}$ and $\|\dot{\mathbf{z}}\|_{\mc{D}_\zhet^\ast}$ is at least an element of $L_{\text{loc}}^{1}(\mathbb{I})$, see~\cite[Ch.~23.5]{Zei90}. Note that in general the weak formulation is the sum of all equations of~\eqref{eq:weak_mixture_energy}. But, since the test function $\boldsymbol{\varphi}$ can be chosen arbitrarily, one can vary one entry of $\boldsymbol{\varphi}$ while setting the other ones to zero.

Since $C^\infty(\Omega) \cap W^{1,3}(\Omega)$ is dense in $W^{1,3}(\Omega)$, see~\cite[Th.~3.17]{Ada75}, system~\eqref{eq:weak_mixture_energy} is also satisfied for arbitrary $\boldsymbol{\varphi}\in W^{1,3}(\Omega)^{\nu + 4}$ if the coefficient functions behave well in $\zhet$. Therefore, one can reinterpret the weak formulation as an operator equation stated in the dual space of $\mc{D}_{\mathbf{z}}$ tested with an arbitrary $\boldsymbol{\varphi} \in \mc{D}_{\mathbf{z}}$.  For further details we refer to~\cite[Ch.~23.1]{Zei90}.

The aim is now to develop GENERIC formulations~\eqref{eq:operator_equation_isolated} and~\eqref{eq:operator_equation_open} such that among other things they encode the weak formulation \eqref{eq:weak_mixture_energy} for an open and isolated system, respectively. 
However, in both cases we need an energy functional~$H$ for the Hamiltonian part and a total entropy functional~$S$ for the dissipative part. Note that the internal energy~$u$ is here a thermodynamic potential and the entropy~$s$ a state variable. For the case that the roles of $u$ and $s$ are interchanged we refer to Section~\ref{sec:mixture_entropy}.

We choose $H$ as the physical energy function over the domain~$\Omega$ of a fluid mixture given by the sum of kinetic and internal energy, 
\begin{equation}
\label{eq:total_energy_mult_u}
H(\mathbf{z}) = \int_\Omega \frac{\mathbf{M}(x)\cdot\mathbf{M}(x)}{2 \sum_{\alpha =1}^\nu \rho_\alpha(x)}\, \dx+ \int_\Omega u(\rho_1,\ldots,\rho_\nu, s)(x) \, \dx,
\end{equation}
and the total entropy functional $S$ is chosen as spatial integral of the entropy 
\begin{equation}
\label{eq:entropy_mult_u}
S(\mathbf{z}) = \int_\Omega s(x)\, \dx.
\end{equation}
The functional derivative of $H$ and $S$ can be calculated with the help of~\eqref{eq:partInt} such that
\begin{equation}
\label{eq:vardif_HS_mult_u}
\vardif{H}{\zhet} = \begin{bmatrix}
	-\frac{\mathbf{v}\cdot\mathbf{v}}{2}   + \mu_1 & \ldots & -\frac{\mathbf{v}\cdot\mathbf{v}}{2} + \mu_\nu& \mathbf{v}^\top  & T 
\end{bmatrix}^\top\!\! \text{ and }~ \vardif{S}{\zhet} = \begin{bmatrix}
	0 & \ldots & 0  & \,\,\mathbf{0}^\top\!\! & 1
\end{bmatrix}^\top,
\end{equation}
where we again have used $\mathbf{v} = \mathbf{M}/ \sum_{\alpha=1}^\nu \rho_\alpha$. 

\subsection{Isolated Systems}\label{sec:mixture_isolated}
We consider at first the case of an isolated system such that there is no interaction between the system and its environment, i.e., convective and non-convective fluxes through the boundary~$\partial \Omega$ are assumed to be zero. In this situation, the normal components of the barycentric velocity~$\mathbf{v}$, the non-convective heat flux~$\mathbf{q}$, the diffusion fluxes~$\mathbf{J}_\alpha$, and the viscosity part of the stress tensor~$\mathbf{S}$ have to vanish at the boundary.  Note that if $\mathbf{v}\cdot \mathbf{n}$ vanishes at the boundary, then also the product of $\mathbf{v}$ and an arbitrary $W^{1,3}$-function, especially $\mathbf{M} = \sum_{\alpha=1}^\nu \rho_\alpha \mathbf{v}$. Consequently, we choose $\mathbf{M}\in \mathbold{W}^{1,3}_N(\Omega) := \{ \boldsymbol{\phi} \in W^{1,3}(\Omega)^3\,|\, \boldsymbol{\phi} \cdot \mathbf{n}|_{\partial \Omega} = 0\}$ such that the boundary condition for the mixture linear momentum is fulfilled automatically. As underlying space for the unknowns associated with~$\mathbf{z}$ where $\zhet(t)\in \mc{Z}$ almost every time we choose
\begin{equation}\label{eq:Dz_isolated}
\mc{D}_{\mathbf{ z}} := W^{1,3}(\Omega)^\nu \times \mathbold{W}^{1,3}_N(\Omega) \times W^{1,3}(\Omega).
\end{equation}
Note that under the condition of smooth data, $\tvardif{H}{\mathbf{z}}$ and $\tvardif{S}{\mathbf{z}}$ are also elements of $\mc{D}_{\mathbf{ z}}$.

For the description of the dynamics for an isolated system we have to define the linear operators $\mc{J}^{(E)}(\mathbf{ z})$ as well as $\mc{R}^{(E)}(\mathbf{ z})$. 
We assume that the coefficients appearing in the operators $\mc{J}^{(E)}(\zhet)$ and $\mc{R}^{(E)}(\zhet)$ behave well in $\zhet \in \mc{D}_\zhet$ such that both operators map continuously from $W^{1,3}(\Omega)^{\nu+4}$ in its dual space. Note that, $\mc{D}_\zhet$ is a subspace of $W^{1,3}(\Omega)^{\nu+4}$ and therefore  $\mc{J}^{(E)}(\zhet), \mc{R}^{(E)}(\zhet)$ are defined in a more general setting. The operator $\mc{J}^{(E)}(\zhet)$ associated to the conservative part is given by
\begin{equation}\label{eq:operator_J_multi_energy}
		\mathcal{J}^{(E)}(\zhet)= 
	\left[\begin{array}{ccc|cc}
		0 & \ldots & 0 & \mathcal{J}^{(E)}_{\rho_1,\mathbf{M}} & 0\\
		\vdots & \ddots & \vdots & \vdots & \vdots\\
		0 & \ldots & 0 & \mathcal{J}^{(E)}_{\rho_\nu,\mathbf{M}} & 0\\
		\hline
		\mathcal{J}^{(E)}_{\mathbf{M},\rho_1} & \ldots & \mathcal{J}^{(E)}_{\mathbf{M},\rho_\nu} & \mathcal{J}^{(E)}_{\mathbf{M},\mathbf{M}} & \mathcal{J}^{(E)}_{\mathbf{M},s}\\
		0 & \ldots & 0 & \mathcal{J}^{(E)}_{s,\mathbf{M}} & 0
		\end{array}\right],
\end{equation} 
where the components of $\mc{J}^{(E)}(\mathbf{z})$ are defined by
\begin{subequations}\label{eq:operator_J_multi_energy_parts}
	\begin{alignat}{3}
	\tweak{\varphi\indi{_{\rho_\alpha}}}{\mc{J}^{(E)}_{\rhoAlpha, \mathbf{M}} \psi\indi{_{\mathbf{M}}}} &=  -  \tweak{\psi\indi{_{\mathbf{M}}}}{\mc{J}^{(E)}_{\mathbf{M}, \rhoAlpha} \varphi\indi{_{\rhoAlpha}}} \\
	 &= \int_\Omega \rhoAlpha  (\psi\indi{_{\mathbf{M}}} \cdot\nabla) \varphi\indi{_{\rhoAlpha}}\!\! -  (\psi\indi{_{\mathbf{M}}} \cdot\nabla)(\varphi\indi{_\rhoAlpha}\mathbb{L}\indi{_\alpha})\, \dx,\notag\\
	\tweak{\varphi\indi{_{\mathbf{M}}}}{\mc{J}^{(E)}_{\mathbf{M} , \mathbf{M}} \psi\indi{_{\mathbf{M}}}} &=  -  \tweak{\psi\indi{_{\mathbf{M}}}}{\mc{J}^{(E)}_{\mathbf{M},\mathbf{M}} \varphi\indi{_{\mathbf{M}}}}\\
	&= \int_\Omega \mathbf{M} \cdot \left[(\psi_{\mathbf{M}}\cdot\nabla)\varphi_{\mathbf{M}} - (\varphi_{\mathbf{M}}\cdot\nabla)\psi\indi{_{\mathbf{M}}}\right]\dx,\notag\\
	\tweak{\varphi\indi{_s}}{\mc{J}^{(E)}_{s,\mathbf{M}}\psi\indi{_{\mathbf{M}}}} &=  -  \tweak{\psi\indi{_{\mathbf{M}}}}{\mc{J}^{(E)}_{\mathbf{M},s} \varphi\indi{_s}} = \int_\Omega s  (\psi\indi{_{\mathbf{M}}} \cdot\nabla) \varphi_{s}\, \dx,
	\end{alignat}
\end{subequations}
with $ \alpha = 1, \ldots ,\nu$ and $\boldsymbol{\varphi}, \boldsymbol{\psi} \in W^{1,3}(\Omega)^{\nu+4}$. The coefficient functions $\mathbb{L}_\alpha=\mathbb{L}_\alpha(\rho_1,\ldots,\rho_\nu,s)$, $\alpha=1,\ldots,\nu$, are defined in~\eqref{eq:newcoff} and are assumed to behave well in $\zhet$ such that $\mathbb{L}_\alpha \in W^{1,p}(\Omega)$, $p>1$, which would lead to $(\psi\indi{_{\mathbf{M}}} \cdot\nabla)(\varphi\indi{_\rhoAlpha}\mathbb{L}\indi{_\alpha}) \in L^q(\Omega)$ with $q>1$, cf.~\cite[Lem.~5.12]{Ada75}. To show that the Hamiltonian part can be described by $\mc{J}^{(E)}$, we prove its skew-adjointness and also the non-interacting condition in the following lemma.
\begin{lemma}\label{lem:mixture_J_energy}
The operator $\mc{J}^{(E)}(\mathbf{z})$ from~\eqref{eq:operator_J_multi_energy} and~\eqref{eq:operator_J_multi_energy_parts} is skew-adjoint on $W^{1,3}(\Omega)^{\nu+4}$. Furthermore, it satisfies the non-interacting condition
\begin{equation}
\label{eq:noninteracting_J_multi_energy}
	\mc{J}^{(E)}(\mathbf{z}) \mvardif{S}{\mathbf{z}}=0.
\end{equation}
\end{lemma}
\begin{proof}
Since the bilinear form associated with $\mc{J}^{(E)}(\mathbf{z})$ can be written as
\begin{align*}
&\tweak{\boldsymbol{\varphi}}{\mc{J}^{(E)} \boldsymbol{\psi}}=\\
\int_{\Omega}&
-\sum_{\alpha=1}^{\nu} \rho_\alpha \left[ \left( \varphi_{\mathbf{M}}\cdot\nabla\right)\psi_{\rho_\alpha} -\left(\psi_{\mathbf{M}}\cdot\nabla\right)\varphi_{\rho_\alpha}\right]\\ &+\sum_{\alpha=1}^{\nu}\left[\left(\varphi_{\mathbf{M}}\cdot\nabla\right)\left(\psi_{\rho_\alpha}\mathbb{L}_\alpha\right) -\left(\psi_{\mathbf{M}}\cdot\nabla\right)\left(\varphi_{\rho_\alpha}\mathbb{L}_\alpha\right)\right]\\
&-\mathbf{M}\cdot\left[\left(\varphi_{\mathbf{M}}\cdot\nabla\right)\psi_{\mathbf{M}} -\left(\psi_{\mathbf{M}}\cdot\nabla\right)\varphi_{\mathbf{M}}\right]
-s\left[\left(\varphi_{\mathbf{M}}\cdot\nabla\right)\psi_{s}-\left(\psi_{\mathbf{M}}\cdot\nabla\right)\varphi_{s}\right]\,\dx\notag,
\end{align*}
the first statement follows by Lemma~\ref{lem:equi_skew} and the second by the form of $\tvardif{S}{\mathbf{z}}$ given in~\eqref{eq:vardif_HS_mult_u}.
\end{proof}
The Hamiltonian part $\mc{J}^{(E)}(\mathbf{z})\tvardif{H}{\mathbf{z}}$ describes the reversible dynamics or in different words the lossless transformation (conservation) of energy, whereas the dissipative part $\mathcal{R}^{(E)}(\zhet)$ given by 
\begin{equation}\label{eq:operator_R_multi_energy}
\begin{split}
		\mathcal{R}^{(E)}(\mathbf{z})= 
	\left[\begin{array}{ccc|cc}
		\mathcal{R}^{(E)}_{\rho_1,\rho_1} & \ldots & \mathcal{R}^{(E)}_{\rho_1,\rho_\nu} & 0 & \mathcal{R}^{(E)}_{\rho_1,s}\\
		\vdots & \ddots & \vdots & \vdots & \vdots\\
		\mathcal{R}^{(E)}_{\rho_\nu,\rho_1} & \ldots & \mathcal{R}^{(E)}_{\rho_\nu,\rho_\nu} & 0 & \mathcal{R}^{(E)}_{\rho_\nu,s}\\[2pt]
		\hline
		0 & \ldots & 0 & \mathcal{R}^{(E)}_{\mathbf{M},\mathbf{M}} & \mathcal{R}^{(E)}_{\mathbf{M},s}\\[2pt]
		\mathcal{R}^{(E)}_{s,\rho_1} & \ldots & \mathcal{R}^{(E)}_{s,\rho_\nu} & \mathcal{R}^{(E)}_{s,\mathbf{M}} & \mathcal{R}^{(E)}_{s,s}
		\end{array}\right].
		\end{split}
\end{equation} 
reflects the irreversible dynamics. Here the non-zero components of $\mc{R}^{(E)}(\mathbf{z})$ are given by
\begin{subequations}\label{eq:operator_R_multi_energy_parts}
	\begin{align}
	&\tweak{\varphi\indi{_\rhoAlpha}}{\mathcal{R}^{(E)}_{\rhoAlpha,{\rho\indi{_\beta}}}\psi_{\rho\indi{_\beta}}}=	 \int_{\Omega}\begin{aligned}[t]
	T\, \mathbb{L}\indi{_\alpha_\beta} \varphi_{\rhoAlpha}
	\psi_{\rho\indi{_\beta}} + B\indi{_\alpha_\beta} \nabla \varphi_{\rhoAlpha}\cdot \nabla\psi_{\rho\indi{_\beta}} 
	\,\dx,\end{aligned}\\
	&\tweak{\varphi\indi{_\rhoAlpha}}{\mathcal{R}^{(E)}_{\rhoAlpha,s}\psi\indi{_s}}
	=\tweak{\psi_s }{\mathcal{R}^{(E)}_{s, \rhoAlpha} \varphi\indi{_{\rhoAlpha}}}\\
	&\begin{aligned}[t] \int_{\Omega} -\sum_{\beta=1}^{\nu} \mathbb{L}\indi{_\alpha_\beta} \mu\indi{_\beta} \varphi\indi{_\rhoAlpha}\psi\indi{_s} + \nabla \varphi\indi{_\rhoAlpha}\!\cdot \Big[B\indi{_\alpha} \nabla\Big(\frac{1}{T}\psi\indi{_s}\Big) -\sum_{\beta=1}^{\nu} B\indi{_\alpha_\beta} \nabla\Big(\frac{\mu_\beta}{T}\psi\indi{_s}\Big)\Big] \,\dx,\end{aligned}\notag\\
	&\tweak{\varphi_{\mathbf{M}}}{\mathcal{R}^{(E)}_{\mathbf{M},\mathbf{M}}\psi_{\mathbf{M}}}=\\
	& \int_{\Omega} \frac{\zeta T}{2}\!\trace\big[(\nabla \varphi_{\mathbf{M}} \!+\! \nabla \varphi_{\mathbf{M}}^\top)\cdot (\nabla \psi_{\mathbf{M}}\!+\! \nabla \psi_{\mathbf{M}}^\top)\big] \!+\! \Big(\lambda-\frac{2\zeta}{3}\Big)T \diver(\varphi_{\mathbf{M}}) \diver(\psi_{\mathbf{M}})\,\dx,\notag\\
	&\tweak{\varphi_{\mathbf{M}}}{\mathcal{R}^{(E)}_{\mathbf{M},s}\psi_{s}}=\tweak{\psi_{s}}{\mathcal{R}^{(E)}_{s,\mathbf{M}}\varphi_{\mathbf{M}}}=\\
	&\begin{aligned}[t]\int_{\Omega}
	-\Big(\zeta \trace\big[ (\nabla \varphi_{\mathbf{M}} + \nabla \varphi_{\mathbf{M}}^\top)\cdot \mathbf{D}\big] + \Big(\lambda-\frac{2\zeta}{3}\Big) \diver(\varphi_{\mathbf{M}}) \diver(\mathbf{v})\Big) \psi_{s}\,\dx,\end{aligned}\notag\\
	&\tweak{\varphi_{s}}{\mathcal{R}^{(E)}_{s,s}\psi_{s}}=\\
	&\int_{\Omega}
	\frac{1}{T}\Big(2\zeta \trace[\mathbf{D} \cdot \mathbf{D}]  + \Big(\lambda-\frac{2\zeta}{3}\Big) (\diver(\mathbf{v}))^2 + \sum_{\alpha,\beta=1}^{\nu}\mu_\alpha \mathbb{L}_{\alpha\beta}\mu_\beta\Big) \varphi_{s}\psi_{s} \notag \\
	&+ \nabla\Big(\frac{1}{T}\varphi_{s}\Big)\cdot \Big[ \kappa T^2 \nabla\Big(\frac{1}{T}\psi_{s}\Big)
	-\sum_{\beta=1}^{\nu} B_{\beta}\nabla\Big(\frac{\mu_\beta}{T}\psi_{s}\Big)\Big]\notag \\
	&-\sum_{\alpha=1}^{\nu}  \nabla\Big(\frac{\mu_\alpha}{T}\varphi_{s}\Big)\cdot \Big(B_\alpha  \nabla\Big(\frac{1}{T}\psi_{s}\Big)\Big)
	 +\sum_{\alpha,\beta=1}^{\nu} \nabla\Big(\frac{\mu_\alpha}{T}\varphi_{s}\Big)\cdot \Big( B_{\alpha\beta} \nabla\Big(\frac{\mu_\beta}{T}\psi_{s}\Big)\Big)\,\dx,\notag	
	\end{align}
\end{subequations}%
with $ \alpha, \beta = 1, \ldots ,\nu$ and $\boldsymbol{\varphi}, \boldsymbol{\psi} \in W^{1,3}(\Omega)^{\nu+4}$. The definition and explanation of the coefficient functions appearing in~\eqref{eq:operator_R_multi_energy_parts} can be found in Section~\ref{sec:TIP}. Again, we assume that these coefficient functions behave well in $\zhet$, such that the operator $\mc{R}^{(E)}(\zhet)$ is continuous, i.e. $\mathbb{L}_{\alpha \beta} \in L^{1+\varepsilon}(\Omega)$ and $\lambda$, $\kappa$, $\zeta$, $B_{\alpha}$, $B_{\alpha \beta} \in L^{3+\varepsilon}(\Omega)$ with an $\varepsilon > 0$, $\alpha, \beta = 1, \ldots, \nu$. Since the second law of thermodynamics has to be satisfied, the operator $\mc{R}^{(E)}$ has to be self-adjoint and semi-elliptic. Furthermore, as we show in the following lemma, it satisfies
\begin{equation}
\label{eq:noninteracting_R_multi_energy}
	\mc{R}^{(E)}(\mathbf{z}) \mvardif{H}{\mathbf{z}}=0.
\end{equation}
\begin{lemma}\label{lem:mixture_R_energy}
The operator $\mc{R}^{(E)}(\mathbf{z})$ from~\eqref{eq:operator_R_multi_energy} and~\eqref{eq:operator_R_multi_energy_parts} is self-adjoint and semi-elliptic in $W^{1,3}(\Omega)^{\nu+4}$. Further, the non-interacting condition~\eqref{eq:noninteracting_R_multi_energy} is satisfied.
\end{lemma}
\begin{proof}
By the  definition of $\mc{R}^{(E)}$  its associated bilinear form is given by
\begin{align}
\label{eq:mixture_dissipation}
&\tweak{\boldsymbol{\varphi}}{\mc{R}^{(E)}(\zhet)\boldsymbol{\psi}} =\\
\int_{\Omega}&  \frac{\zeta T}{2} \trace \Big[\Big(\nabla\varphi_{\mathbf{M}} + \nabla\varphi_{\mathbf{M}}^\top - \frac{1}{T}(\nabla \mathbf{v}  + \nabla \mathbf{v}^\top)\varphi_{s}\Big) \notag\\
& \qquad \qquad \qquad \qquad \qquad \qquad \qquad \cdot \Big(\nabla\psi_{\mathbf{M}} + \nabla\psi_{\mathbf{M}}^\top - \frac{1}{T}(\nabla \mathbf{v}  + \nabla \mathbf{v}^\top)\psi_{s}\Big)\Big] \notag \\
&+ T\Big(\lambda-\frac{2\zeta}{3}\Big) \Big(\diver(\varphi_{\mathbf{M}})- \frac{1}{T} \diver(\mathbf{v})\varphi_{s}\Big) \Big(\diver(\psi_{\mathbf{M}})- \frac{1}{T} \diver(\mathbf{v})\psi_{s}\Big) \notag \\
&+ \begin{bmatrix}
\nabla\big(\mfrac{1}{T}\varphi_{s}\big)\\
\nabla\big(\varphi_{\rho_1}-\mfrac{\mu_1}{T}\varphi_{s}\big)\\
\vdots\\
\nabla\big(\varphi_{\rho_\nu}-\mfrac{\mu_\nu}{T}\varphi_{s}\big)\\
\end{bmatrix}\! \odot\!
\left(\!
  \begin{bmatrix}
  \kappa T^2 & B\indi{_1} & \cdots & B\indi{_{\nu}}\\ 
  B\indi{_1} & B\indi{_1_{,1}} & \cdots & B\indi{_1_{,\nu}}\\
  \vdots & \vdots & \ddots & \vdots\\
  B\indi{_\nu} & B\indi{_{\nu}_{,1}} & \cdots  & B\indi{_{\nu}_{,\nu}}
  \end{bmatrix} 
\! \otimes_{\text{kron}} I_3 \!\right)\!\begin{bmatrix}
\nabla\big(\mfrac{1}{T}\psi_{s}\big)\\
\nabla\big(\psi_{\rho\indi{_1}}-\mfrac{\mu\indi{_1}}{T}\psi_{s}\big)\\
\vdots\\
\nabla\big(\psi_{\rho_\nu}-\mfrac{\mu_\nu}{T}\psi_{s}\big)\\
\end{bmatrix} \notag\\
&+\sum_{\alpha,\beta=1}^{\nu} T \Bigr(\varphi_{\rho_\alpha}-\frac{\mu_\alpha}{T}\varphi_{s}\Bigr)
\mathbb{L}_{\alpha \beta} \Bigr(\psi_{\rho_\beta}-\frac{\mu_\beta}{T}\psi_{s}\Bigr)\, \dx, \notag
\end{align}
where $\otimes_{\text{kron}}$ is the Kronecker product of two matrices, \cite[Def.~4.2.1]{HorJ91} and the term with $\odot$ should be read as 
\begin{align*}
&\nabla\Big(\frac{1}{T}\varphi_{s}\Big)\cdot \Big( \kappa T^2 \nabla\Big(\frac{1}{T}\psi_{s}\Big) + \sum_{\beta = 1}^\nu B_\beta \nabla\Big(\psi_{\rho\indi{_\beta}}-\frac{\mu\indi{_\beta}}{T}\psi_{s}\Big)\Big)\\
&\quad + \sum_{\alpha = 1}^\nu \nabla\Big(\psi_{\rho\indi{_\alpha}}-\frac{\mu\indi{_\alpha}}{T}\psi_{s}\Big)\cdot \Big( B_\alpha \nabla\Big(\frac{1}{T}\psi_{s}\Big) + \sum_{\beta = 1}^\nu B_{\alpha,\beta} \nabla\Big(\psi_{\rho\indi{_\beta}}-\frac{\mu\indi{_\beta}}{T}\psi_{s}\Big)\Big).
\end{align*}
The self-adjointness can  be seen directly by the symmetry of $B_{\alpha \beta}$ and $\mathbb{L}_{\alpha \beta}$, see p.~\pageref{eq:const}. For the semi-ellipticity we note that the matrices appearing in the third and fourth summand of the integrand are positive semi-definite,~\cite[p.~425~f.]{MeiR59}, and that $\lambda$ as well as $T$ are positive. For the parts 
with $\zeta \geq 0$ as prefactor we use that the trace is independent of the choice of basis. Therefore, we may choose Cartesian coordinates. Then the terms with~$\zeta$ are in sum non-negative, since by the Cauchy-Schwarz inequality it holds that $|\sum_{i=1}^3 A_{ii}|\leq \sqrt{\sum_{i=1}^3 1}\sqrt{\sum_{i=1}^3 A^2_{ii}}$ for every $A\in \R^{3\times 3}$, and therefore,
\begin{align*}
&\frac{1}{2}\text{trace}[(A+A^\top)^2]-\frac{2}{3}\big(\sum_{i=1}^3 A_{ii}\big)^2\\
 = & \frac{1}{2}\sum_{\substack{i,k=1\\i\neq k}}^3(A_{ik}+A_{ki})^2+2\sum_{i=1}^3 A_{ii}^2 -\frac{2}{3}\big(\sum_{i=1}^3 A_{ii}\big)^2 \geq \frac{1}{2}\sum_{\substack{i,k=1\\i\neq k}}^3(A_{ik}+A_{ki})^2.
\end{align*}
Hence, the integrand of~\eqref{eq:mixture_dissipation} is non-negative almost everywhere and thus $\tweak{\boldsymbol{\varphi}}{\mc{R}^{(E)}(\zhet)\boldsymbol{\varphi}}\geq 0$. The non-interaction condition~\eqref{eq:noninteracting_R_multi_energy} follows by
$\sum_{\beta=1}^\nu B_\beta = \sum_{\beta=1}^\nu B_{\alpha \beta}= \sum_{\beta=1}^\nu \mathbb{L}_{\alpha \beta}=0$ for $\alpha=1,\ldots,\nu$, see \cite[p.~427~f.]{MeiR59}, and a straight-forward calculation.
\end{proof}

We can now prove the main result for isolated systems of reactive fluid mixtures.

\begin{theorem}[Isolated system of fluid mixture]\label{th:mixture_isolated}
Let the vector of unknowns $\mathbf{z}(t) \in \mc{Z} \subset \mc{D}_{\mathbf{z}}$ be smooth enough
such that the functional derivatives~$\tvardif{H}{\mathbf{z}}$ and~$\tvardif{S}{\mathbf{z}}$ are elements of~$\mc{D}_{\mathbf{z}}$ for almost every time. Assume that the coefficients of the linear operator~$\mc{J}^{(E)}(\mathbf{z})$ and~$\mc{R}^{(E)}(\mathbf{z})$ given in~\eqref{eq:operator_J_multi_energy} and~\eqref{eq:operator_R_multi_energy}, respectively, behave well in $\zhet$, such that
$\mathbb{L}_\alpha \in W^{1,1+\varepsilon}(\Omega)$, $\mathbb{L}_{\alpha \beta} \in L^{1+\varepsilon}(\Omega)$, and $\lambda$, $\kappa$, $\zeta$, $B_{\alpha}$, $B_{\alpha \beta} \in L^{3+\varepsilon}(\Omega)$ with an $\varepsilon >0$ uniformly in time, $\alpha, \beta = 1, \ldots, \nu$.
Suppose that the barycentric velocity~$\mathbf{v}$, the non-convective heat flux~$\mathbf{q}$, the diffusion fluxes~$\mathbf{J}_\alpha$, and the viscosity part of the stress tensor $\mathbf{S}$ vanish at the boundary $\partial \Omega$ in normal direction.

Then the GENERIC formulation~\eqref{eq:operator_equation_isolated} encodes the weak formulation~\eqref{eq:weak_mixture_energy}, where
the operator $\mc{J}=\mc{J}^{(E)}$ is skew-adjoint and $\mc{R}=\mc{R}^{(E)}$ is self-adjoint and semi-elliptic. Furthermore, both non-interaction conditions~\eqref{eq:noninteracting_J_multi_energy} and~\eqref{eq:noninteracting_R_multi_energy} are satisfied, the system is energy preserving, and the second law of thermodynamics is fulfilled, i.e., 
$\frac{\mathrm{d}}{\mathrm{d}t} H(\mathbf{z}) = 0$ and $\frac{\mathrm{d}}{\mathrm{d}t} S(\mathbf{z}) \geq 0.$
\end{theorem}
\begin{proof}
Let us first consider the Hamiltonian part. Since $\sum_{\alpha=1}^\nu \L_\alpha =0$, see \cite[p.~427]{MeiR59}, as well as  $\sum_{\alpha=1}^\nu \rho_\alpha (\varphi_{\mathbf{M}}\cdot \nabla) \mfrac{\mathbf{v}\cdot\mathbf{v}}{2} = \sum_{\alpha=1}^\nu  \rho_\alpha\mathbf{v} \cdot (\varphi_{\mathbf{M}}\cdot \nabla )\mathbf{v}= \mathbf{M}\cdot(\varphi_{\mathbf{M}}\cdot\nabla)\mathbf{v}$, we get 
\begin{align*}
&\tweak{\boldsymbol{\varphi}}{\mc{J}^{(E)}\mvardif{H}{\mathbf{z}}}\\
=&\int_{\Omega}\begin{aligned}[t]
&-\sum_{\alpha=1}^{\nu}\rho_\alpha \left[ \left( \varphi_{\mathbf{M}} \cdot\nabla\right) \Big(\! -\frac{\mathbf{v}\cdot\mathbf{v}}{2}   + \mu_\alpha\Big)- \left(\mathbf{v} \cdot\nabla\right) \varphi_{\rho_\alpha}\right]\\
&+ \sum_{\alpha=1}^{\nu}( \varphi_{\mathbf{M}}\cdot\nabla) \Big(\Big(\!-\frac{\mathbf{v}\cdot\mathbf{v}}{2}   + \mu_\alpha\Big)\mathbb{L}_\alpha\Big) -\left( \mathbf{v}\cdot\nabla \right) \left( \varphi_{\rho_\alpha} \mathbb{L}_\alpha\right)\\
&-\mathbf{M}\cdot\left[\left(\varphi_{\mathbf{M}}\cdot\nabla\right)\mathbf{v}-\left(\mathbf{v}\cdot\nabla\right)\varphi_{\mathbf{M}}\right] -  s\left[\left(\varphi_{\mathbf{M}}\cdot\nabla\right)T-\left(\mathbf{v}\cdot\nabla\right)\varphi_{s}\right]\,\dx\end{aligned}\\
=&\int_{\Omega}\begin{aligned}[t] &\sum_{\alpha=1}^{\nu}\!-\rho_\alpha [ ( \varphi_{\mathbf{M}} \cdot\nabla)  \mu_\alpha- (\mathbf{v} \cdot\nabla) \varphi_{\rho_\alpha}] 
+ ( \varphi_{\mathbf{M}}\cdot\nabla ) ( \mu_\alpha \mathbb{L}_\alpha) -( \mathbf{v}\cdot\nabla ) ( \varphi_{\rho_\alpha} \mathbb{L}_\alpha) \\
& +\mathbf{M}\cdot(\mathbf{v}\cdot\nabla)\varphi_{\mathbf{M}} 
  -  s[(\varphi_{\mathbf{M}}\cdot\nabla)T -(\mathbf{v}\cdot\nabla)\varphi_{s}]\,\dx.\end{aligned}
\end{align*}
For the dissipation part we use the fact that $2\trace[\mathbf{A}\cdot(\mathbf{B}+\mathbf{B}^\top)]=\trace[(\mathbf{A}+\mathbf{A}^\top)\cdot(\mathbf{B}+\mathbf{B}^\top)]$, see~\eqref{eq:tensor_contraction_identity_relation}. With the definition of $\mathbf{T}^\mathrm{d}$, $\mathbf{q}$, $\mathbf{J}_\alpha$, $\tau_\alpha$, and $-\pi$ in~\eqref{eq:massprod} and~\eqref{eq:const},
we then have
\begin{align*}
&\tweak{\boldsymbol{\varphi}}{\mc{R}^{(E)}\mvardif{S}{\zhet}}\\
=&\int_{\Omega}
-\frac{\zeta}{2} \trace \left[\left(\nabla\varphi_{\mathbf{M}}  - \frac{1}{T}(\nabla\mathbf{v}) \varphi_{s}\right)\cdot (\nabla \mathbf{v}  + \nabla \mathbf{v}^\top)\right]\\
&\hphantom{\int_{\Omega}} -\Big(\lambda-\frac{2\zeta}{3}\Big) \left(\diver(\varphi_{\mathbf{M}})- \frac{1}{T} \diver(\mathbf{v})\varphi_{s}\right) \diver(\mathbf{v})\\
&\hphantom{\int_{\Omega}} + \begin{bmatrix}
\nabla\big(\mfrac{1}{T}\varphi_{s}\big)\\
\nabla\big(\varphi_{\rho\indi{_1}}-\mfrac{\mu\indi{_1}}{T}\varphi\indi{_s}\big)\\
\vdots\\
\nabla\big(\varphi_{\rho\indi{_\nu}}-\mfrac{\mu\indi{_\nu}}{T}\varphi\indi{_s}\big)\\
\end{bmatrix}\odot
\left(
\begin{bmatrix}
\kappa T^2 & B\indi{_1} & \cdots &  B\indi{_{\nu}}\\ 
B\indi{_1} & B\indi{_1_{,1}} & \cdots & B\indi{_1_{,\nu}}\\
\vdots & \vdots & \ddots &  \vdots\\
B\indi{_\nu} & B\indi{_{\nu}_{,1}} & \cdots  & B\indi{_{\nu}_{,\nu}}
\end{bmatrix} 
\otimes_{\text{kron}} I_3 \right) \begin{bmatrix}
\nabla\big(\mfrac{1}{T}\big)\\
\nabla\big(-\mfrac{\mu\indi{_1}}{T}\big)\\
\vdots\\
\nabla\big(-\mfrac{\mu\indi{_\nu}}{T}\big)\\
\end{bmatrix}\\
&\hphantom{\int_{\Omega}}-\sum_{\alpha,\beta=1}^{\nu} \Big(\varphi_{\rho_\alpha}-\frac{\mu_\alpha}{T}\varphi_{s}\Big)
\mathbb{L}_{\alpha \beta} \mu_\beta \dx\\
=&\int_{\Omega} \trace\Big[ (\mathbf{T}^\mathrm{d}-\pi \mathbf{I})\cdot\Big(\frac{\varphi_s}{T}\nabla\mathbf{v}\Big) 
- \mathbf{T}^\mathrm{d}\cdot \nabla \varphi_{\mathbf{M}}\Big] 
-\lambda\diver(\mathbf{v}) \diver(\varphi_{\mathbf{M}}) 
+  \mathbf{q} \cdot\nabla\Big(\frac{\varphi_{s}}{T}\Big)\\
&\hphantom{\int_{\Omega}} +\sum_{\alpha=1}^{\nu}\mathbf{J}\indi{_\alpha}\left(\nabla\varphi_{\rho_\alpha}-\nabla\Big(\frac{\mu_\alpha}{T}\varphi_{s}\Big)\right) +\sum_{\alpha=1}^{\nu} \tau_\alpha \frac{-\mu_\alpha}{T}\varphi_s 
-\sum_{\alpha,\beta=1}^{\nu}\varphi_{\rho_\alpha} \mathbb{L}\indi{_\alpha_\beta}\mu\indi{_\beta}\,\dx,
\end{align*}
where we add  the zero
$\sum_{\beta=1}^{\nu}\L_\beta \mu_\beta \diver(\mathbf{v})\frac{\varphi_s}{T} + \sum_{\alpha=1}^{\nu} \L_\alpha \diver(\mathbf{v}) \frac{-\mu_\alpha}{T} \varphi_s$ in the second equality.
If we use that $\mathbf{v}$ and $\varphi_{\mathbf{M}}$ vanish at the boundary in normal direction, by the boundary conditions  and by the choice of $\mc{D}_{\zhet}$, respectively, we get by partial integration that
\allowdisplaybreaks
\begin{align}\label{eq:mixture_isolated_help}
\int_{\Omega} &( \varphi_{\mathbf{M}}\cdot\nabla ) \Big( \sum_{\alpha=1}^{\nu}\mu_\alpha \mathbb{L}_\alpha\Big)-( \mathbf{v}\cdot\nabla ) \Big(\sum_{\alpha=1}^{\nu} \varphi_{\rho_\alpha} \mathbb{L}_\alpha\Big) -\lambda\diver \mathbf{v} \diver \varphi_{\mathbf{M}} \\
 & 
-\!\!\!\sum_{\alpha,\beta=1}^{\nu}\!\! \varphi_{\rho_\alpha} \mathbb{L}_{\alpha \beta} \mu_\beta\,\dx- \int_{\Omega} \sum_{\alpha=1}^{\nu} \tau_\alpha \varphi_{\rho_\alpha}  - \trace(-\pi \mathbf{I}\cdot \nabla \varphi_{\mathbf{M}})\,\dx\notag\\
=  \int_{\partial \Omega}& \sum_{\alpha=1}^{\nu}(\L_\alpha\mu_\alpha \varphi_{\mathbf{M}} - \varphi_{\rho_\alpha} \L_\alpha  \mathbf{v})\cdot\mathbf{n} \,\dS = 0.\notag
\end{align}
By these calculations, the operator equation tested with $\boldsymbol{\varphi}$ equals  the weak formulation~\eqref{eq:weak_mixture_energy} without the boundary terms. Since the boundary integrals are zero by the assumptions on $\mathbf{v}$, $\mathbf{q}$, $\mathbf{J}_\alpha$, and~$\mathbf{S}$, this shows the equality of the weak formulation and the operator equation for isolated systems.

The properties of $\mc{J}^{(E)}$ and $\mc{R}^{(E)}$ are proven in Lemma~\ref{lem:mixture_J_energy} and~\ref{lem:mixture_R_energy}, respectively. Since $\mc{J}^{(E)}(\mathbf{z})$ is skew-adjoint and $\mc{R}^{(E)}(\mathbf{z})$ is self-adjoint and semi-elliptic, it follows with the non-interaction conditions~\eqref{eq:noninteracting_J_multi_energy} and~\eqref{eq:noninteracting_R_multi_energy} that
\begin{gather*}
\mfrac{\mathrm{d}}{\mathrm{d}t} H = \weak{\vardif{H}{\mathbf{z}}}{\dot{\mathbf{z}}}= \weak{\vardif{H}{\mathbf{z}}}{\mc{J}^{(E)}(\zhet)\vardif{H}{\mathbf{z}}} = 0,\\
 \frac{\mathrm{d}}{\mathrm{d}t} S = \weak{\vardif{S}{\mathbf{z}}}{\dot{\mathbf{z}}}= \weak{\vardif{S}{\mathbf{z}}}{\mc{R}^{(E)}(\zhet)\vardif{S}{\mathbf{z}}} \geq 0. \qedhere
\end{gather*}
\end{proof}

\begin{remark}\normalfont
We point out that the term $\int_{\Omega} \sum_{\alpha=1}^{\nu} [( \varphi_{\mathbf{M}}\cdot\nabla)( \psi_{\rho_\alpha}\mathbb{L}_\alpha) - ( \psi_{\mathbf{M}}\cdot\nabla)(\varphi_{\rho_\alpha}\mathbb{L}_\alpha)]\,\dx$ from the Poisson bracket~\eqref{eq:operator_J_multi_energy} has neither an effect on the total energy nor on the entropy. Therefore, it could be shifted from the operator $\mc{J}^{(E)}$ to the dissipation operator $\mc{R}^{(E)}$ via the additional term
\begin{align*}
\tweak{\boldsymbol{\varphi}}{\mc{R}^{(E)}_{\text{add}} \boldsymbol{\psi}}=
\int_{\Omega} &T\sum_{\beta=1}^\mu \Big(\diver(\varphi_{\mathbf{M}})- \frac{1}{T} \diver(\mathbf{v})\varphi_{s}\Big)\mathbb{L}_{\beta} \Big(\psi_{\rho_\beta}-\frac{\mu_\beta}{T}\psi_{s}\Big)\\
-&T\sum_{\alpha=1}^{\nu} \Big(\varphi_{\rho_\alpha}-\frac{\mu_\alpha}{T}\varphi_{s}\Big)
\mathbb{L}_{\alpha} \Big(\diver(\psi_{\mathbf{M}})- \frac{1}{T} \diver(\mathbf{v})\psi_{s}\Big)\, \dx,
\end{align*}
which may seem natural since this term is connected to the reaction rate densities~$\Lambda^k$ and the dynamic pressure~$\left(-\pi\right)$ which are both thermodynamic fluxes. Note that this additional term vanishes if evaluated at~$\vardif{H}{\zhet}$, and if evaluated at~$\vardif{S}{\zhet}$ it gives $\int_{\Omega}\sum_{\alpha=1}^{\nu}[( \varphi_{\mathbf{M}}\cdot\nabla ) ( \mu_\alpha \mathbb{L}_\alpha) -( \mathbf{v}\cdot\nabla ) ( \varphi_{\rho_\alpha} \mathbb{L}_\alpha)]\dx$
under integration by parts. However, if one would add this term to~$\mc{R}^{(E)}$, then the operator $\tilde{\mc{R}}^{(E)}(\zhet):= \mc{R}^{(E)}(\zhet)+\mc{R}^{(E)}_{\text{add}}(\zhet)$ would be still semi-elliptic but no longer self-adjoint. Furthermore, at equilibrium, the new dissipation operator~$\tilde{\mc{R}}^{(E)}(\zhet)$ would couple the mass densities~$\rho_\alpha$ and the entropy density~$s$ with the momentum density~$\mathbf{M}$, which have different parities. This violates the Onsager-Casimir reciprocal relations, cf.~\cite[Ch.~II]{PavKG14}.
\end{remark}

\begin{remark}[Euler equations of fluid dynamics for reactive mixture] \normalfont
The operator equation~\eqref{eq:operator_equation_isolated} allows us to define an analogue of the Euler equations of fluid dynamics for reactive fluid mixtures, i.e., a description of reactive fluids with a constant total entropy, by neglecting the operator~$\mc{R}^{(E)}$. Using the equivalence of the operator equation and its weak formulation, and integration by parts we obtain under the assumption of a smooth solution 
\allowdisplaybreaks\\[2pt]
\textit{Hamiltonian part of full dynamics}
	\begin{align*}
		\hspace*{1cm}&\partial_t\rho\alp+\diver\left(\rho\indi{_\alpha}\mathbf{v}\right)=\L\indi{_\alpha}\diver\left(\mathbf{v}\right), \hspace*{1.1cm}(\alpha=1,\dots,\nu)\\
		&\partial_t\mathbf{M} + \diver\left(\mathbf{M}\otimes\mathbf{v}\right)		
		=-\nabla \Big(p -\sum_{\alpha=1}^{\nu}\L\indi{_\alpha}\mu\indi{_\alpha}\Big), \\
		&\partial_t s + \diver\!\left(s\mathbf{v}\right) =0.
		\end{align*}
Therein, we have used $\mathbf{v}= \mathbf{M}/(\sum_{\alpha=1}^\nu \rho_\alpha)$ and $\nabla p=s\nabla T+\sum_{\alpha=1}^{\nu}\rho\indi{_\alpha}\nabla\mu\indi{_\alpha}$ in combination with the Gibbs equation~\eqref{eq:Gibbs} and the thermodynamic constitutive equation~\eqref{cr:thermoeqpressure}.
\end{remark}
\subsection{Open Systems}\label{sec:mixture_open}
In Section~\ref{sec:mixture_isolated} we made the restrictions that the barycentric velocity~$\mathbf{v}$, the non-convective heat flux~$\mathbf{q}$, the diffusion fluxes~$\mathbf{J}_\alpha$, and the viscosity part of the stress tensor $\mathbf{S}$ as well as the linear momentum density $\mathbf{M}$ should vanish at the boundary in normal direction. These restrictions led us to the desired operator differential equation~\eqref{eq:operator_equation_isolated}. In the case of non-vanishing boundary terms, this kind of description is no longer valid, since the environment can for example influence the total energy. Therefore, we add an input-port variable $\mathbf{u}$ and two output-port variables  $\mathbf{y}_H$, $\mathbf{y}_S$ to describe the interaction of the system with its environment. 
For the space~$\mc{D}_\zhet$ we have to keep in mind that we now consider an open system. The necessary boundary conditions that the
linear momentum density~$\mathbf{M}$ had to satisfy in case of an isolated system is dropped in the following considerations of an open system.  Hence we now choose 
\begin{equation}\label{eq:Dz_open}
\mc{D}_\zhet:= W^{1,3}(\Omega)^{\nu+4}
\end{equation}
as the underlying space for the unknowns associated with $\zhet$ where $\zhet(t) \in \mc{Z} \subset \mc{D}_\zhet$ almost every time with $\mc{Z}$ given in~\eqref{eq:set_Z}.
The operators $\mc{J}^{(E)}$ and $\mc{R}^{(E)}$ are the same as in Section~\ref{sec:mixture_isolated}. We define the operator
$\mc{B}^{(E)}(\zhet)[\,\cdot\,]\colon \mc{D}_{\mathbf{u}} \to \mc{D}_\zhet^\ast$ via the pairing
\begin{align}\label{eq:operator_B_mixture}
&\tweak{\boldsymbol{\varphi}}{\mc{B}^{(E)}(\zhet)\mathbf{u}}:= \\
\int_{\partial \Omega} &\sum_{\alpha=1}^{\nu}\varphi_{\rho_\alpha}((\L_\alpha-\rho_\alpha) \mathrm{u}\indi{_2} - \mathrm{u}\indi{_{3+\alpha}}) + \varphi_{\mathbf{M}}\cdot(\mathrm{u}\indi{_{[\nu+4:\nu+6]}} - \mathbf{M}\mathrm{u}\indi{_2} - \mathbf{n} \mathrm{u}\indi{_1})\notag\\
 &- \varphi_s\Big(s\mathrm{u}\indi{_2} + \frac{1}{T}\mathrm{u}\indi{_3} -\sum\limits_{\alpha=1}^{\nu} \frac{\mu_\alpha}{T}\mathrm{u}\indi{_{3+\alpha}}\Big) \, \dS.\notag
\end{align}
For the continuity of $\mc{B}^{(E)}(\zhet)[\,\cdot\,]$ we have to find an appropriate space~$\mc{D}_{\mathbf{u}}$ for $\mathbf{u}$. We notice that the normal vector $\mathbf{n}$ is bounded almost everywhere on $\partial \Omega$ and that the restriction of an arbitrary $W^{1,3}(\Omega)$-function onto the boundary is an element of $L^q(\partial \Omega)$ for every $q \in [1,\infty)$, \cite[Ch.~2, Th.~4.6]{Nec12}. Furthermore, we assume again that the terms~$\mathbb{L}_\alpha$ behave well in $\zhet$ such that $\mathbb{L}_\alpha \in W^{1,1+\varepsilon}(\Omega)$ with $\varepsilon > 0$, see Section~\ref{sec:mixture_isolated}. It is well known, that then $\mathbb{L}_\alpha|_{\partial \Omega} \in L^{\tilde{p}}(\partial \Omega)$ for every $\tilde{p} \leq \frac{2 + 2\varepsilon}{2-\varepsilon}$, see \cite[Ch.~2, Th.~4.2]{Nec12}. Therefore, an appropriate space for $\mathbf{u}$ is given by
\begin{equation}
\label{eq:Du}
\mc{D}_{\mathbf{u}}  := 
L^{p_1}(\partial \Omega) \times 
L^{p_2}(\partial \Omega) \times 
L^{p_3}(\partial \Omega) \times 
L^{p_4}(\partial \Omega)^\nu \times 
L^{p_5}(\partial \Omega)^3, 
\end{equation}
where $p_2 > \frac{2+2\varepsilon}{3\varepsilon}$, $p_1,p_3,p_4,p_5 > 1$. In addition, $p_1$ is bounded by $p_1 < \frac{2 + 2\varepsilon}{2-\varepsilon}$ such that the later choice of $\mathrm{u}_1= \sum_{\beta=1}^{\nu}\L\indi{_\beta} \mu\indi{_\beta}|_{\partial \Omega}$ is well-defined.
The adjoint operator of $\mc{B}^{(E)}(\zhet)[\,\cdot\,]$  is again defined as the linear operator ${\mc{B}^{(E)}}^{\ast}(\zhet)[\,\cdot\,]\colon \mc{D}_\zhet \to \mc{D}_{\mathbf{u}}^\ast$ which fulfills $\tweak{\boldsymbol{\varphi}}{{\mc{B}^{(E)}}(\zhet) \mathbf{u}} = \tweak{ \mathbf{u}}{{\mc{B}^{(E)}}^{\ast}\!(\zhet)\boldsymbol{\varphi}}$. 
Note that, $\mc{D}_{\mathbf{u}}^\ast$ can be identified with
$$\mc{D}_{\mathbf{u}}^\ast \cong L^{q_1}(\partial \Omega) \times 
L^{q_2}(\partial \Omega) \times 
L^{q_3}(\partial \Omega) \times 
L^{q_4}(\partial \Omega)^\nu \times 
L^{q_5}(\partial \Omega)^3$$
where $q_i= \frac{p_i}{p_i-1}$, $i=1,\ldots,5$.
 
The following theorem
proves that the description of the weak formulation~\eqref{eq:weak_mixture_energy} in its operator setting can be written in the form of~\eqref{eq:operator_equation_open_I}.

\begin{theorem}[Open system of fluid mixture]\label{th:mixture_open}
Let the vector of unknowns $\zhet(t)\in \mc{Z} \subset \mc{D}_\zhet$ be smooth enough such that the functional derivatives~$\vardif{H}{\zhet}$ and~$\vardif{S}{\zhet}$ are elements of $\mc{D}_\zhet$ at almost every time point. Suppose that the coefficient function in the linear operators $\mc{J}=\mc{J}^{(E)}$,  $\mc{R}=\mc{R}^{(E)}$, and $\mc{B}=\mc{B}^{(E)}$ given in~\eqref{eq:operator_J_multi_energy}, \eqref{eq:operator_R_multi_energy}, and~\eqref{eq:operator_B_mixture} behave well in~$\zhet$, such that 
$\mathbb{L}_\alpha \in W^{1,1+\varepsilon}(\Omega)$, $\mathbb{L}_{\alpha \beta} \in L^{1+\varepsilon}(\Omega)$, and $\lambda$, $\kappa$, $\zeta$, $B_{\alpha}$, $B_{\alpha \beta} \in L^{3+\varepsilon}(\Omega)$ with an $\varepsilon >0$ uniformly in time, $\alpha, \beta = 1, \ldots, \nu$. 
Then the representation of the weak formulation~\eqref{eq:weak_mixture_energy} as operator equation is given by~\eqref{eq:operator_equation_open_I}
and the corresponding port~$\mathbf{u}$ is specified through
\begin{equation}\label{eq:u_mixture}
\mathbf{u}=\begin{bmatrix} {\textstyle\sum_{\beta=1}^{\nu}}\L\indi{_\beta} \mu\indi{_\beta}|_{\partial \Omega} &
\mathbf{v}|_{\partial \Omega}\cdot\mathbf{n}   & \mathbf{q}|_{\partial \Omega}\cdot \mathbf{n} &  \mathbf{J}_1|_{\partial \Omega}\cdot \mathbf{n} & \ldots & \mathbf{J}_\nu|_{\partial \Omega} \cdot \mathbf{n} & (\mathbf{S}|_{\partial \Omega}\cdot \mathbf{n})^\top
\end{bmatrix}^\top.
\end{equation}
The operator $\mc{J}^{(E)}(\mathbf{z})$ is skew-adjoint,  $\mc{R}^{(E)}(\mathbf{z})$ is self-adjoint and semi-elliptic, and satisfies the non-interaction conditions~\eqref{eq:noninteracting_J_multi_energy} and~\eqref{eq:noninteracting_R_multi_energy}, respectively.
\end{theorem}
\begin{proof}
By the proof of Theorem~\ref{th:mixture_isolated} it is enough to show that $\mc{B}^{(E)}(\mathbf{z})\mathbf{u}$ describes the boundary terms of the weak formulation~\eqref{eq:weak_mixture_energy} which are left out in $\mc{J}^{(E)}\vardif{H}{\mathbf{z}}+ \mc{R}^{(E)}\vardif{S}{\mathbf{z}}$. Recall that the partial integration step at~\eqref{eq:mixture_isolated_help} resulted in the boundary term
$\int_{\partial \Omega} \sum_{\alpha=1}^{\nu}(\L_\alpha\mu_\alpha \varphi_{\mathbf{M}} - \varphi_{\rho_\alpha} \L_\alpha  \mathbf{v})\cdot\mathbf{n} \,\dS,$
 which were created by $\mc{J}^{(E)}\vardif{H}{\mathbf{z}}+ \mc{R}^{(E)}\vardif{S}{\mathbf{z}}$ or more accurately by $\mc{J}^{(E)}\vardif{H}{\mathbf{z}}$. In case of an isolated system these boundary terms vanish by assumption, whereas for an open system these boundary contributions do not vanish and have to be considered as well. Using definition~\eqref{eq:u_mixture} for the port~$\mathbf{u}$ in~\eqref{eq:operator_B_mixture} we obtain
\begin{equation}
\begin{split}
&\tweak{\boldsymbol{\varphi}}{\mc{B}^{(E)}(\mathbf{z})\mathbf{u}} + \int_{\partial \Omega} \sum_{\alpha=1}^{\nu}(\L_\alpha\mu_\alpha \varphi_{\mathbf{M}}\cdot\mathbf{n} - \varphi_{\rho_\alpha} \L_\alpha  \mathbf{v}\cdot\mathbf{n}) \,\dS \\
		=& \int_{\partial \Omega}\Big[-\sum_{\alpha=1}^{\nu}\varphi_{\rho_\alpha}(\rho_\alpha \mathbf{v} + \mathbf{J}_\alpha ) + \varphi_{\mathbf{M}}\cdot\mathbf{S}\\
		&\qquad \qquad \qquad \qquad - (\varphi_{\mathbf{M}}\cdot\mathbf{M})\mathbf{v} - \varphi_s\Big(s\mathbf{v} + \frac{1}{T}\mathbf{q} - \sum_{\alpha=1}^{\nu}\frac{\mu_\alpha}{T}\mathbf{J}_\alpha \Big)\Big]\cdot \mathbf{n}\, \dS.  \qedhere
\end{split}
\end{equation}	
\end{proof}
For an isolated thermodynamic system whose state space representation in the operator setting is given by~\eqref{eq:operator_equation_isolated}, the total energy is conserved and the entropy can only increase as shown in Theorem~\ref{th:mixture_isolated} in accordance with the second law of thermodynamics. 
This corresponds to the properties of the GENERIC formalism for isolated systems as stated in Section~\ref{sec:GENERIC}. Since for open thermodynamic systems convective and non-convective transport over the boundary have to be taken into account, the statements related to the conservation of energy and the production of entropy for isolated systems have to be replaced by the more general statements. These are motivated by the conservation laws of continuum mechanics and thermodynamics, in particular by the first and second law of thermodynamics written for open systems in the sense of~\eqref{eq:energyintbal2} and~\eqref{ineq:entbal2}, respectively. 
Expressed with the total energy given by the Hamiltonian $H$ and the entropy functional $S$ with respect to the time-independent domain $\Omega$ considered in this section, these two balance laws become \\[10pt]
\begin{subequations}\textit{Energy conservation} 
	\begin{alignat}{4}
	\label{HS:1}
	\frac{\mathrm{d}}{\mathrm{d}t} H(\mathbf{z}) &=
	 -\int_{\partial \Omega}\,\Big[\Big(\frac{\mathbf{M}\cdot\mathbf{v}}{2}+ u\Big)\mathbf{v}
	+\Big(\mathbf{q}-\mathbf{T}^\top\!\cdot\mathbf{v}\Big)\Big]\cdot\mathbf{n}\,\dS,\\
	\intertext{\textit{Entropy Inequality}}
	\label{HS:2}
	\frac{\mathrm{d}}{\mathrm{d}t} S(\mathbf{z}) &\geq
	-\int_{\partial \Omega}\Big[s\mathbf{v}+\frac{1}{T}\Big(\mathbf{q}-\sum\limits_{\alpha=1}^{\nu}\mathbf{J}\alp\mu\alp\Big)\Big]\cdot\mathbf{n}\,\dS\,,
	\end{alignat}
\end{subequations}
with  $\mathbf{v}= \mathbf{M}/(\sum_{\alpha=1}^\nu \rho_\alpha)$. 

These two important properties are encoded in the bracket formulation induced by the system of operator equations~\eqref{eq:operator_equation_open}. The interaction of the system with its environment is described through ports, given by the~\emph{combined input port} $\mathbf{u}$ and the~\emph{output ports} $\mathbf{y}_H$ and $\mathbf{y}_S$ of the operator equation. The pairing between combined input port and output port $\mathbf{y}_H$ represents the interaction of the system with the environment related to the change of the Hamiltonian, whereas the pairing between the combined input port and output port $\mathbf{y}_S$ describes the interaction of the system with the environment related to the change of the total entropy.
Since the mathematical expression for the change of the total energy is obtained when the time-evolution equation~\eqref{eq:GENERIC_open} is evaluated with the Hamiltonian $H$, the change of the total energy should only depend on the combined input port~$\mathbf{u}$ and~$\mathbf{y}_H$. Analogously, the change of the total entropy is obtained if the time-evolution equation is evaluated with the entropy functional $S$. Due to the considerations made in Section~\ref{sec:FirstSecond}, we expect the change of the entropy $S$ to be bounded by the duality pairing of combined input port~$\mathbf{u}$ and~$\mathbf{y}_S$. Hence the pairing of the combined input $\mathbf{u}$ and the output ports $\mathbf{y}_H$ (w.r.t. the Hamiltonian part) and $\mathbf{y}_S$ (w.r.t. the entropic part) should result in an expression that is in accordance with the first and the second law of thermodynamics, respectively. More precisely, the result of the pairing of combined input $\mathbf{u}$ and the output port  $\mathbf{y}_H$ and $\mathbf{y}_S$ should have the form of the right-hand side of~\eqref{HS:1} and~\eqref{HS:2}, respectively.

\begin{corollary}\label{cor:mixture_isolated}
Under the assumptions of Theorem~\ref{th:mixture_open}, the total energy~$H$ and total entropy~$S$ satisfy the balance relation
\[\mfrac{\mathrm{d}}{\mathrm{d}t} H(\mathbf{z}) = \tweak{\mathbf{y}_H}{\mathbf{u}} \quad \text{ and } \quad \mfrac{\mathrm{d}}{\mathrm{d}t} S(\mathbf{z}) \geq \tweak{\mathbf{y}_S}{\mathbf{u}}.\]
\end{corollary}
\begin{proof}
Since $\mc{J}^{(E)}$ is skew-adjoint and the non-interaction condition~\eqref{eq:noninteracting_J_multi_energy} is satisfied, it holds that
\begin{align*}
\frac{\mathrm{d}H}{\mathrm{d}t} = \weak{\vardif{H}{\zhet}}{\dot{\zhet}} =& \weak{\vardif{H}{\zhet}}{\mc{J}^{(E)}(\zhet)\vardif{H}{\mathbf{z}}+\mc{R}^{(E)}(\zhet)\vardif{S}{\mathbf{z}}+ \mc{B}^{(E)}(\zhet)\mathbf{u}}\\
 =& \weak{\vardif{H}{\zhet}}{\mc{B}^{(E)}(\zhet)\mathbf{u}}= \weak{{\mc{B}^{(E)}}^\ast(\zhet) \vardif{H}{\zhet}}{\mathbf{u}}= \tweak{\mathbf{y}_H}{\mathbf{u}}.
\end{align*}
The proof for the time evolution of the total entropy follows analogously by the semi-ellipticness of~$\mc{R}^{(E)}$ and non-interaction condition~\eqref{eq:noninteracting_R_multi_energy}.
\end{proof}

The ports $\mathbf{y}_H$ and $\mathbf{y}_S$ from Corollary~\ref{cor:mixture_isolated} can be calculated explicitly by the equations~\eqref{eq:operator_equation_open_II} and~\eqref{eq:operator_equation_open_III}. With the functional derivatives~\eqref{eq:vardif_HS_mult_u} we get 
\begin{equation}\label{eq:expression_yH_mixture}
\mathbf{y}_H=\begin{bmatrix} - \mathbf{v}|_{\partial \Omega}\cdot\mathbf{n}   \\
\Big({\textstyle\sum_{\alpha=1}^{\nu}}\mu\indi{_\alpha}( \L\indi{_\alpha} - \rho\indi{_\alpha}) - \mfrac{\mathbf{M}\cdot \mathbf{v}}{2} - Ts\Big)\Big|_{\partial \Omega} \\ -1 \\ \mfrac{\mathbf{v}\cdot \mathbf{v}}{2}\Big|_{\partial \Omega} \\ \vdots \\ \mfrac{\mathbf{v}\cdot \mathbf{v}}{2}\Big|_{\partial \Omega} \\ \mathbf{v}|_{\partial \Omega}
\end{bmatrix}.
\end{equation}
We mention that $\mathbf{y}_H$ is the conjugate variable of $\mathbf{u}$, i.e., the
pairing $\tweak{\mathbf{y}_H}{\mathbf{u}}$ between these two has the physical dimension of power. The associated energy balance has the form 
\begin{align}\label{eq:energy_balance_mixture}
\mfrac{\mathrm{d}}{\mathrm{d}t} H
 =&
 \tweak{\mathbf{y}_H}{\mathbf{u}}\\
=& -\int_{\partial \Omega} \Big[\Big(\sum_{\beta=1}^{\nu}\L_\beta \mu_\beta - \sum_{\alpha=1}^{\nu}(\L_\alpha \mu_\alpha + \mu_\alpha \rho_\alpha) + \frac{\mathbf{M}\cdot \mathbf{v}}{2} + Ts\Big)\mathbf{v}\notag\\
 &\qquad \qquad \qquad \qquad \qquad \qquad  \qquad + \mathbf{q} - \sum_{\alpha=1}^\nu \mathbf{J}_\alpha  \frac{\mathbf{v}\cdot \mathbf{v}}{2} -  \mathbf{v}\cdot \mathbf{S}\Big]\cdot \mathbf{n} \, \dx\notag\\
=&  -\int_{\partial \Omega} \Big[\Big(\frac{\mathbf{M}\cdot \mathbf{v}}{2} + u\Big)\mathbf{v} + \mathbf{q}  -  \mathbf{v}\cdot (\mathbf{S}-p\mathbf{I})\Big]\cdot \mathbf{n}\, \dx, \notag
\end{align}
where we have used equation~\eqref{cr:thermoeqpressure} and $\sum_{\alpha=1}^\nu \mathbf{J}_\alpha =0$, which follows from $\sum_{\alpha=1}^\nu B_\alpha = \sum_{\alpha=1}^\nu B_{\alpha \beta} =0$. Taking into account that the influx of internal energy is neglected in our operator formulation, the mixture related energy balance~\eqref{eq:energy_balance_mixture} has the form of the integral total energy balance~\eqref{HS:1} for open systems, cf.~\cite[p.~6]{Oet06}.
For the change of the total entropy we get as port
\begin{equation}\label{eq:expression_yS_mixture}
\mathbf{y}_S=\begin{bmatrix} 0 &
-s|_{\partial \Omega} & -\frac{1}{T}|_{\partial \Omega} & \frac{\mu_1}{T}|_{\partial \Omega} & \ldots & \frac{\mu_\nu}{T}|_{\partial \Omega} & \boldsymbol{0}^\top
\end{bmatrix}^\top.
\end{equation}
Note that  the pairing $\tweak{\mathbf{y}_S}{\mathbf{u}}$ has the physical unit of Joule per Kelvin second and it gives the correct lower bound for the temporal change of the entropy, cf.~\cite[p.~6]{Oet06}, by 
\begin{align*}
\mfrac{\mathrm{d}}{\mathrm{d}t} S 
 &\geq  \tweak{\mathbf{y}_S}{\mathbf{u}}  = - \int_{\partial \Omega} \Big[s\mathbf{v} + \frac{1}{T}\Big(\mathbf{q} - \sum_{\alpha=1}^{\nu}\mathbf{J}\indi{_\alpha}\mu\indi{_\alpha}\Big)\Big]\cdot \mathbf{n} \, \dS,
\end{align*}
which is in accordance to~\eqref{HS:2}.
\subsection{Formulation with Entropy as Thermodynamic Potential} \label{sec:mixture_entropy}

In the previous sections~\ref{sec:mixture_isolated} and~\ref{sec:mixture_open} we have investigated the case where the entropy density~$s$ is an independent state variable and energy is the thermodynamic potential. In this subsection we consider the case where the internal energy density~$u$ is amongst the independent state variables associated with $\zhet$ and where entropy constitutes the thermodynamic potential. The state evolution equations given as a collection of field equations describing time evolution and spatial distribution of the fields associated with the state $\zhet$ through
\begin{align}\label{stateVar:Entpot}
	\zhet = \begin{bmatrix}
	\rho_1 & \ldots & \rho_\nu & \mathbf{M}^\top & u
	\end{bmatrix}^\top,
\end{align}
are summarized in~\eqref{eq:stateEvoEntPot}. Again, $\zhet$ maps from a bounded time interval~$\mathbb{I}$ into the open subset~$\mc{Z}$ of the space~$\mc{D}_\zhet$, where $\mc{Z}$ is defined in~\eqref{eq:set_Z} and~$\mc{D}_\zhet$ in~\eqref{eq:Dz_isolated} or~\eqref{eq:Dz_open} depending on whether an isolated or an open system is considered. The energy functional $H$ for this case is given by
\begin{equation}
\label{eq:total_energy_s}
H(\mathbf{z}) = \int_\Omega \frac{\mathbf{M}(x)\cdot\mathbf{M}(x)}{2 \sum_{\alpha =1}^\nu \rho_\alpha(x)}\, \dx+ \int_\Omega u(x) \, \dx,
\end{equation}
and the entropy functional has the form
\begin{equation}
\label{eq:entropy_mult_s}
S(\mathbf{z}) = \int_\Omega s(\rho_1,\dots,\rho_\nu,u)(x)\, \dx.
\end{equation}
Note that in this case entropy is the thermodynamic potential and the internal energy density~$u$ is chosen to be one of the independent functions associated with the state $\zhet$ through~\eqref{stateVar:Entpot}. Therefore the definition of the physical energy~\eqref{eq:total_energy_s} and the total entropy~\eqref{eq:entropy_mult_s} differ from the case before, see~\eqref{eq:total_energy_mult_u} and~\eqref{eq:entropy_mult_u}. The corresponding functional derivatives are given
\begin{equation}
\label{eq:vardif_HS_mult_s}
\vardif{H}{\zhet} = \begin{bmatrix}
	-\frac{\mathbf{v}\cdot\mathbf{v}}{2} & \ldots & -\frac{\mathbf{v}\cdot\mathbf{v}}{2}& \mathbf{v}^\top  & 1 
\end{bmatrix}^\top \text{ and }~~ \vardif{S}{\zhet} = \begin{bmatrix}
	-\frac{\mu_1}{T} & \ldots & -\frac{\mu_\nu}{T} & \mathbf{0}^\top   & \frac{1}{T}
\end{bmatrix}^\top,
\end{equation}
where we have used the relations in~\eqref{rel:entpot} to replace the directional derivatives appearing in~$\tvardif{S}{\zhet}$. The transformations required to obtain the operator formulation associated to the differential balance laws~\eqref{eq:stateEvoEntPot} are analogous to the steps performed in Section~\ref{sec:mixture_isolated} and~\ref{sec:mixture_open}. Therefore in the remainder of this section we skip the details in case they are analogous to the steps done in the subsections before.

Let $\zhet$ be smooth enough, then a weak formulation of~\eqref{eq:stateEvoEntPot} is given by
\begin{subequations}
\label{eq:weak_mixture_entropy}
\begin{alignat}{2}
\tweak{\varphi_{\rhoAlpha}}{\partial\indi{_t} \rhoAlpha} &= \int_\Omega (\rhoAlpha\mathbf{v} + \mathbf{J}\indi{_\alpha})\cdot \nabla \varphi_{\rhoAlpha}  + \tau\indi{_\alpha}\varphi_{\rhoAlpha}\, \dx- \int_{\partial \Omega} \!\mathbf{n}\cdot\left({\rhoAlpha}\mathbf{v} + \mathbf{J}\indi{_\alpha}\right) \varphi_{\rhoAlpha}\, \dS, \\
		\tweak{\lvarphi\indi{_{\mathbf{M}}}}{\partial\indi{_t}\mathbf{M}} &= \int_\Omega	(\mathbf{M}\otimes\mathbf{v})\dcont\nabla\lvarphi\indi{_{\mathbf{M}}}
+\lvarphi\indi{_{\mathbf{M}}}\cdot \nabla\Big(\!-p+\sum_{\alpha=1}^{\nu}\mathbb{L}\indi{_\alpha}\mu\indi{_\alpha}\Big)
+\mathbf{S}\dcont\nabla\lvarphi\indi{_{\mathbf{M}}} \, \dx\\
& - \int_{\partial \Omega}\mathbf{n}\cdot\big(\mathbf{M}\otimes\mathbf{v}-\mathbf{S}\big)\cdot\lvarphi\indi{_{\mathbf{M}}}\, \dS,\notag\\
\tweak{\varphi\indi{_u}}{\partial\indi{_t}u} &=\int_\Omega
(u\mathbf{v}+\mathbf{q})\cdot\nabla\varphi\indi{_u}
-\mathbf{v}\cdot\nabla\Big(\Big(-p+\sum_{\alpha=1}^{\nu}\mathbb{L}\indi{_\alpha}\mu\indi{_\alpha}\Big)\varphi\indi{_u}\Big)\\
&\quad -(\mathbf{S}\dcont\nabla\mathbf{v})\,\varphi\indi{_u}\,\dx-\int_{\partial\Omega}\mathbf{n} \cdot\Big(\Big(u+p-\sum_{\alpha=1}^{\nu}\mathbb{L}\indi{_\alpha}\mu\indi{_\alpha}\Big)\mathbf{v}+\mathbf{q}\Big)\varphi\indi{_u}\,\dS,\notag
\end{alignat}
\end{subequations}
 where $\boldsymbol{\varphi}$ is an arbitrary element of $W^{1,3}(\Omega)^{\nu+4}$. 

Again we are aiming for  a skew-adjoint operator $\mc{J}^{(S)}(\zhet)$ and a self-adjoint, semi-elliptic operator $\mc{R}^{(S)}(\zhet)$, both mapping from $W^{1,3}(\Omega)^{\nu+4}$ into is dual space, such that the weak formulation for an isolated system can be obtained by an operator equation of the form~\eqref{eq:operator_equation_isolated}. These operators should also satisfy non-interaction conditions equal to~\eqref{eq:noninteracting_J_multi_energy} and~\eqref{eq:noninteracting_R_multi_energy}. The operator~$\mc{J}^{(S)}(\mathbf{ z})$ is given by
\begin{equation}\label{eq:operator_J_multi_entropy}
\begin{split}
		\mathcal{J}^{(S)}(\mathbf{z})= 
	\left[\begin{array}{ccc|cc}
		0 & \ldots & 0 & \mathcal{J}^{(S)}_{\rho_1,\mathbf{M}} & 0\\
		\vdots & \ddots & \vdots & \vdots & \vdots\\
		0 & \ldots & 0 & \mathcal{J}^{(S)}_{\rho_\nu,\mathbf{M}} & 0\\
		\hline
		\mathcal{J}^{(S)}_{\mathbf{M},\rho_1} & \ldots & \mathcal{J}^{(S)}_{\mathbf{M},\rho_\nu} & \mathcal{J}^{(S)}_{\mathbf{M},\mathbf{M}} & \mathcal{J}^{(S)}_{\mathbf{M},u}\\
		0 & \ldots & 0 & \mathcal{J}^{(S)}_{u,\mathbf{M}} & 0
		\end{array}\right]
\end{split}
\end{equation}
where the single blocks in the operator are given by
\begin{align*}
\tweak{\varphi\indi{_{\rho_\alpha}}}{\mc{J}^{(S)}_{\rhoAlpha, \mathbf{M}} \psi\indi{_{\mathbf{M}}}} &=  -  \tweak{\psi\indi{_{\mathbf{M}}}}{\mc{J}^{(S)}_{\mathbf{M}, \rhoAlpha} \varphi\indi{_{\rhoAlpha}}}\\
&= \int_\Omega \rhoAlpha  (\psi\indi{_{\mathbf{M}}} \cdot\nabla) \varphi\indi{_{\rhoAlpha}}\!\! -  (\psi\indi{_{\mathbf{M}}} \cdot\nabla)(\varphi\indi{_\rhoAlpha}\mathbb{L}\indi{_\alpha})\, \dx,\\
	\tweak{\varphi\indi{_{\mathbf{M}}}}{\mc{J}^{(S)}_{\mathbf{M} , \mathbf{M}} \psi\indi{_{\mathbf{M}}}} &=  -  \tweak{\psi\indi{_{\mathbf{M}}}}{\mc{J}^{(S)}_{\mathbf{M},\mathbf{M}} \varphi\indi{_{\mathbf{M}}}}\\
	 &= \int_\Omega \mathbf{M} \cdot \left[(\psi_{\mathbf{M}}\cdot\nabla)\varphi_{\mathbf{M}} - (\varphi_{\mathbf{M}}\cdot\nabla)\psi\indi{_{\mathbf{M}}}\right]\dx,\\
	\tweak{\varphi\indi{_u}}{\mc{J}^{(S)}_{u,\mathbf{M}}\psi\indi{_{\mathbf{M}}}} &=  -  \tweak{\psi\indi{_{\mathbf{M}}}}{\mc{J}^{(S)}_{\mathbf{M},u} \varphi\indi{_u}} \\
	&= \int_\Omega u  (\psi\indi{_{\mathbf{M}}} \cdot\nabla) \varphi_{u}+ \left(\psi_{\mathbf{M}}\cdot\nabla\right)(\varphi_{u}(p-\textstyle\sum_{\beta=1}^{\nu} \mathbb{L}_\beta\mu_\beta))]\,\dx
\end{align*}
with $\boldsymbol{\varphi}, \boldsymbol{\psi} \in W^{1,3}(\Omega)^{\nu+4}$.
The dissipation operator  $\mc{R}^{(S)}(\mathbf{ z})$ is defined via

\begin{equation}\label{eq:operator_R_multi_entropy}
\begin{split}
		\mathcal{R}^{(S)}(\mathbf{ z})= 
	\left[\begin{array}{ccc|cc}
		\mathcal{R}^{(S)}_{\rho_1,\rho_1} & \ldots & \mathcal{R}^{(S)}_{\rho_1,\rho_\nu} & 0 & \mathcal{R}^{(S)}_{\rho_1,u}\\
		\vdots & \ddots & \vdots & \vdots & \vdots\\
		\mathcal{R}^{(S)}_{\rho_\nu,\rho_1} & \ldots & \mathcal{R}^{(S)}_{\rho_\nu,\rho_\nu} & 0 & \mathcal{R}^{(S)}_{\rho_\nu,u}\\
		\hline
		0 & \ldots & 0 & \mathcal{R}^{(S)}_{\mathbf{M},\mathbf{M}} & \mathcal{R}^{(S)}_{\mathbf{M},u}\\
		\mathcal{R}^{(S)}_{u,\rho_1} & \ldots & \mathcal{R}^{(S)}_{u,\rho_\nu} & \mathcal{R}^{(S)}_{u,\mathbf{M}} & \mathcal{R}^{(S)}_{u,u}
		\end{array}\right]
		\end{split}
\end{equation} 
\allowdisplaybreaks
	\begin{align*}
	&\tweak{\varphi\indi{_\rhoAlpha}}{\mathcal{R}^{(S)}_{\rhoAlpha,{\rho\indi{_\beta}}}\psi_{\rho\indi{_\beta}}}=	 \int_{\Omega}\begin{aligned}[t]
	\mathbb{L}\indi{_\alpha_\beta} \varphi_{\rhoAlpha}
	\psi_{\rho\indi{_\beta}} + B\indi{_\alpha_\beta} \nabla \varphi_{\rhoAlpha}\cdot \nabla\psi_{\rho\indi{_\beta}} 
	\,\dx,\end{aligned}\\
	&\tweak{\varphi\indi{_\rhoAlpha}}{\mathcal{R}^{(S)}_{\rhoAlpha,u}\psi\indi{_u}}
	=\tweak{\psi_u }{\mathcal{R}^{(S)}_{u, \rhoAlpha} \varphi\indi{_{\rhoAlpha}}}
	=\int_{\Omega} B\indi{_\alpha} \nabla \varphi\indi{_\rhoAlpha}\!\cdot \nabla\psi\indi{_u}\,\dx,\\
	&\tweak{\varphi_{\mathbf{M}}}{\mathcal{R}^{(S)}_{\mathbf{M},\mathbf{M}}\psi_{\mathbf{M}}}=\\
	\int_{\Omega}& \frac{\zeta T}{2}\! \trace\!\big[(\nabla \varphi_{\mathbf{M}} \!+\! \nabla \varphi_{\mathbf{M}}^\top)\cdot (\nabla \psi_{\mathbf{M}} + \nabla \psi_{\mathbf{M}}^\top)\big] + \Big(\!\lambda-\frac{2\zeta}{3}\Big)T \diver(\varphi_{\mathbf{M}}) \diver(\psi_{\mathbf{M}})\,\dx,\notag\\
	&\tweak{\varphi_{\mathbf{M}}}{\mathcal{R}^{(S)}_{\mathbf{M},u}\psi_{u}}
	=	\tweak{\psi_{u}}{\mathcal{R}^{(S)}_{u,\mathbf{M}}\varphi_{\mathbf{M}}}=\\
	\int_{\Omega}&
	-\Big(\zeta T \trace\big[ (\nabla \varphi_{\mathbf{M}} + \nabla \varphi_{\mathbf{M}}^\top)\cdot \mathbf{D}\big] + \Big(\lambda-\frac{2\zeta}{3}\Big)T \diver(\varphi_{\mathbf{M}}) \diver(\mathbf{v})\Big) \psi_{u}\,\dx,\\
	&\tweak{\varphi_{u}}{\mathcal{R}^{(S)}_{u,u}\psi_{u}}=\\
	\int_{\Omega}&
	\Big(2\zeta T  \trace[\mathbf{D} \cdot \mathbf{D}]  + \Big(\lambda-\frac{2\zeta}{3}\Big)T \diver(\mathbf{v})^2\Big) \varphi_{u}\psi_{u}+ \kappa T^2 \nabla(\varphi_{u})\cdot \nabla(\psi_{u})\, \dx.
	\end{align*}
\allowdisplaybreaks
We assume that the coefficient functions in the operators~$\mc{J}^{(S)}$ and~$\mc{R}^{(S)}$ behave well in $\zhet$, such that the operators define linear continuous mappings, i.e.  $\mathbb{L}_\alpha, p \in W^{1,1+\varepsilon}(\Omega)$, $\mathbb{L}_{\alpha \beta} \in L^{1+\varepsilon}(\Omega)$  and $\lambda$, $\kappa$, $\zeta$, $B_{\alpha}$, $B_{\alpha \beta} \in L^{3+\varepsilon}(\Omega)$ with $\varepsilon >0$, $\alpha, \beta = 1, \ldots, \nu$, cf.~Section~\ref{sec:mixture_isolated}. Then the next lemma shows that the operators defined in~\eqref{eq:operator_J_multi_entropy} and~\eqref{eq:operator_R_multi_entropy} satisfy the asserted properties. 
\begin{lemma}\label{lem:mixture_entropy}
Suppose the coefficients in the operator $\mc{J}^{(S)}(\zhet)$ and $\mc{R}^{(S)}(\zhet)$ defined via~\eqref{eq:operator_J_multi_entropy} and~\eqref{eq:operator_R_multi_entropy} behave well in $\zhet$. Then $\mc{J}^{(S)}(\zhet)$ is skew-adjoint and $\mc{R}^{(S)}(\zhet)$ is self-adjoint, semi-elliptic on $W^{1,3}(\Omega)^{\nu+4}$. Furthermore, $\mc{J}^{(S)}(\zhet)$ and $\mc{R}^{(S)}(\zhet)$ satisfy the non-interaction conditions
\begin{equation}
\label{eq:noninteracting_multi_entropy}
	\mc{J}^{(S)}(\mathbf{z}) \mvardif{S}{\mathbf{z}}=0 \qquad \text{and} \qquad	\mc{R}^{(S)}(\mathbf{z}) \mvardif{H}{\mathbf{z}}=0.
\end{equation}
\end{lemma}
\begin{proof}
The skew-adjointness of $\mc{J}^{(S)}$ and  the self-adjointness of $\mc{R}^{(S)}$ are obvious. The proof that $\mc{R}^{(S)}$ is semi-elliptic and that $\mc{R}^{(S)}(\zhet) \vardif{H}{\mathbf{z}}=0$ holds can be done analogously to the proof of Lemma~\ref{lem:mixture_R_energy}. Finally, the non-interaction condition of $\mc{J}^{(S)}$ follows by
\begin{equation*}
\begin{split}
&\weak{\boldsymbol{\varphi}}{\mc{J}^{(S)} \mvardif{S}{\zhet}}\\
\overset{\hphantom{\eqref{cr:thermoeqpressure}}}{=}
&\int_{\Omega} \!\varphi_{\mathbf{M}}\cdot \Big[\sum\limits_{\alpha=1}^{\nu} \rho_\alpha\nabla\frac{\mu_\alpha}{T}- \nabla\Big(\frac{\mu_\alpha}{T}\mathbb{L}_\alpha\Big) - u \nabla \frac{1}{T}  - \nabla\Big(\frac{1}{T}\Big(p-\sum\limits_{\beta=1}^{\nu}\mu_\beta \mathbb{L}_\beta\Big)\Big)\Big]\dx\\
\overset{\eqref{cr:thermoeqpressure}}{=}&\int_{\Omega} \!\varphi_{\mathbf{M}}\cdot \Big[-\sum\limits_{\alpha=1}^{\nu}\frac{\mu_\alpha}{T}\nabla\rho_\alpha +  \frac{1}{T}  \nabla u - \nabla s \Big]\,\dx\overset{\eqref{eq:Gibbs}}{=} 0.\\[-1.6em] \qedhere
\end{split}
\end{equation*}
\end{proof}
The proof that $\mc{J}^{(S)}$ and $\mc{R}^{(S)}$ describe the dynamics of an isolated system will be given later together with the proof for an open system. As we saw before, for the description of an open system ports $\mathbf{u}$, $\mathbf{y}_H$, and $\mathbf{y}_S$ have been introduced.  The linear operator $\mc{B}^{(S)}(\zhet)[\,\cdot\,]\colon \mc{D}_{\mathbf{u}} \to \mc{D}_\zhet^\ast$ which specifies the influence of $\mathbf{u}$ on the time evolution of the state~$\zhet$ is given by
\begin{align}\label{eq:operator_B_mixture_entropy}
\tweak{\boldsymbol{\varphi}}{\mc{B}^{(S)}\mathbf{u}}:=  \int_{\partial \Omega}&\sum\limits_{\alpha=1}^{\nu} \varphi_{\rho_\alpha}((\L_\alpha-\rho_\alpha) \mathrm{u}\indi{_2} - \mathrm{u}\indi{_{3+\alpha}})-\varphi_{\mathbf{M}}\cdot( \mathbf{M}\mathrm{u}\indi{_2} + \mathbf{n} \mathrm{u}\indi{_1})\\ &   + \varphi_{\mathbf{M}}\cdot\mathrm{u}\indi{_{[\nu+4:\nu+6]}} - \varphi_u\Big(\Big(u+p-\sum\limits_{\beta=1}^{\nu}\mu\indi{_\beta} \L\indi{_\beta}\Big) \mathrm{u}\indi{_2} + \mathrm{u}\indi{_3} \Big) \, \dS,\notag
\end{align}
where $\mc{D}_{\mathbf{u}}$ is defined as in~\eqref{eq:Du}. 
With the definition of $\mc{B}^{(S)}$, we can show the connection between the weak formulation~\eqref{eq:weak_mixture_entropy} and the operator equations.
\begin{theorem}[Systems with entropy as thermodynamic potential]
\label{th:mixture_entropy}
Let the vector of unknowns~$\zhet$ contain the internal energy~$u$ and suppose that exactly one of the following assumption is satisfied. 

\begin{enumerate}
\item The set $\mc{D}_{\mathbf{z}}$ is defined as in~\eqref{eq:Dz_isolated} and  the barycentric velocity~$\mathbf{v}$, the non-convective heat flux~$\mathbf{q}$, the diffusion fluxes~$\mathbf{J}_\alpha$, as well as viscosity stress tensor~$\mathbf{S}$ vanish at the boundary $\partial \Omega$ in normal direction. \label{item:isolated}
\item The set $\mc{D}_{\mathbf{z}}$ is defined as in~\eqref{eq:Dz_open}.\label{item:open}
\end{enumerate}
Let $\mathbf{z}(t) \in \mc{Z} \subset \mc{D}_{\mathbf{z}} $ be smooth enough such that the functional derivatives~$\vardif{H}{\mathbf{z}}$ and $\vardif{S}{\mathbf{z}}$ are elements of $\mc{D}_{\mathbf{z}}$ at almost every time point. Assume that the coefficients appearing in the linear operators $\mc{J}(\mathbf{z})=\mc{J}^{(S)}(\mathbf{z})$, $\mc{R}(\mathbf{z})=\mc{R}^{(S)}(\mathbf{z})$ behave well in $\zhet$ such that 
$\mathbb{L}_\alpha, p \in W^{1,1+\varepsilon}(\Omega)$, $\mathbb{L}_{\alpha \beta} \in L^{1+\varepsilon}(\Omega)$, and $\lambda$, $\kappa$, $\zeta$, $B_{\alpha}$, $B_{\alpha \beta} \in L^{3+\varepsilon}(\Omega)$ with an $\varepsilon >0$ uniformly in time, $\alpha, \beta = 1, \ldots, \nu$. 

If the system is isolated, i.e. assumption~\ref{item:isolated}. is satisfied, then the weak formulation~\eqref{eq:weak_mixture_entropy} in its operator formulation is given by~\eqref{eq:operator_equation_isolated}.
The system is energy preserving and the second law of thermodynamics is fulfilled, i.e., 
$\frac{\mathrm{d}}{\mathrm{d}t} H(\mathbf{z}) = 0$ and $\frac{\mathrm{d}}{\mathrm{d}t} S(\mathbf{z}) \geq 0.$

If the system is open, i.e. assumption~\ref{item:open}. is satisfied, then the operator equation~\eqref{eq:operator_equation_open_I}
encodes the weak formulation~\eqref{eq:weak_mixture_entropy}, where $\mathbf{u}$ is given by~\eqref{eq:u_mixture} and $\mc{B}=\mc{B}^{(S)}$. Furthermore, the time evolution of the total energy~$H$ and the total entropy~$S$ fulfill $\frac{\mathrm{d}}{\mathrm{d}t} H(\mathbf{z}) = \tweak{\mathbf{y}_H}{\mathbf{u}}$  and $\frac{\mathrm{d}}{\mathrm{d}t} S(\mathbf{z}) \geq \tweak{\mathbf{y}_S}{\mathbf{u}}$, where the ports are  $\mathbf{y}_{H} = {\mc{B}^{(S)}}^\ast(\mathbf{z}) \vardif{H}{\mathbf{z}}$ and $\mathbf{y}_{S} = {\mc{B}^{(S)}}^\ast(\mathbf{z}) \vardif{S}{\mathbf{z}}$.
\end{theorem}
\begin{proof}
The assertions can be proven following the steps of the proofs of Theorem~\ref{th:mixture_isolated},~\ref{th:mixture_open}, and Corollary~\ref{cor:mixture_isolated} under the consideration of
\begin{align*}
&\tweak{\boldsymbol{\varphi}}{\mc{J}^{(S)} \mvardif{H}{\zhet}}\\
=&\int_{\Omega}
\!\!\!\!\begin{aligned}[t]&\sum\limits_{\alpha=1}^{\nu}\rho_\alpha[ (\mathbf{v}\cdot\nabla)
+\mathbb{L}_\alpha \diver(\mathbf{v})] \varphi_{\rho_\alpha} + \mathbf{M}\cdot(\mathbf{v}\cdot\nabla)\varphi_{\mathbf{M}} - (\varphi_{\mathbf{M}}\cdot\nabla)p \\
 -&\sum\limits_{\beta=1}^{\nu}\diver(\varphi_{\mathbf{M}})\mu_\beta \mathbb{L}_\beta
 + u(\mathbf{v}\cdot\nabla)\varphi_{u} +(\mathbf{v}\cdot\nabla)(p\varphi_{u}) 
 +\sum\limits_{\beta=1}^{\nu}\diver (\mathbf{v}) \mu_\beta \mathbb{L}_\beta \varphi_{u} \,\dx\end{aligned}\\
+&\int_{\partial \Omega}
	\sum\limits_{\beta=1}^{\nu}[ (\varphi_{\mathbf{M}}\cdot\mathbf{n}) \mu_\beta \mathbb{L}_\beta 
	- (\mathbf{v}\cdot\mathbf{n})(\mu_\beta \mathbb{L}_\beta)\varphi_{u}]
	- \sum\limits_{\alpha=1}^{\nu}(\mathbf{v}\cdot \mathbf{n})(\varphi_{\rho_\alpha}\mathbb{L}_\alpha) \, \dS,
\end{align*}
where we have used integration by parts.
\end{proof}

As for the system with the entropy as state variable it is possible to express the ports~$\mathbf{y}_{H} = {\mc{B}^{(S)}}^\ast(\zhet) \vardif{H}{\mathbf{z}}$ and~$\mathbf{y}_{S} = {\mc{B}^{(S)}}^\ast(\zhet) \vardif{S}{\mathbf{z}}$ explicitly. This will again result in the expressions~\eqref{eq:expression_yH_mixture} and~\eqref{eq:expression_yS_mixture}. 
Therefore, the pairing $\tweak{\mathbf{y}_H}{\mathbf{u}}$ and $\tweak{\mathbf{y}_S}{\mathbf{u}}$ will be the same as in~\ref{sec:mixture_open} and thus equal to the change of the total energy and also justify the introduction of a lower bound for the change of entropy, respectively.

\subsection{GENERIC Formulation}\label{sec:mixture_GENERIC}
In the previous sections~\ref{sec:mixture_isolated}-\ref{sec:mixture_entropy} we have investigated the description of the dynamic of an reactive fluid mixture by the theory of operator equations. Depending on the choice of thermodynamic potential, we have defined a skew- and a self-adjoint operator~$\mc{J}$ and~$\mc{R}$ for an isolated system as well as an operator $\mc{B}$ describing the connection between the system and its environment. These operators are now used to define the Poisson and dissipation bracket as well as the boundary contribution of the GENERIC formalism~\eqref{eq:GENERIC_open}. Instead of test functions as in the operator formulation, the GENERIC formulation works with test observables or rather their functional derivatives. Therefore, the brackets are easy to define by formally replacing the test functions by functionals. This approach is well-defined if the functional derivatives are smooth enough, cf.~Section~\ref{sec:GENERIC}. However, in general, the resulting expressions are unbounded and should be understood in a distributional sense, see e.g. \cite{GrmOet97I,OetGrm97II}.
Starting with the case where energy constitutes the thermodynamic potential, we obtain the Poisson bracket 
\begin{equation}
\label{eq:poisson_bracket_mixture_energy}
\begin{aligned}
\{A,B\}^{(E)} =\int_{\Omega}
&-\sum_{\alpha=1}^{\nu}\rho_\alpha\Big[\Big(\vardif{A}{\mathbf{M}}\cdot\nabla\Big)\vardif{B}{\rho_\alpha}-\Big(\vardif{B}{\mathbf{M}}\cdot\nabla\Big)\vardif{A}{\rho_\alpha}\Big]\\
&+\sum_{\alpha=1}^{\nu}\Big[\Big(\vardif{A}{\mathbf{M}}\cdot\nabla\Big)\Big(\vardif{B}{\rho_\alpha}\mathbb{L}_\alpha\Big)-\Big(\vardif{B}{\mathbf{M}}\cdot\nabla\Big)\Big(\vardif{A}{\rho_\alpha}\mathbb{L}_\alpha\Big)\Big]\\
&- \mathbf{M}\cdot\Big[\Big(\vardif{A}{\mathbf{M}}\cdot\nabla\Big)\vardif{B}{\mathbf{M}}-\Big(\vardif{B}{\mathbf{M}}\cdot\nabla\Big)\vardif{A}{\mathbf{M}}\Big]\\
&- s\Big[\Big(\vardif{A}{\mathbf{M}}\cdot\nabla\Big)\vardif{B}{s}-\Big(\vardif{B}{\mathbf{M}}\cdot\nabla\Big)\vardif{A}{s}\Big]\dx,
\end{aligned}
\end{equation}
where the formal expression $\{A,B\}^{(E)}=\tweak{\tvardif{A}{\zhet}}{\mc{J}^{(E)} \tvardif{B}{\zhet}}$ has been used. 
For the case where entropy constitutes the thermodynamic potential and the internal energy density $u$ represents an independent state variable, we define
\allowdisplaybreaks
\begin{align}
\label{eq:poisson_bracket_mixture_entropy}
&\{A,B\}^{(S)} =\\
&\int_{\Omega}
-\sum_{\alpha=1}^{\nu}\rho_\alpha\Big[\Big(\vardif{A}{\mathbf{M}}\cdot\nabla\Big)\vardif{B}{\rho_\alpha}-\Big(\vardif{B}{\mathbf{M}}\cdot\nabla\Big)\vardif{A}{\rho_\alpha}\Big]\notag\\
&+\sum_{\alpha=1}^{\nu}\Big[\Big(\vardif{A}{\mathbf{M}}\cdot\nabla\Big)\Big(\vardif{B}{\rho_\alpha}\mathbb{L}_\alpha\Big)-\Big(\vardif{B}{\mathbf{M}}\cdot\nabla\Big)\Big(\vardif{A}{\rho_\alpha}\mathbb{L}_\alpha\Big)\Big]\notag\\
&- \mathbf{M}\cdot\Big[\!\Big(\vardif{A}{\mathbf{M}}\cdot\nabla\Big)\vardif{B}{\mathbf{M}}-\Big(\vardif{B}{\mathbf{M}}\cdot\nabla\Big)\vardif{A}{\mathbf{M}}\Big]
- u\Big[\!\Big(\vardif{A}{\mathbf{M}}\cdot\nabla\Big)\vardif{B}{u}-\Big(\vardif{B}{\mathbf{M}}\cdot\nabla\Big)\vardif{A}{u}\Big] \notag\\
&- \Big[\Big(\vardif{A}{\mathbf{M}}\cdot\nabla\Big)\Big(\vardif{B}{u}\big(p-\sum_{\beta=1}^{\nu}\mu_\beta \mathbb{L}_\beta\big)\Big)-\Big(\vardif{B}{\mathbf{M}}\cdot\nabla\Big)\Big(\vardif{A}{u}\big(p-\sum_{\beta=1}^{\nu}\mu_\beta \mathbb{L}_\beta\big)\Big)\Big]\dx.\notag
\end{align}
The Poisson brackets then satisfy the properties claimed in Section~\ref{sec:GENERIC}.
\begin{lemma}\label{lem:poisson_mixture}
Let the bracket~$\{\cdot,\cdot\}$ be defined as~\eqref{eq:poisson_bracket_mixture_energy} if the entropy density $s$ is a state variable and the internal energy~$u$ the thermodynamic potential. In case entropy constitutes the thermodynamic potential and internal energy an independent state variable, then let the bracket~$\{\cdot,\cdot\}$ be defined by~\eqref{eq:poisson_bracket_mixture_entropy}.\newline
Then~$\{\cdot,\cdot\}$ is anti-symmetric and satisfies the Leibniz rule and the Jacobi identity. Furthermore, the non-interaction condition
\[ \{A,S\} = 0 \quad \text{for all } A\in C^\infty(\mc{Z}) \]
is fulfilled.
\end{lemma}
\begin{proof}
The anti-symmetry and the non-interaction condition can be proven in the same way as the skew-symmetry and the non-interaction condition of $\mc{J}^{(E)}$ and $\mc{J}^{(S)}$. The Leibniz rule was shown in Section~\ref{sec:GENERIC}. Finally, the Jacobi identity follows by Theorem~\ref{th:jacobi_identity} with $\tilde{z}_\gamma=\rho_\gamma$,  $f_\gamma=\mathbb{L}_\gamma$, $\gamma =1,\ldots,\nu$, and $\tilde{z}_{\nu+1}=s$, $f_{\nu+1}=0$ for the case with the entropy $s$ as state variable or $\tilde{z}_{\nu+1}=u$, $f_{\nu+1}=\sum_{\beta=1}^{\nu}\mu_\beta \mathbb{L}_\beta - p$ for the internal energy $u$ as state variable.
\end{proof}
As for the Poisson bracket, the operator $\mc{R}$ allows us to define the dissipation bracket of the GENERIC formulation. For this, we define
\begin{align}
\label{eq:dissipation_bracket_mixture_energy}
&[A ,B]^{(E)}=\\
&\begin{aligned}[t]&\int_{\Omega} 
\frac{\zeta T}{2}\left[\nabla\vardif{A}{\mathbf{M}} + \nabla\vardif{A}{\mathbf{M}}^\top\!\! -\, \mathbf{D}\frac{1}{T}\vardif{A}{s}\right]\dcont \left[\nabla\vardif{B}{\mathbf{M}} + \nabla\vardif{B}{\mathbf{M}}^\top\!\! -\, \mathbf{D}\frac{1}{T}\vardif{B}{s}\right]\\
+ &T\Big(\lambda-\frac{2\zeta}{3}\Big) \Big(\!\diver\Big(\vardif{A}{\mathbf{M}}\Big)- \frac{1}{2}\trace(\mathbf{D})\frac{1}{T} \vardif{A}{s}\Big) \Big(\!\diver\Big(\vardif{B}{\mathbf{M}}\Big)-  \frac{1}{2}\trace(\mathbf{D})\frac{1}{T}\vardif{B}{s}\Big)
\\
+& \begin{bmatrix}
\nabla\Big(\mfrac{1}{T}\mvardif{A}{s}\Big)\\
\nabla\Big(\mvardif{A}{\rho_1}-\mfrac{\mu_1}{T}\mvardif{A}{s}\Big)\\
\vdots\\
\nabla\Big(\mvardif{A}{\rho_\nu}-\mfrac{\mu_\nu}{T}\mvardif{A}{s}\Big)\\
\end{bmatrix}\odot
\left(
\begin{bmatrix}
\kappa T^2 & B_1 &\dots &B_\nu\\[8pt]
B_1 & B\indi{_1_1} & \dots & B\indi{_1_\nu}\\[4pt]
\vdots & \vdots & \ddots & \vdots\\[4pt]
B_\nu & B\indi{_\nu_1} & \dots & B\indi{_\nu_\nu}\\
\end{bmatrix} \otimes_{\text{kron}} I_3 \right) \begin{bmatrix}
\nabla\Big(\mfrac{1}{T}\mvardif{B}{s}\Big)\\
\nabla\Big(\mvardif{B}{\rho_1}-\mfrac{\mu_1}{T}\mvardif{B}{s}\Big)\\
\vdots\\
\nabla\Big(\mvardif{B}{\rho_\nu}-\mfrac{\mu_\nu}{T}\mvardif{B}{s}\Big)\\
\end{bmatrix}\\
+& T \sum_{\alpha,\beta=1}^{\nu}\Big(\vardif{A}{\rho_\alpha}-\frac{\mu_\alpha}{T}\vardif{A}{s}\Big)
\mathbb{L}_{\alpha \beta} \Big(\vardif{B}{\rho_\beta}-\frac{\mu_\beta}{T}\vardif{B}{s}\Big)\dx,\end{aligned}\notag
\end{align}
for the case where energy constitutes the thermodynamic potential, and
\begin{align}
\label{eq:dissipation_bracket_mixture_entropy}
 [A,B]^{(S)}=
&\int_{\Omega}\phantom{+}
\frac{\zeta T}{2} \left[\nabla\vardif{A}{\mathbf{M}} + \nabla\vardif{A}{\mathbf{M}}^\top\!\!  -\, \mathbf{D}\vardif{A}{u}\right]\dcont\left[\nabla\vardif{B}{\mathbf{M}} + \nabla\vardif{B}{\mathbf{M}}^\top\!\!  -\, \mathbf{D}\vardif{B}{u}\right] \\
&\hphantom{\int_{\Omega}}+ T\Big(\lambda-\frac{2\zeta}{3}\Big) \Big(\!\diver\Big(\vardif{A}{\mathbf{M}}\Big)-  \frac{1}{2}\trace(\mathbf{D})\vardif{A}{u}\Big) \Big(\!\diver\Big(\vardif{B}{\mathbf{M}}\Big) - \frac{1}{2}\trace(\mathbf{D}) \vardif{B}{u} \Big)
\notag\\
&\hphantom{\int_{\Omega}}+ \begin{bmatrix}
\nabla \mvardif{A}{u}\\[4pt]
\nabla \mvardif{A}{\rho_1}\\
\vdots\\[1pt]
\nabla \mvardif{A}{\rho_\nu}\\
\end{bmatrix}\odot
\left(
\begin{bmatrix}
\kappa T^2 & B_1 &\dots &B_\nu\\[5pt]
B_1 & B\indi{_1_1} & \dots & B\indi{_1_\nu}\\[4pt]
\vdots & \vdots & \ddots & \vdots\\[4pt]
B_\nu & B\indi{_\nu_1} & \dots & B\indi{_\nu_\nu}\\
\end{bmatrix} \otimes_{\text{kron}} I_3 \right) \begin{bmatrix}
\nabla \mvardif{B}{u}\\[4pt]
\nabla \mvardif{B}{\rho_1}\\
\vdots\\[1pt]
\nabla \mvardif{B}{\rho_\nu}\\
\end{bmatrix} \notag\\
&\hphantom{\int_{\Omega}}+ T\sum_{\alpha,\beta=1}^{\nu} \vardif{A}{\rho_\alpha}
\mathbb{L}_{\alpha \beta} \vardif{B}{\rho_\alpha}\, \dx,\notag
\end{align}
in case entropy constitutes the thermodynamic potential. In the Sections~\ref{sec:mixture_isolated} and~\ref{sec:mixture_entropy} we showed that the associated operators~$\mc{R}^{(E)}$ and~$\mc{R}^{(S)}$ inducing the dissipation brackets above are self-adjoint, semi-elliptic, and satisfy the non-interaction conditions~\eqref{eq:noninteracting_R_multi_energy} and~\eqref{eq:noninteracting_multi_entropy}. Therefore, the brackets~$[\,\cdot,\cdot\,]^{(E)}$ and~$[\,\cdot,\cdot\,]^{(S)}$ are symmetric, non-negative and fulfill the degeneracy (or non-interacting) condition
\begin{equation}
\label{eq:degeneracy_condition}
[A,H]^{(E)}=[A,H]^{(S)}=0 \quad \text{for all } A\in C^\infty(\mc{Z}).
\end{equation}
The Leibniz rule follows, since the associated operators are linear spatial-differential operators, see Section~\ref{sec:GENERIC}. We summarize all properties of the dissipation brackets in the following lemma.
\begin{lemma}\label{lem:dissipation_mixture}
In case energy constitutes the thermodynamic potential, let the dissipation bracket be defined as in~\eqref{eq:dissipation_bracket_mixture_energy}. If entropy constitutes the thermodynamic potential, then choose the dissipation bracket defined by~\eqref{eq:dissipation_bracket_mixture_entropy}.\newline
Then the bracket is symmetric as well as non-negative and satisfies the Leibniz rule as well as the non-interaction condition~\eqref{eq:degeneracy_condition}.
\end{lemma}
As stated in Section~\ref{sec:GENERIC}, the Poisson bracket and dissipation bracket should be enough to describe the evolution of every observable for isolated systems. These observables have to depend upon the state variable~$\zhet$ and only implicitly on time.
\begin{theorem}[GENERIC for isolated systems of fluid mixture]\label{th:GENRERIC_mixture_isolated}
Suppose the system  of fluid mixture is isolated, i.e., the barycentric velocity~$\mathbf{v}$, the non-convective heat flux~$\mathbf{q}$, the diffusion fluxes~$\mathbf{J}_\alpha$, and the viscosity part of the stress tensor~$\mathbf{S}$ vanish at the boundary $\partial \Omega$ in normal direction. Let the Poisson bracket be given by~\eqref{eq:poisson_bracket_mixture_energy} and the dissipation bracket by~\eqref{eq:dissipation_bracket_mixture_energy}, if the entropy $s$ is a state variable, and by~\eqref{eq:poisson_bracket_mixture_entropy} and~\eqref{eq:dissipation_bracket_mixture_entropy}, if the internal energy $u$ is a state variable. \newline
Then the evolution for every smooth observable $A$ depending on the state~$\zhet$ is given by the GENERIC formulation~\eqref{eq:GENERIC_isolated}. 
\end{theorem}
\begin{proof} This follows by $\mfrac{\mathrm{d}A}{\mathrm{d}t}=\tweak{\mvardif{A}{\zhet}}{\dot \zhet}$ and Theorems~\ref{th:mixture_isolated} and~\ref{th:mixture_entropy}, respectively.
\end{proof}
Theorem~\ref{th:GENRERIC_mixture_isolated} shows that for reactive fluid mixtures the dynamics of every observable in an isolated system can be described with the GENERIC formulation~\eqref{eq:GENERIC_isolated}. For the formulation of open systems we have to consider the dynamics determined by the bulk related contribution of Poisson and dissipation brackets, or in case the full brackets are used, subtract the contribution of the boundary brackets from the full brackets, see~\eqref{eq:GENERIC_open}. In Sections~\ref{sec:mixture_open} and~\ref{sec:mixture_entropy} we proved that the boundary contribution is given by $\mc{B}\mathbf{u}$. Therefore, we look for associated boundary brackets which satisfy
\begin{equation*}
	\weak{\vardif{A}{\zhet}}{\mc{B}\mathbf{u}}= -\{A,H\}_{\text{boundary}} - [A,S]_{\text{boundary}}\,,
\end{equation*}
for every observable $A$ with smooth enough functional derivatives. In addition, the boundary brackets have to fulfill the non-interaction conditions, since the full brackets satisfy it by Lemma~\ref{lem:poisson_mixture} and~\ref{lem:dissipation_mixture} and the GENERIC formulation for open system claims that the bulk contribution does so. Therefore, for the case where the entropy density~$s$ is a state variable and the energy functional contains the thermodynamic potential we define
\begin{align}
\label{eq:poisson_bracket_mixture_energy_boundary}
\{A,B\}^{(E)}_{\text{boundary}}&=\int_{\partial\Omega} \Big(\sum_{\alpha=1}^{\nu}\vardif{A}{\rho_\alpha}\bigr(\rho_\alpha - \mathbb{L}_\alpha\bigr)   + \vardif{A}{\mathbf{M}}\cdot \mathbf{M} +  \vardif{A}{s} s\Big) \vardif{B}{\mathbf{M}}\cdot\mathbf{n}\, \dS,\\
\label{eq:dissipation_bracket_mixture_energy_boundary}
[A,B]_{\text{boundary}}^{(E)}&=
\int_{\partial\Omega} 
\zeta T \vardif{A}{\mathbf{M}}\cdot\Big(\nabla\vardif{B}{\mathbf{M}} + \nabla\vardif{B}{\mathbf{M}}^\top\!\! -\,\mathbf{D} \frac{1}{T}\vardif{B}{s}\Big)\cdot \mathbf{n}\, \\
&\hspace*{-1cm}+\Big(\vardif{A}{\mathbf{M}}\cdot \mathbf{n}\Big) T\Big(\lambda -\frac{2\zeta}{3}\Big) \Big(\diver\Big(\vardif{B}{\mathbf{M}}\Big)-  \frac{1}{2}\trace(\mathbf{D})\frac{1}{T}\vardif{B}{s}\Big)\notag\\
&\hspace*{-1cm} + \begin{bmatrix}
\mfrac{1}{T}\mvardif{A}{s}\\
\mvardif{A}{\rho_1}-\mfrac{\mu_1}{T}\mvardif{A}{s} \\
\vdots\\
\mvardif{A}{\rho_\nu}-\mfrac{\mu_\nu}{T}\mvardif{A}{s} \\
\end{bmatrix}^\top
\begin{bmatrix}
\kappa T^2 & B_1 &\dots &B_\nu\\[8pt]
B_1 & B\indi{_1_1} & \dots & B\indi{_1_\nu}\\[4pt]
\vdots & \vdots & \ddots & \vdots\\[4pt]
B_\nu & B\indi{_\nu_1} & \dots & B\indi{_\nu_\nu}\\
\end{bmatrix} \begin{bmatrix}
\nabla\Big(\mfrac{1}{T}\mvardif{B}{s}\Big)\cdot \mathbf{n}\\
\nabla\Big(\mvardif{B}{\rho_1}-\mfrac{\mu_1}{T}\mvardif{B}{s}\Big)\cdot \mathbf{n}\\
\vdots\\
\nabla\Big(\mvardif{B}{\rho_\nu}-\mfrac{\mu_\nu}{T}\mvardif{B}{s}\Big)\cdot \mathbf{n}\\
\end{bmatrix}
\dS.\notag
\end{align}
In case the internal energy density $u$ is a state variable and entropy constitutes the thermodynamic potential, we define the boundary contributions to be given by
\begin{align}
\label{eq:poisson_bracket_mixture_entropy_boundary}
\{A,B\}^{(S)}_{\text{boundary}}&=\\
&\hspace*{-2cm}\int_{\partial\Omega}\! \Big(\!\sum_{\alpha=1}^{\nu}\vardif{A}{\rho_\alpha}\bigr(\rho_\alpha - \mathbb{L}_\alpha\bigr)   +  \vardif{A}{\mathbf{M}}\cdot\mathbf{M} +\vardif{A}{u} \bigr( u + p - \sum_{\beta=1}^{\nu}\mu_\beta \mathbb{L}_\beta\bigr)  \Big)\vardif{B}{\mathbf{M}}\cdot\mathbf{n}\,\dS,\notag\\
\label{eq:dissipation_bracket_mixture_entropy_boundary}
[A,B]_{\text{boundary}}^{(S)}&=
\int_{\partial\Omega} 
\zeta T \vardif{A}{\mathbf{M}}\cdot\Big(\nabla\vardif{B}{\mathbf{M}} + \nabla\vardif{B}{\mathbf{M}}^\top \!\! -\, \mathbf{D}\vardif{B}{u}\Big)\cdot \mathbf{n} \\
&\hphantom{= \int_{\partial\Omega}}+\Big(\vardif{A}{\mathbf{M}}\cdot \mathbf{n}\Big) T\Big(\lambda -\frac{2\zeta}{3}\Big) \Big(\diver\Big(\vardif{B}{\mathbf{M}}\Big)- \frac{1}{2}\trace(\mathbf{D})\vardif{B}{u}\Big) 
\notag \\
&\hphantom{= \int_{\partial\Omega}} + \begin{bmatrix}
\mvardif{A}{u}\\[4pt]
\mvardif{A}{\rho_1}\\
\vdots\\
\mvardif{A}{\rho_\nu}\\
\end{bmatrix}^\top
\begin{bmatrix}
\kappa T^2 & B_1 &\dots &B_\nu\\
B_1 & B\indi{_1_1} & \dots & B\indi{_1_\nu}\\
\vdots & \vdots & \ddots & \vdots\\
B_\nu & B\indi{_\nu_1} & \dots & B\indi{_\nu_\nu}\\
\end{bmatrix} \begin{bmatrix} 
\nabla\mvardif{B}{u}\cdot \mathbf{n}\\[4pt]
\nabla\mvardif{B}{\rho_1}\cdot \mathbf{n} \\
\vdots\\
\nabla\mvardif{B}{\rho_\nu}\cdot \mathbf{n}\\
\end{bmatrix}\dS.\notag 
\end{align}
With these boundary contributions we finally can prove that the dynamics of the considered fluid mixture modeled as open system can be described by the GENERIC formulation~\eqref{eq:GENERIC_open}.
\begin{theorem}[GENERIC for open systems of fluid mixture]\label{th:GENRERIC_mixture_open}
Let one of the following assumptions be satisfied.

\begin{enumerate}
\item The entropy density~$s$ is a state variable and energy~constitutes the thermodynamical potential. The Poisson bracket is given by~\eqref{eq:poisson_bracket_mixture_energy} and the dissipation bracket by~\eqref{eq:dissipation_bracket_mixture_energy} with boundary contribution~\eqref{eq:poisson_bracket_mixture_energy_boundary} and~\eqref{eq:dissipation_bracket_mixture_energy_boundary}, respectively. \label{item:GENERIC_mixture_open_energy}
\item The internal energy density~$u$ is a state variable and entropy constitutes the thermodynamical potential. The Poisson bracket is given by~\eqref{eq:poisson_bracket_mixture_entropy} and the dissipation bracket by~\eqref{eq:dissipation_bracket_mixture_entropy} with boundary contribution~\eqref{eq:poisson_bracket_mixture_entropy_boundary} and~\eqref{eq:dissipation_bracket_mixture_entropy_boundary}, respectively.\label{item:GENERIC_mixture_open_entropy}
\end{enumerate}
Then the Poisson bracket is anti-symmetric, satisfies the Leibniz rule and the Jacobi identity. The dissipation bracket is symmetric, non-negative and satisfies the Leibniz rule as well. The bulk contributions of both brackets, which are defined as 
\[ \{A,B\}_{\text{\normalfont bulk}} = \{A,B\} - \{A,B\}_{\text{\normalfont boundary}}, \quad [A,B]_{\text{\normalfont bulk}} = [A,B] - [A,B]_{\text{\normalfont boundary}},\]
fulfill the associated non-interaction conditions~\eqref{eq:noninteracting_bulk}. Furthermore, the evolution for every observable $A$ depending only on the state~$\zhet$ is given by the GENERIC formulation~\eqref{eq:GENERIC_open}. 
\end{theorem}
\begin{proof}
The properties of the Poisson bracket and the dissipation bracket are shown in Lemma~\ref{lem:poisson_mixture} and~\ref{lem:dissipation_mixture}. Since these lemmas prove also the non-inter\-action condition for the whole bracket, it is enough to show that $\{\cdot,S\}_{\text{boundary}}$ and $[\,\cdot,H]_{\text{boundary}}$ vanish. For the Poisson brackets this follows immediately by $\tvardif{S}{\mathbf{M}}=\mathbf{0}$ and for the dissipation brackets one uses again $\sum_{\beta=1}^\nu B_{\beta } = \sum_{\beta=1}^\nu B_{\alpha \beta } = 0$, $\alpha=1,\ldots,\nu$, and a straight forward calculation. For the evolution equation we notice that under assumption~\ref{item:GENERIC_mixture_open_energy}. it holds that
\begin{align*}
 &\{A,H\}^{(E)}_{\text{boundary}} +  [A,S]^{(E)}_{\text{boundary}} \\
= &
\int_{\partial \Omega} \!\Big(\! \sum_{\alpha=1}^{\nu} \vardif{A}{\rho_\alpha}\big(\rho_\alpha - \mathbb{L}_\alpha\big)  + \vardif{A}{\mathbf{M}}\cdot\mathbf{M} + \vardif{A}{s} s\Big)\mathbf{n}\cdot \mathbf{v}  
- \zeta  \vardif{A}{\mathbf{M}}\cdot\mathbf{D}\cdot \mathbf{n} \\
& - \Big(\vardif{A}{\mathbf{M}}\cdot \mathbf{n}\Big)\Big(\lambda -\frac{2\zeta}{3}\Big) \frac{1}{2}\trace(\mathbf{D}) 
+\frac{1}{T}\vardif{A}{s} \Big[\kappa T^2 \nabla\Big(\frac{1}{T}\Big) +\sum_{\beta=1}^{\nu} B_\beta \nabla\Big(\!\!-\frac{\mu_\beta}{T}\Big)\Big]\cdot \mathbf{n} \\
& +\sum_{\alpha=1}^{\nu}\Big(\vardif{A}{\rho_\alpha}-\frac{\mu_\alpha}{T}\vardif{A}{s}\Big)\Big[B_\alpha \nabla\Big(\frac{1}{T}\Big)
+\sum_{\beta=1}^{\nu} B_{\alpha \beta} \nabla\Big(\!\!-\frac{\mu_\beta}{T}\Big)\Big]\cdot \mathbf{n}\, \dS \\
= &\int_{\partial \Omega}\! \Big(\! \sum_{\alpha=1}^{\nu}\vardif{A}{\rho_\alpha}(\rho_\alpha - \mathbb{L}_\alpha)   +  \vardif{A}{\mathbf{M}}\cdot \mathbf{M} + \vardif{A}{s} s\Big)\mathbf{n}\cdot \mathbf{v} - \vardif{A}{\mathbf{M}} \cdot \mathbf{T}^\mathrm{d} \cdot \mathbf{n} + \frac{1}{T}\vardif{A}{s} \mathbf{q}\cdot \mathbf{n}\\
&  +\sum_{\alpha=1}^{\nu}\Big(\vardif{A}{\rho_\alpha}-\frac{\mu_\alpha}{T}\vardif{A}{s}\Big)\mathbf{J}_\alpha\cdot \mathbf{n}
- \vardif{A}{\mathbf{M}} \cdot (-\pi\mathbf{I}) \cdot \mathbf{n}
+\sum_{\beta=1}^{\nu}\mathbb{L}_\beta \mu_\beta  \vardif{A}{\mathbf{M}}\cdot \mathbf{n}\, \dS
\\
= & -\weak{\vardif{A}{\zhet}}{\mc{B}^{(E)} \mathbf{u}},
\end{align*}
where we used the definitions of the constitutive relations,~$\mc{B}^{(E)}$, and~$\mathbf{u}$ in~\eqref{eq:operator_B_mixture} and~\eqref{eq:u_mixture}. Analogously, under assumption~\ref{item:GENERIC_mixture_open_entropy}.~and with~$\mc{B}^{(S)}$ given by~\eqref{eq:operator_B_mixture_entropy} one proves that
\begin{align}
	\{A,H\}^{(S)}_{\text{boundary}} +  [A,S]^{(S)}_{\text{boundary}} = - \weak{\vardif{A}{\zhet}}{\mc{B}^{(S)} \mathbf{u}}.
\end{align}
The description of the evolution for an observable~$A$ by formulation~\eqref{eq:GENERIC_open} follows then by  Theorems~\ref{th:mixture_open},~\ref{th:mixture_entropy}, and~\ref{th:GENRERIC_mixture_isolated}.
\end{proof}

\begin{remark}\normalfont
Lemma~\ref{lem:poisson_mixture} and~\ref{lem:dissipation_mixture} as well as the proof of Theorem~\ref{th:GENRERIC_mixture_open} show that not only the bulk contributions fulfill the non-interaction conditions~\eqref{eq:noninteracting_bulk} but also the whole brackets and therefore also the boundary contributions. This property was shown before for hydro-dynamical systems in \cite{Oet06} and is extended here to homogeneous mixtures of heat-conducting Newtonian fluids consisting of a finite number of reactive constituents. 
\end{remark}

\section{Conclusions}\label{sec:conclusion}
In the first part of this work we introduced operator based state space representations of abstract nonlinear dissipative dynamical systems which interact
with their environment in a system theoretic sense. This was achieved through the reinterpretation of the GENERIC framework for open non-equilibrium thermodynamic systems. Any of these Operator-GENERIC formulations \eqref{eq:operator_equation_open} and \eqref{eq:operator_equation_isolated} is a combination of a generalized Hamiltonian system and a gradient system. 
As concrete example for an Operator-GENERIC formulation we considered homogeneous mixtures of heat-conducting compressible Newtonian fluids consisting of reactive constituents. 
Therefore we motivated the differential balance laws and complementary closure relations that represent the mathematical model of the fluid mixture in the framework of classical continuum physics. As closure relations we chose the constitutive equations of TIP. We presented the Operator-GENERIC formulations for fluid mixtures in an operator setting for the cases whether energy or entropy represents the thermodynamic potential. We proved that these Operator-GENERIC formulations encode a weak formulation of the field equations in an operator setting. We presented a new mixture related Poisson bracket and a dissipation bracket such that the Onsager-Casimir reciprocal relations could be fully incorporated.

In future work we plan to investigate the relation between the operators of the Operator-GENERIC formulations with underlying Dirac structures. Furthermore, Lie-Poisson integrators and structure preserving discretization methods will be in focus.

\subsection*{Acknowledgments}

The authors were supported by the Einstein Foundation Berlin through the project "Model reduction for complex transport-dominated phenomena and reactive flows". In addition, the first author is supported by the German Federal Ministry of Education and Research (BMBF) via BMBF-project "Verbundprojekt 05M2018 - EiFer: Energieeffizienz durch intelligente Fernw\"{a}rmenetze - Teilprojekt 3: Regelung von gekoppelten port-Hamiltonischen W\"{a}rme-Stom-Systemen", and the second author by the DFG Collaborative Research Center 910 through the project "Control of self-organizing nonlinear
systems: Theoretical methods and concepts of application".
The authors would like to thank Christopher Beattie, Volker Mehrmann, Robert Altmann, and Philipp Schulze for helpful discussions.
\endgroup

\bibliography{references}

\begin{thebibliography}{10}

\bibitem{Abr87}
R.~Abraham and J.~E. Marsden.
\newblock {\em Foundations of {M}echanics}.
\newblock Addison-Wesley Publishing Company, Inc., Redwood City, CA, second
  edition, 1987.

\bibitem{AbrMarRat88}
R.~Abraham, J.~E. Marsden, and T.~Ratiu.
\newblock {\em Manifolds, {T}ensor {A}nalysis, and {A}pplications}.
\newblock Springer, New York, second edition, 1988.

\bibitem{Ada75}
R.~A. Adams.
\newblock {\em Sobolev {S}paces}.
\newblock Academic Press, New York-London, 1975.

\bibitem{AltD14}
H.~W. Alt.
\newblock Distributions.
\newblock \url{http://www-m6.ma.tum.de/~alt/alt-distributions.pdf}.
\newblock Accessed 07/07/2018, Version: 20161020.

\bibitem{AltCM17}
H.~W. Alt.
\newblock Lectures on mathematical continuum mechanics.
\newblock \url{http://www-m6.ma.tum.de/~alt/alt-continuum.pdf}.
\newblock Accessed 07/07/2018, Version: 2017116.

\bibitem{Alt16}
H.~W. Alt.
\newblock {\em Linear Functional Analysis: An Application-Oriented
  Introduction}.
\newblock Springer London, London, 2016.

\bibitem{AmaEschII08}
H.~Amann and J.~Escher.
\newblock {\em Analysis II}.
\newblock Birkh{\"a}user Basel, 2008.
\newblock Translated from the German.

\bibitem{AmaEschIII09}
H.~Amann and J.~Escher.
\newblock {\em Analysis III}.
\newblock Birkh{\"a}user Basel, 2009.
\newblock Translated from the German.

\bibitem{AGS08}
L.~Ambrosio, N.~Gigli, and G.~Savar{\'e}.
\newblock {\em Gradient Flows: in {M}etric {S}paces and in the {S}pace of
  {P}robability {M}easures}.
\newblock Birkh{\"a}user, Basel, second edition, 2008.

\bibitem{AveSmo67}
V.~I. Averbukh and O.~G. Smolyanov.
\newblock The theory of differentiation in linear topological spaces.
\newblock {\em Russ. Math. Surv+}, 22(6):201--258, 1967.

\bibitem{BerE94}
A.~N. Beris and B.~J. Edwards.
\newblock {\em Thermodynamics of {F}lowing {S}ystems: with {I}nternal
  {M}icrostructure}.
\newblock Oxford University Press, Oxford, 1994.

\bibitem{Bl13}
A.~M. Bloch, P.~J. Morrison, and T.~S. Ratiu.
\newblock Gradient flows in the normal and {K}{\"a}hler metrics and triple
  bracket generated metriplectic systems.
\newblock In {\em Recent {T}rends in {D}ynamical {S}ystems: {P}roceedings of a
  {C}onference in {H}onor of {J}{\"u}rgen {S}cheurle}, pages 371--415.
  Springer, Basel, 2013.

\bibitem{BotDre15}
D.~Bothe and W.~Dreyer.
\newblock Continuum thermodynamics of chemically reacting fluid mixtures.
\newblock {\em Acta Mech.}, 226(6):1757--1805, 2015.

\bibitem{BAW91}
C.~I. Byrnes, A.~Isidori, and J.~C. Willems.
\newblock Passivity, feedback equivalence, and the global stabilization of
  minimum phase nonlinear systems.
\newblock {\em IEEE Trans. Automat. Contr.}, 36(11):1228--1240, 1991.

\bibitem{Cas45}
H.~B.~G. Casimir.
\newblock On {O}nsager's principle of microscopic reversibility.
\newblock {\em Rev. Mod. Phys.}, 17:343--350, 1945.

\bibitem{CZT09}
G.-Q. Chen, W.~P. Ziemer, and M.~Torres.
\newblock Gauss-{G}reen theorem for weakly differentiable vector fields, sets
  of finite perimeter, and balance laws.
\newblock {\em Commun. Pure. Appl. Math.}, 62(2):242--304, 2009.

\bibitem{Daf93}
C.~M. Dafermos.
\newblock Equivalence of referential and spatial field equations in continuum
  physics.
\newblock In {\em Nonlinear Hyperbolic Problems: Theoretical, Applied, and
  Computational Aspects}, pages 179--183. Vieweg+Teubner Verlag,
  Braunschweig/Wiesbaden, 1993.

\bibitem{Daf16}
C.~M. Dafermos.
\newblock Introduction to continuum physics.
\newblock In {\em Hyperbolic {C}onservation {L}aws in {C}ontinuum {P}hysics},
  pages 25--51. Springer, Berlin Heidelberg, 2016.

\bibitem{Mazur}
S.~R. {de Groot} and P.~Mazur.
\newblock {\em Non-Equilibrium Thermodynamics}.
\newblock Dover Publications, Mineola, NY, 1984.

\bibitem{Dui09}
V.~Duindam, A.~Macchelli, S.~Stramigioli, and H.~Bruyninckx.
\newblock {\em Modeling and {C}ontrol of {C}omplex {P}hysical {S}ystems: the
  {P}ort-{H}amiltonian {A}pproach}.
\newblock Springer, Berlin Heidelberg, 2009.

\bibitem{EMV05}
D.~Eberard, B.~M. Maschke, and A.~J. van~der Schaft.
\newblock Port contact systems for irreversible thermodynamical systems.
\newblock In {\em Decision and Control, 2005 and 2005 European Control
  Conference. 44th IEEE Conference on}, pages 5977--5982, 2005.

\bibitem{EMV07}
D.~Eberard, B.~M. Maschke, and A.~J. van~der Schaft.
\newblock An extension of {H}amiltonian systems to the thermodynamic phase
  space: Towards a geometry of nonreversible processes.
\newblock {\em Rep. Math. Phys.}, 60(2):175--198, 2007.

\bibitem{Eck1}
C.~Eckart.
\newblock The thermodynamics of irreversible processes. {I}. {T}he simple
  fluid.
\newblock {\em Phys. Rev.}, 58:267--269, 1940.

\bibitem{Eck2}
C.~Eckart.
\newblock The thermodynamics of irreversible processes. {II}. {F}luid mixtures.
\newblock {\em Phys. Rev.}, 58:269--275, 1940.

\bibitem{Eck3}
C.~Eckart.
\newblock The thermodynamics of irreversible processes. {III}. {R}elativistic
  theory of the simple fluid.
\newblock {\em Phys. Rev.}, 58:919--924, 1940.

\bibitem{Ed98}
B.~J. Edwards.
\newblock An analysis of single and double generator thermodynamic formalisms
  for the macroscopic description of complex fluids.
\newblock {\em J. Non-Equil. Thermody.}, 23(4):301--333, 1998.

\bibitem{BriOet97}
B.~J. Edwards and H.~C. {\"O}ttinger.
\newblock Time-structure invariance criteria for closure approximations.
\newblock {\em Phys. Rev. E}, 56:4097--4103, 1997.

\bibitem{FeiN17}
E.~Feireisl and A.~Novotn{\'y}.
\newblock {\em Singular Limits in Thermodynamics of Viscous Fluids}.
\newblock Birkh{\"a}user, Cham, second edition, 2017.

\bibitem{GrmOet97I}
M.~Grmela and H.~C. {\"O}ttinger.
\newblock Dynamics and thermodynamics of complex fluids. {I}. {D}evelopment of
  a general formalism.
\newblock {\em Phys. Rev. E}, 56(6):6620--6632, 1997.

\bibitem{Gur87}
M.~E. Gurtin, W.~O. Williams, and W.~P. Ziemer.
\newblock {\em Geometric Measure Theory and the Axioms of Continuum
  Thermodynamics}, pages 379--400.
\newblock Springer Berlin Heidelberg, Berlin, Heidelberg, 1987.

\bibitem{Ham82}
E.~P. Hamilton and M.~Z. Nashed.
\newblock Global and local variational derivatives and integral representations
  of {G}{\^a}teaux differentials.
\newblock {\em J. Funct. Anal.}, 49(1):128--144, 1982.

\bibitem{HorJ91}
R.~A. Horn and C.~R. Johnson.
\newblock {\em Topics in Matrix Analysis}.
\newblock Cambridge University Press, Cambridge, 1991.

\bibitem{Hut09}
K.~Hutter.
\newblock {\em Solid-Fluid Mixtures of Frictional Materials in Geophysical and
  Geotechnical Context}.
\newblock Springer, Berlin Heidelberg, 2009.

\bibitem{IVM16}
E.~A. Ivanova, E.~N. Vilchevskaya, and W.~H. M{\"u}ller.
\newblock {\em Time Derivatives in Material and Spatial Description---What Are
  the Differences and Why Do They Concern Us?}, pages 3--28.
\newblock Springer, Singapore, 2016.

\bibitem{Kau84}
A.~N. Kaufman.
\newblock Dissipative {H}amiltonian systems: A unifying principle.
\newblock {\em Phys. Lett. A}, 100(8):419--422, 1984.

\bibitem{KauMor82}
A.~N. Kaufman and P.~J. Morrison.
\newblock Algebraic structure of the plasma quasilinear equations.
\newblock {\em Phys. Lett. A}, 88(8):405--406, 1982.

\bibitem{KosM97}
A.~I. Kostrikin and Y.~I. Manin.
\newblock {\em Linear {A}lgebra and {G}eometry}.
\newblock CRC Press, Boca Raton, FL, 1997.

\bibitem{Leb08}
G.~Lebon, D.~Jou, and J.~Casas-V{\'a}zquez.
\newblock {\em Understanding Non-equilibrium Thermodynamics: Foundations,
  Applications, Frontiers}.
\newblock Springer, Berlin Heidelberg, 2008.

\bibitem{Liu02}
I.-S. Liu.
\newblock {\em Continuum Mechanics}.
\newblock Springer, Berlin Heidelberg, 2002.

\bibitem{Liu83}
I.-S. Liu and I.~M{\"u}ller.
\newblock Extended thermodynamics of classical and degenerate ideal gases.
\newblock {\em Arch. Ration. Mech. Anal.}, 83(4):285--332, 1983.

\bibitem{MarHug94}
J.~E. Marsden and T.~J.~R. Hughes.
\newblock {\em Mathematical Foundations of Elasticity}.
\newblock Dover Publiciations, Mineola, NY, 1994.

\bibitem{MarR99}
J.~E. Marsden and T.~S. Ratiu.
\newblock {\em Introduction to {M}echanics and {S}ymmetry: {A} {B}asic
  {E}xposition of {C}lassical {M}echanical {S}ystems}.
\newblock Springer, New York, 1999.

\bibitem{Mar02}
A.~Marzocchi and A.~Musesti.
\newblock On the measure-theoretic foundations of the second law of
  thermodynamics.
\newblock {\em Math. Models Methods Appl. Sci.}, 12(05):721--736, 2002.

\bibitem{Mas96}
B.~M. Maschke and A.~J. van~der Schaft.
\newblock Interconnection of systems: the network paradigm.
\newblock In {\em Decision and Control, 1996., Proceedings of the 35th IEEE
  Conference on}, pages 207--212, 1996.

\bibitem{Mei41}
J.~Meixner.
\newblock Zur {T}hermodynamik der {T}hermodiffusion.
\newblock {\em Ann. Phys.}, 431(5):333--356, 1941.
\newblock In German.

\bibitem{Mei73}
J.~Meixner.
\newblock Consistency of the {O}nsager-{C}asimir reciprocal relations.
\newblock {\em Adv. Mol. Relax. Int. Pr.}, 5(4):319--331, 1973.

\bibitem{MeiR59}
J.~Meixner and H.~G. Reik.
\newblock Thermodynamik der irreversiblen {P}rozesse.
\newblock In {\em Handbuch der {P}hysik, {B}d. 3/2}, pages 413--523. Springer,
  Berlin G{\"o}ttingen Heidelberg, 1959.
\newblock In German.

\bibitem{MeyS64}
N.~G. Meyers and J.~Serrin.
\newblock ${H}={W}$.
\newblock {\em Proc. Natl. Acad. Sci. U.S.A.}, 51:1055--1056, 1964.

\bibitem{Mie11}
A.~Mielke.
\newblock Formulation of thermoelastic dissipative material behavior using
  {GENERIC}.
\newblock {\em Continuum Mech. Therm.}, 23(3):233--256, 2011.

\bibitem{Mie15}
A.~Mielke.
\newblock On thermodynamical couplings of quantum mechanics and macroscopic
  systems.
\newblock In {\em Mathematical Results in Quantum Mechanics: Proceedings of the
  QMath12 Conference}, pages 331--348, 2015.

\bibitem{Mis17}
C.~W. Misner, K.~S. Thorne, and J.~A. Wheeler.
\newblock {\em Gravitation}.
\newblock Princeton University Press, 2017.

\bibitem{MiM17}
M.~Mittnenzweig and A.~Mielke.
\newblock An entropic gradient structure for {L}indblad equations and couplings
  of quantum systems to macroscopic models.
\newblock {\em J. Stat. Phys.}, 167(2):205--233, 2017.

\bibitem{Mor82}
P.~J. Morrison.
\newblock Poisson brackets for fluids and plasmas.
\newblock {\em AIP Conference Proceedings}, 88(1):13--46, 1982.

\bibitem{Mor84}
P.~J. Morrison.
\newblock Bracket formulation for irreversible classical fields.
\newblock {\em Phys. Lett. A}, 100(8):423--427, 1984.

\bibitem{Mor86}
P.~J. Morrison.
\newblock A paradigm for joined {H}amiltonian and dissipative systems.
\newblock {\em Physica D}, 18(1-3):410--419, 1986.

\bibitem{Mor09}
P.~J. Morrison.
\newblock Thoughts on brackets and dissipation: Old and new.
\newblock {\em J. Phys. Conf. Ser.}, 169(1):{ }012006, 2009.

\bibitem{Mue85}
I.~M{\"u}ller.
\newblock {\em Thermodynamics}.
\newblock Pitman, London, 1985.

\bibitem{IngoM}
I.~M{\"u}ller and T.~Ruggeri.
\newblock {\em Rational {E}xtended {T}hermodynamics}.
\newblock Springer, New York, second edition, 1998.

\bibitem{Mueller2012}
I.~M{\"u}ller and W.~Weiss.
\newblock Thermodynamics of irreversible processes --- past and present.
\newblock {\em Eur. Phys. J. H}, 37(2):139--236, 2012.

\bibitem{Nec12}
J.~Ne{\v c}as.
\newblock {\em Direct {M}ethods in the {T}heory of {E}lliptic {E}quations}.
\newblock Springer, Heidelberg, 2012.

\bibitem{Noll1973}
W.~Noll.
\newblock Lectures on the foundations of continuum mechanics and
  thermodynamics.
\newblock {\em Arch. Ration. Mech. Anal.}, 52(1):62--92, 1973.

\bibitem{Ons1}
L.~Onsager.
\newblock Reciprocal relations in irreversible processes. {I}.
\newblock {\em Phys. Rev.}, 37:405--426, 1931.

\bibitem{Ons2}
L.~Onsager.
\newblock Reciprocal relations in irreversible processes. {II}.
\newblock {\em Phys. Rev.}, 38:2265--2279, 1931.

\bibitem{Oet05}
H.~C. {\"O}ttinger.
\newblock {\em Beyond {E}quilibrium {T}hermodynamics}.
\newblock John Wiley \& Sons, Hoboken, NJ, 2005.

\bibitem{Oet06}
H.~C. {\"O}ttinger.
\newblock Nonequilibrium thermodynamics for open systems.
\newblock {\em Phys. Rev. E}, 73(3):036126, 2006.

\bibitem{Oet11}
H.~C. {\"O}ttinger.
\newblock The geometry and thermodynamics of dissipative quantum systems.
\newblock {\em Europhys Lett.}, 94(1):10006, 2011.

\bibitem{OetGrm97II}
H.~C. {\"O}ttinger and M.~Grmela.
\newblock Dynamics and thermodynamics of complex fluids. {II}. {I}llustrations
  of a general formalism.
\newblock {\em Phys. Rev. E}, 56(6):6633 -- 6655, 1997.

\bibitem{PavKG14}
M.~Pavelka, V.~Klika, and M.~Grmela.
\newblock Time reversal in nonequilibrium thermodynamics.
\newblock {\em Phys. Rev. E}, 90:062131, 2014.

\bibitem{Pfe85}
W.~F. Pfeffer.
\newblock On the continuity of the volterra variational derivative.
\newblock {\em J. Funct. Anal.}, 71(1):195--197, 1987.

\bibitem{Rod95}
W.~A. Rodrigues, Q.~A.~G. de~Souza, and Y.~Bozhkov.
\newblock The mathematical structure of {N}ewtonian spacetime: Classical
  dynamics and gravitation.
\newblock {\em Found. Phys.}, 25(6):871--924, 1995.

\bibitem{Rug08}
T.~Ruggeri.
\newblock The entropy principle from continuum mechanics to hyperbolic systems
  of balance laws: {T}he modern theory of extended thermodynamics.
\newblock {\em Entropy}, 10(3):319--333, 2008.

\bibitem{Sch06}
F.~Schwabl.
\newblock {\em Statistische Mechanik}.
\newblock Springer, Berlin Heidelberg, 2006.
\newblock In German.

\bibitem{Sil97}
M.~{\v S}ilhav{\'y}.
\newblock {\em The Mechanics and Thermodynamics of Continuous Media}.
\newblock Springer, Berlin Heidelberg, 1997.

\bibitem{Sur03}
Y.~B. Suris.
\newblock {\em The Problem of Integrable Discretization: Hamiltonian Approach}.
\newblock Birkh{\"a}user, Basel, 2003.

\bibitem{Tal02}
Y.~R. Talpaert.
\newblock {\em Tensor Analysis and Continuum Mechanics}.
\newblock Springer Netherlands, Dordrecht, 2002.

\bibitem{Tra70}
A.~Trautman.
\newblock Fibre bundles associated with space-time.
\newblock {\em Rep. Math. Phys.}, 1(1):29--62, 1970.

\bibitem{Trues}
C.~Truesdell.
\newblock {\em Rational Thermodynamics}.
\newblock Springer, New York, second edition, 1984.

\bibitem{TruesRM}
C.~Truesdell.
\newblock {\em A First Course in Rational Continuum Mechanics -- Volume 1}.
\newblock Academic Press, Boston, MA, second edition, 1997.

\bibitem{TruesdNoll}
C.~Truesdell and W.~Noll.
\newblock {\em The Non-Linear Field Theories of Mechanics}.
\newblock Springer, Berlin Heidelberg, third edition, 2004.

\bibitem{SchM01}
A.~J. van~der Schaft and B.~M. Maschke.
\newblock Fluid dynamical systems as {H}amiltonian boundary control systems.
\newblock In {\em Proceedings of the 40th IEEE Conference on Decision and
  Control}, volume~5, pages 4497--4502, 2001.

\bibitem{SchM02}
A.~J. van~der Schaft and B.~M. Maschke.
\newblock Hamiltonian formulation of distributed-parameter systems with
  boundary energy flow.
\newblock {\em J. Geom. Phys.}, 42(1-2):166--194, 2002.

\bibitem{Wil72I}
J.~C. Willems.
\newblock Dissipative dynamical systems part {I}: {G}eneral theory.
\newblock {\em Arch. Ration. Mech. Anal.}, 45(5):321--351, 1972.

\bibitem{Wil72II}
J.~C. Willems.
\newblock Dissipative dynamical systems part {II}: {L}inear systems with
  quadratic supply rates.
\newblock {\em Arch. Ration. Mech. Anal.}, 45(5):352--393, 1972.

\bibitem{Zei86}
E.~Zeidler.
\newblock {\em Nonlinear {F}unctional {A}nalysis and its {A}pplications. {I}:
  {F}ixed-{P}oint {T}heorems}.
\newblock Springer, New York, 1986.

\bibitem{ZeidIV}
E.~Zeidler.
\newblock {\em Nonlinear {F}unctional {A}nalysis and its {A}pplications: {IV}:
  {A}pplications to {M}athematical {P}hysics}.
\newblock Springer New York, 1988.

\bibitem{Zei90}
E.~Zeidler.
\newblock {\em Nonlinear {F}unctional {A}nalysis and its {A}pplications.
  {II}/{A}: {L}inear {M}onotone {O}perators}.
\newblock Springer, New York, 1990.

\bibitem{ZeiQFT3}
E.~Zeidler.
\newblock {\em Quantum Field Theory III: Gauge Theory: A Bridge between
  Mathematicians and Physicists}.
\newblock Springer, Berlin Heidelberg, 2011.

\end{thebibliography}
\bibliographystyle{plain}


	\appendix
	\appendixpageoff
	\counterwithin*{equation}{subsection}
	\section*{Appendix}

	\begin{appendices}
		\section*{Proof of the Jacobi Identity, Theorem~\ref{th:jacobi_identity}}
\label{sec:proof_Jacobi_identity}
\setcounter{section}{1}
In this section we prove the Jacobi identity for brackets of the general form~\eqref{eq:jacobi_identity_bracket}. We assume that the set~$\mc{Z}$  where the state lives is a open subset of $\mc{D}_\zhet$ which is a closed subspace of $[W^{1,p}(\Omega)]^{n}$ (not necessarily with zero boundaries). Further, $\zhet$ should be smooth enough. Note that, $\mc{Z}$ as a open subset of a linear space is a manifold with a trivial bundle. For the readability of the proof, we adopt in the following the~\emph{summation convention} for coordinate representations on physical space. This means that, if an index appears twice (\emph{and only twice}) in the same term, \emph{once as a subscript and once as a superscript} it implies summation of that term over all the values of the index. The corresponding summation symbol $\sum$ is omitted.

Before we prove Theorem~\ref{th:jacobi_identity} we have to calculate formulas for the partial functional derivatives and their spatial derivatives 
when they are acting on elements of $W^{1,p}(\Omega)$. This can be done by the following Lemma.
\begin{lemma}\label{lem:functional_derivative}
Let the functions $F, G\in C^\infty(\mc{Z})$ be given by 
\begin{align*}
F(\zhet) &=\int_\Omega a_k(x,\zhet(x)) \g\indi{^k^\ell}(x) \nabla_\ell \chi(x,\zhet(x))\, \dx,\\
G(\zhet) &= \int_\Omega A_{ik}(x,\zhet(x)) \g\indi{^k^\ell}(x)  \nabla_\ell (\g\indi{^i^j}(x)a_j(x,\zhet(x)))  \, \dx,
\end{align*}
where $\chi(\cdot,\zhet(\cdot))$, $a\indi{_j}(\cdot,\zhet(\cdot))$, and $A\indi{_i_k}(\cdot,\zhet(\cdot))$ are coordinate representations of a scalar, a co-vector and a two-covariant tensor field, respectively, which map the domain $\Omega$ into the real numbers and depend smoothly enough on $x$. Let the mappings of $\mc{Z}$ to these fields be smooth enough as well, and 
$\alpha=1,\ldots,\mu+1$ be arbitrary but fixed. Then for every $\phi\in W^{1,p}(\Omega)\setminus \{0\}$ and $\mathbf{h}\in W^{1,p}(\Omega)^d \setminus \{0\}$, such that the vector with $\phi$ in the $\alpha^{\text{th}}$ position and otherwise zero-function, as well as $[0,\ldots,0,\mathbf{h}]$ are elements of $\cD_{\zhet}$, it holds that
\begin{subequations}\label{eq:functional_derivative}
\begin{alignat}{2}
\int_\Omega \phi \mvardif{F}{\tilde{z}_\alpha}  \,\dx &= \int_\Omega \phi \mpd{a_k}{\tilde{z}_\alpha} \g\indi{^k^\ell} \nabla_\ell \chi + a_k \g\indi{^k^\ell} \nabla_\ell (\phi \mpd{\chi}{\tilde{z}_\alpha}  )\,\dx, \label{eq:functional_derivative_a}\\
\int_\Omega \phi \mvardif{G}{\tilde{z}_\alpha}\,\dx &= \int_\Omega \phi\mpd{A_{ik}}{\tilde{z}_\alpha} \g\indi{^k^\ell} \nabla_\ell (\g\indi{^i^j}a_j)  + A_{ik} \g\indi{^k^\ell} \nabla_\ell (\phi\g\indi{^i^j}\mpd{a_j}{\tilde{z}_\alpha} ) \,\dx, \label{eq:functional_derivative_b}\\
\int_\Omega h_{p} \g\indi{^p^q} \mvardif{F}{M_q}\,\dx &= \int_\Omega h_p \g\indi{^p^q}\mpd{a_k}{M_q} \g\indi{^k^\ell}  \nabla_\ell \chi + a_k \g\indi{^k^\ell}  \nabla_\ell (h_p \g\indi{^p^q}\mpd{\chi}{M_q})\,\dx, \label{eq:functional_derivative_c}\\
\int_\Omega h_{p} \g\indi{^p^q} \mvardif{G}{M_q}\,\dx &= \int_\Omega h_p \g\indi{^p^q}\mpd{A_{ik}}{M_q} \g\indi{^k^\ell} \nabla_\ell (\g\indi{^i^j} a_j)  + A_{ik} \g\indi{^k^\ell} \nabla_\ell (h_p \g\indi{^p^q}\g\indi{^i^j}\mpd{a_j}{M_q} ) \,\dx.\label{eq:functional_derivative_d}
\end{alignat}
\end{subequations}
Furthermore, for the covariant derivative of the functional derivatives of $F$ and $G$ we have 
\allowdisplaybreaks
\begin{align*}
&\int_\Omega \phi M_p \g\indi{^p^q} \nabla_q \mvardif{F}{\tilde{z}_\alpha}  \,\dx\\
 =& \int_\Omega \phi M_p \g\indi{^p^q} \nabla_q \Big( \mpd{a_k}{\tilde{z}_\alpha} \g\indi{^k^\ell} \nabla_\ell \chi + a_k \g\indi{^k^\ell} \nabla_\ell \mpd{\chi}{\tilde{z}_\alpha}\Big) +  \nabla_\ell( \phi M_p \g\indi{^p^q})\nabla_q ( a_k \g\indi{^k^\ell} \mpd{\chi}{\tilde{z}_\alpha}) \,\dx, \\
&\int_\Omega \phi M_p \g\indi{^p^q} \nabla_q \mvardif{G}{\tilde{z}_\alpha}  \,\dx\\
=&  \int_\Omega \phi M_p \g\indi{^p^q} \nabla_q \Big(\mpd{A_{ik}}{\tilde{z}_\alpha} \g\indi{^k^\ell} \nabla_\ell (\g\indi{^i^j}a_j)  + A_{ik} \g\indi{^k^\ell} \nabla_\ell (\g\indi{^i^j}\mpd{a_j}{\tilde{z}_\alpha}) \Big)  + \nabla_\ell(\phi M_p \g\indi{^p^q}) \nabla_q (A_{ik} \g\indi{^k^\ell}  \g\indi{^i^j}\mpd{a_j}{\tilde{z}_\alpha} )\,\dx, \\
&\int_\Omega h_{s}  M_p \g\indi{^p^q}\nabla_q ( \g\indi{^s^t}\mvardif{F}{M_t})\,\dx\\ 
=& \int_\Omega h_{s}  M_p \g\indi{^p^q}\nabla_q ( \g\indi{^s^t}\mpd{a_k}{M_t} \g\indi{^k^\ell} \nabla_\ell \chi +\g\indi{^s^t} a_k \g\indi{^k^\ell} \nabla_\ell \mpd{\chi}{M_t}) +  \nabla_\ell( h_{s} \g\indi{^p^q} M_p)\nabla_q ( \g\indi{^s^t}a_k \g\indi{^k^\ell} \mpd{\chi}{M_t}) \,\dx, \\
&\int_\Omega h_{s}  M_p \g\indi{^p^q}\nabla_q (\g\indi{^s^t} \mvardif{G}{M_t}) \,\dx\\ 
=& \int_\Omega  h_{s}  M_p \g\indi{^p^q}\nabla_q \Big(\g\indi{^s^t}\mpd{A_{ik}}{M_t} \g\indi{^k^\ell} \nabla_\ell (\g\indi{^i^j} a_j)  + \g\indi{^s^t}A_{ik} \g\indi{^k^\ell} \nabla_\ell (\g\indi{^i^j}\mpd{a_j}{M_t} )\Big) \\
&\qquad+ \nabla_{\ell}( h_{s}  M_p \g\indi{^p^q})\nabla_q(\g\indi{^s^t}A_{ik} \g\indi{^k^\ell} \g\indi{^i^j}\mpd{a_j}{M_t}) \,\dx.
\end{align*}
\end{lemma}
\begin{proof}
We will only prove the first equation for the integral equations without and with the spatial version of the functional derivatives. The other equation follows with analogous arguments. The first equality follows by the definition of the functional derivative via  
\begin{align*}
&\int_\Omega \phi \mvardif{F}{\tilde{z}_\alpha} \, \dx\\
=& \lim_{\varepsilon \searrow 0} \tfrac{1}{\varepsilon} \{F(\tilde{z}_1,\ldots,\tilde{z}_{\alpha-1},\tilde{z}_\alpha+\varepsilon \phi, z_{\alpha+1},\ldots,z_{\mu+1},\mathbf{M}) - F(\zhet)\}\\
=& \lim_{\varepsilon \searrow 0} \int_\Omega \tfrac{1}{\varepsilon} \Big\{ a_k \g^{k\ell} \nabla_\ell \chi + \varepsilon \Big(\phi \mpd{a}{\tilde{z}_\alpha} a_k \g^{k\ell} \nabla_\ell \chi  + a_k \g^{k\ell} \nabla_\ell\big( \phi \mpd{\chi}{\tilde{z}_\alpha}\big)\Big) + \mathcal{O}(\varepsilon^2) -  a_k \g^{k\ell} \nabla_\ell \chi\Big\} \, \dx\\
=&\int_\Omega \phi \mpd{a_k}{\tilde{z}_\alpha} \g\indi{^k^\ell} \nabla_\ell \chi + a_k \g\indi{^k^\ell} \nabla_\ell (\phi \mpd{\chi}{\tilde{z}_\alpha}  )\,\dx,
\end{align*}
where we have made use of the smoothness of $\mathbf{a}$ and $b$ and that $\zhet + [0,\ldots,0,\varepsilon \phi,0\ldots,0]$ is an element of $\mc{D}_\zhet$ for small enough $\varepsilon > 0$. For the first equation with a covariant derivative, let $\phi\in C^\infty(\Omega)\cap W^{1,p}(\Omega)$. By a formal calculation we get
\begin{equation}
\label{eq:functional_derivative_spatial_a}
\begin{split}
&\int_\Omega \phi M_p \g\indi{^p^q} \nabla_q \mvardif{F}{\tilde{z}_\alpha}  \,\dx \\
=&\int_\Omega \nabla_q \Big(\phi M_p \g\indi{^p^q}  \mvardif{F}{\tilde{z}_\alpha}\Big)- \nabla_q (\phi M_p \g\indi{^p^q}) \mvardif{F}{\tilde{z}_\alpha} \,\dx\\
=&\int_\Omega \nabla_q \Big(\phi M_p \g\indi{^p^q} \mpd{a_k}{\tilde{z}_\alpha} \g\indi{^k^\ell} \nabla_\ell \chi + a_k \g\indi{^k^\ell} \nabla_\ell \Big(\phi M_p \g\indi{^p^q} \mpd{\chi}{\tilde{z}_\alpha}  \Big)\Big)\\
&\qquad - \nabla_q (\phi M_p \g\indi{^p^q})\mpd{a_k}{\tilde{z}_\alpha} \g\indi{^k^\ell} \nabla_\ell \chi - a_k \g\indi{^k^\ell} \nabla_\ell \Big(\nabla_q (\phi M_p \g\indi{^p^q})\mpd{\chi}{\tilde{z}_\alpha}  \Big) \,\dx \\
 =& \int_\Omega \phi M_p \g\indi{^p^q} \nabla_q \Big( \mpd{a_k}{\tilde{z}_\alpha} \g\indi{^k^\ell} \nabla_\ell \chi + a_k \g\indi{^k^\ell} \nabla_\ell \mpd{\chi}{\tilde{z}_\alpha}\Big) +  \nabla_\ell ( \phi M_p \g\indi{^p^q})\nabla_q \Big( a_k \g\indi{^k^\ell} \mpd{\chi}{\tilde{z}_\alpha}\Big) \,\dx.
\end{split}
\end{equation}
Note that the second and third line is well-defined by the smoothness of $\phi$ and $\zhet$. Since $C^\infty(\Omega)\cap W^{1,p}(\Omega)$ is dense in $W^{1,p}(\Omega)$ and since the last line is well-defined for $\phi \in W^{1,p}(\Omega)$ if $\zhet$ is smooth enough, the integral equality~\eqref{eq:functional_derivative_spatial_a} can be extended to $W^{1,p}(\Omega)$. 
\end{proof}

\begin{proof}[Proof of Theorem~\ref{th:jacobi_identity}]
Let us define the brackets 
\begin{align*}
\{A,B\}_1 &:= \int_\Omega  M_i \Big[\mvardif{A}{M_k} \g\indi{^k^\ell} \nabla_\ell (\g\indi{^i^j}\mvardif{B}{M_j}) - \mvardif{B}{M_k} \g\indi{^k^\ell} \nabla_\ell (\g\indi{^i^j}\mvardif{A}{M_j})\Big]\, \dx,\\
\{A,B\}_2 &:= \int_\Omega \sum_{\alpha=1}^{\mu+1} \tilde{z}_\alpha \Big[\mvardif{A}{M_k} \g\indi{^k^\ell} \nabla_\ell \mvardif{B}{\tilde{z}_\alpha} - \mvardif{B}{M_k} \g\indi{^k^\ell} \nabla_\ell \mvardif{A}{\tilde{z}_\alpha}\Big]\, \dx,\\ 
\{A,B\}_3 &:= \int_\Omega \sum_{\alpha=1}^{\mu+1} \Big[\mvardif{A}{M_k} \g\indi{^k^\ell} \nabla_\ell  (f_\alpha \mvardif{B}{\tilde{z}
_\alpha}) - \mvardif{B}{M_k} \g\indi{^k^\ell} \nabla_\ell  (f_\alpha \mvardif{A}{\tilde{z}_\alpha})\Big]\,\dx. 
\end{align*}
Then, it obviously holds that $\{\cdot,\cdot\}= -\{\cdot,\cdot\}_1 - \{\cdot,\cdot\}_2 +  \{\cdot,\cdot\}_3$ for the bracket defined in~\eqref{eq:jacobi_identity_bracket}. For the proof we consider all possible combinations of the brackets and number the terms which arise from applying Lemma~\ref{lem:functional_derivative} in these integral equations. Then, these terms will be split into subterms and we will show which combination of subterms sum up to zero under cyclic summation. Note that, a single term can annul itself under cyclic summation. If two terms only vanish together, we write "together with". 
At first, let us consider the bracket $\{\cdot,\cdot\}_1$ for which
\allowdisplaybreaks
\begin{align*}
 &\{A,\{B,C\}_{1}\}_{1}\\
=& \int_\Omega M_s\mvardif{A}{M_p}\g\indi{^p^q}\nabla_q \Big( \g\indi{^s^t}\mvardif{}{M_t} \int_\Omega  M_i \Big[\mvardif{B}{M_k}\g\indi{^k^\ell}\nabla_\ell (\g\indi{^i^j}\mvardif{C}{M_j}) - \mvardif{C}{M_k}\g\indi{^k^\ell}\nabla_\ell( \g\indi{^i^j}\mvardif{B}{M_j})\Big] \,\dx^\prime\Big)\\
&\qquad \qquad - M_s\nabla_q \Big(\g\indi{^s^t}\mvardif{A}{M_t}\Big) \g\indi{^q^p}\mvardif{}{M_p} \int_\Omega   M_i \Big[\mvardif{B}{M_k}\g\indi{^k^\ell}\nabla_\ell( \g\indi{^i^j}\mvardif{C}{M_j} )- \mvardif{C}{M_k}\g\indi{^k^\ell}\nabla_\ell (\g\indi{^i^j}\mvardif{B}{M_j})\Big] \,\dx^\prime \,\dx\\
=& \int_\Omega M_s\mvardif{A}{M_p}\g\indi{^p^q}\nabla_q  \Big(\g\indi{^s^t}\Big[\mvardif{B}{M_k}\g\indi{^k^\ell}\nabla_\ell \mvardif{C}{M_t} - \mvardif{C}{M_k}\g\indi{^k^\ell}\nabla_\ell \mvardif{B}{M_t}\Big]\Big){}_{(1)}\\
& \phantom{\int}  + M_s\mvardif{A}{M_p}\g\indi{^p^q}\nabla_q \Big( \g\indi{^s^t} M_i \g\indi{^i^j} \Big[\mvardif{^2 B}{M_k \delta M_t}\g\indi{^k^\ell}\nabla_\ell \mvardif{C}{M_j} - \mvardif{^2 C}{M_k \delta M_t}\g\indi{^k^\ell}\nabla_\ell \mvardif{B}{M_j}\Big]\Big){}_{(2)} \\
& \phantom{\int}  +M_s\mvardif{A}{M_p}\g\indi{^p^q}\nabla_q   \Big(\g\indi{^s^t} M_i \g\indi{^i^j} \Big[\mvardif{B}{M_k}\g\indi{^k^\ell}\nabla_\ell \mvardif{^2 C}{M_j \delta M_t} - \mvardif{C}{M_k}\g\indi{^k^\ell}\nabla_\ell \mvardif{^2 B}{M_j \delta M_t}\Big]\Big){}_{(3)}\\
& \phantom{\int}  +  \nabla_\ell  (M_s\mvardif{A}{M_p} )\g\indi{^\ell^k}\g\indi{^p^q}\nabla_q  \Big(\g\indi{^s^t} M_i\Big[\mvardif{B}{M_k}\mvardif{^2 C}{M_t \delta M_j} - \mvardif{C}{M_k}\mvardif{^2 B}{M_t \delta M_j}\Big]\Big) {}_{(4)}\\
& \phantom{\int} -M_s \nabla_q (\g\indi{^s^t}\mvardif{A}{M_t})\g\indi{^q^p} \Big[\mvardif{B}{M_k}\g\indi{^k^\ell}\nabla_\ell \mvardif{C}{M_p} - \mvardif{C}{M_k}\g\indi{^k^\ell}\nabla_\ell \mvardif{B}{M_p}\Big]{}_{(5)}\\
& \phantom{\int}  -  M_sM_i\nabla_q (\g\indi{^s^t}\mvardif{A}{M_t}) \g\indi{^q^p} \g\indi{^i^j} \Big[\mvardif{^2 B}{M_p \delta M_k}\g\indi{^k^\ell}\nabla_\ell \mvardif{C}{M_j} - \mvardif{^2 C}{M_p \delta M_k}\g\indi{^k^\ell}\nabla_\ell \mvardif{B}{M_j}\Big]{}_{(6)}\\
& \phantom{\int} - M_i \g\indi{^i^j} \Big[\mvardif{B}{M_k}\g\indi{^k^\ell}\nabla_\ell ( \mvardif{^2 C}{M_j \delta M_p} M_s \g\indi{^p^q}\nabla_q( \g\indi{^s^t}\mvardif{A}{M_t})) - \mvardif{C}{M_k}\g\indi{^k^\ell}\nabla_\ell (\mvardif{^2 B}{M_j \delta M_p} M_s \g\indi{^p^q}\nabla_q( \g\indi{^s^t}\mvardif{A}{M_t}))\Big]{}_{(7)}\, \dx\\
=& \int_\Omega  M_s\mvardif{A}{M_p}\g\indi{^p^q}\g\indi{^s^t}\Big[\mvardif{B}{M_k}\nabla_q(\g\indi{^k^\ell}\nabla_\ell \mvardif{C}{M_t}) - \mvardif{C}{M_k}\nabla_q(\g\indi{^k^\ell}\nabla_\ell \mvardif{B}{M_t})\Big]{}_{(1.a)}\\
& \phantom{\int} 
 + M_s\mvardif{A}{M_p}\g\indi{^p^q} \g\indi{^s^t} \Big[\nabla_q  (\mvardif{B}{M_k})\g\indi{^k^\ell}\nabla_\ell (\mvardif{C}{M_t}) -\nabla_q(\mvardif{C}{M_k})\g\indi{^k^\ell}\nabla_\ell (\mvardif{B}{M_t})\Big]{}_{(1.b)}\\
& \phantom{\int}  + M_s\mvardif{A}{M_p}\g\indi{^p^q}\nabla_q \Big( \g\indi{^s^t} M_i \g\indi{^i^j} \Big[\mvardif{^2 B}{M_k \delta M_t}\g\indi{^k^\ell}\nabla_\ell \mvardif{C}{M_j} - \mvardif{^2 C}{M_k \delta M_t}\g\indi{^k^\ell}\nabla_\ell \mvardif{B}{M_j}\Big]\Big){}_{(2)} \\
& \phantom{\int} 
+  M_sM_i\mvardif{A}{M_p} \g\indi{^p^q} \g\indi{^s^t} \g\indi{^i^j} \Big[\mvardif{B}{M_k}\nabla_q( \g\indi{^k^\ell}\nabla_\ell \mvardif{^2 C}{M_j \delta M_t}) - \mvardif{C}{M_k}\nabla_q ( \g\indi{^k^\ell}\nabla_\ell \mvardif{^2 B}{M_j \delta M_t})\Big]{}_{(3.a)} \\
& \phantom{\int}  + M_s\mvardif{A}{M_p}\g\indi{^p^q} \g\indi{^s^t} \Big[\nabla_q  (M_i \mvardif{B}{M_k})\g\indi{^i^j} \g\indi{^k^\ell}\nabla_\ell \mvardif{^2 C}{M_j \delta M_t} - \nabla_q  (M_i\mvardif{C}{M_k})\g\indi{^i^j} \g\indi{^k^\ell}\nabla_\ell \mvardif{^2 B}{M_j \delta M_t}\Big]{}_{(3.b)} \\
& \phantom{\int} 
+  M_i\nabla_\ell ( M_s\mvardif{A}{M_p} )\g\indi{^i^j} \g\indi{^\ell^k}\g\indi{^p^q}\g\indi{^s^t}\Big[\mvardif{B}{M_k}\nabla_q\mvardif{^2 C}{M_t \delta M_j} - \mvardif{C}{M_k}\nabla_q\mvardif{^2 B}{M_t \delta M_j}\Big]  {}_{(4.a)} \\
& \phantom{\int} 
+ \nabla_\ell ( M_s\mvardif{A}{M_p} ) \g\indi{^\ell^k}\g\indi{^p^q}\g\indi{^s^t}\Big[\nabla_q ( M_i \mvardif{B}{M_k})\g\indi{^i^j}\mvardif{^2 C}{M_t \delta M_j} - \nabla_q (M_i \mvardif{C}{M_k})\g\indi{^i^j}\mvardif{^2 B}{M_t \delta M_j}\Big]{}_{(4.b)} \\
& \phantom{\int} -M_s \nabla_q (\g\indi{^s^t}\mvardif{A}{M_t})\g\indi{^q^p} \Big[\mvardif{B}{M_k}\g\indi{^k^\ell}\nabla_\ell \mvardif{C}{M_p} - \mvardif{C}{M_k}\g\indi{^k^\ell}\nabla_\ell \mvardif{B}{M_p}\Big]{}_{(5)}\\
& \phantom{\int}  -  M_sM_i\nabla_q (\g\indi{^s^t}\mvardif{A}{M_t}) \g\indi{^q^p} \g\indi{^i^j} \Big[\mvardif{^2 B}{M_p \delta M_k}\g\indi{^k^\ell}\nabla_\ell \mvardif{C}{M_j} - \mvardif{^2 C}{M_p \delta M_k}\g\indi{^k^\ell}\nabla_\ell \mvardif{B}{M_j}\Big]{}_{(6)}\\
& \phantom{\int} - M_i \g\indi{^i^j} \Big[\mvardif{B}{M_k}\g\indi{^k^\ell}\nabla_\ell ( \mvardif{^2 C}{M_j \delta M_p} M_s \g\indi{^p^q}\nabla_q( \g\indi{^s^t}\mvardif{A}{M_t})) - \mvardif{C}{M_k}\g\indi{^k^\ell}\nabla_\ell (\mvardif{^2 B}{M_j \delta M_p} M_s \g\indi{^p^q}\nabla_q( \g\indi{^s^t}\mvardif{A}{M_t}))\Big]{}_{(7)}\, \dx.
\end{align*}
Then the cyclic sum of $(1.a)$; $(1.b)$ together with $(5)$; $(2)$ together with $(7)$; $(3.a)$; $(3.b)$ together with $(4.a)$; $(4.b)$; as well as $(6)$ vanish.
If we consider~$\{A,\{B,C\}_{2}\}_{2}$ then terms equivalent to the ones from $\{A,\{B,C\}_{1}\}_{1}$ will arise with the exception of $(5)$, since $\tvardif{\tilde{z}_\beta}{M_k}=0.$ Therefore, only the term $\int_\Omega \sum_{\alpha=1}^{\mu+1} 
\tilde{z}_\alpha\tvardif{A}{M_p}\g\indi{^p^q} [\nabla_q(\tvardif{B}{M_k})\g\indi{^k^\ell}\nabla_\ell (\tvardif{C}{\tilde{z}_\alpha}) -\nabla_q(\tvardif{C}{M_k})\g\indi{^k^\ell}\nabla_\ell (\tvardif{B}{\tilde{z}_\alpha})]\,\dx$, which is equivalent to $(1.b)$ from above, will not vanish under cyclic summation of~$\{A,\{B,C\}_{2}\}_{2}$.  This term will vanish under cyclic summation with a term of $\{A,\{B,C\}_{1}\}_{2} + \{A,\{B,C\}_{2}\}_{1}$, to be more precise with a part of
$-\int_{\Omega}  \sum_{\alpha=1}^{\mu+1} 
\tilde{z}_\alpha \nabla_\ell(\tvardif{A}{\tilde{z}_\alpha})\g\indi{^\ell^k} \tvardif{\{B,C\}_{2}}{M_k}\,\dx$. The remaining terms of $\{A,\{B,C\}_{1}\}_{2} + \{A,\{B,C\}_{2}\}_{1}$ will also cyclic sum up to zero. To prove this, the splitting of the subterms is similar to these of $\{A,\{B,C\}_{1}\}_{1}$. In the end, $\{A,\{B,C\}_{2}\}_{2} + \{A,\{B,C\}_{1}\}_{2} + \{A,\{B,C\}_{2}\}_{1}$ vanish under cyclic summation.

For the third part of the bracket we derive
\begin{align*}
&\{A,\{B,C\}_{3}\}_{3}\\
=& \int_\Omega \sum_{\alpha=1}^{\mu+1} f_\alpha\mvardif{A}{M_p} \g\indi{^p^q}\nabla_q \Big(\mvardif{}{\tilde{z}_\alpha} \int_\Omega \sum_{\beta=1}^{\mu+1}\Big[ \mvardif{B}{M_k}\g\indi{^k^\ell}\nabla_\ell (f_\beta\mvardif{C}{\tilde{z}_\beta} ) - \mvardif{C}{M_k}\g\indi{^k^\ell}\nabla_\ell(f_\beta\mvardif{B}{\tilde{z}_\beta} ) \Big]\,\dx^\prime\Big) \\
& \phantom{\int} +\sum_{\alpha=1}^{\mu+1} \nabla_q (f_\alpha)  \g\indi{^q^p} \mvardif{A}{M_p}\mvardif{}{\tilde{z}_\alpha} \int_\Omega \sum_{\beta=1}^{\mu+1} \Big[\mvardif{B}{M_k}\g\indi{^k^\ell}\nabla_\ell (f_\beta \mvardif{C}{\tilde{z}_\beta} ) - \mvardif{C}{M_k}\g\indi{^k^\ell}\nabla_\ell(f_\beta \mvardif{B}{\tilde{z}_\beta})\Big] \,\dx^\prime\\
& \phantom{\int}   - \sum_{\alpha=1}^{\mu+1} \nabla_q (f_\alpha \mvardif{A}{\tilde{z}_\alpha})\g\indi{^q^p}\mvardif{ }{M_p} \int_\Omega \sum_{\beta=1}^{\mu+1}  \Big[\mvardif{B}{M_k}\g\indi{^k^\ell}\nabla_\ell (f_\beta \mvardif{C}{\tilde{z}_\beta}) - \mvardif{C}{M_k}\g\indi{^k^\ell}\nabla_\ell(f_\beta \mvardif{B}{\tilde{z}_\beta})\Big] \,\dx^\prime\,\dx\\
=& \int_\Omega \sum_{\alpha,\beta=1}^{\mu+1} \mvardif{A}{M_p} \g\indi{^p^q}\nabla_q \Big(f_\alpha [\mvardif{^2 B}{M_k \delta \tilde{z}_\alpha}\g\indi{^k^\ell}\nabla_\ell (f_\beta \mvardif{C}{\tilde{z}_\beta} ) - \mvardif{^2 C}{M_k \delta \tilde{z}_\alpha}\g\indi{^k^\ell}\nabla_\ell(f_\beta \mvardif{B}{\tilde{z}_\beta} )]\Big){}_{(1)}\\
& \phantom{\int} + \sum_{\alpha,\beta=1}^{\mu+1} \mvardif{A}{M_p}\g\indi{^p^q}\nabla_q  \Big( f_\alpha[\mvardif{B}{M_k}\g\indi{^k^\ell}\nabla_\ell (\mpd{f_\beta}{\tilde{z}_\alpha}\mvardif{C}{\tilde{z}_\beta} + f_\beta \mvardif{^2 C}{\tilde{z}_\beta \delta \tilde{z}_\alpha}) - \mvardif{C}{M_k}\g\indi{^k^\ell}\nabla_\ell(\mpd{f_\beta}{\tilde{z}_\alpha}\mvardif{B}{\tilde{z}_\beta} + f_\beta \mvardif{^2 B}{\tilde{z}_\beta \delta \tilde{z}_\alpha})]\Big){}_{(2)} \\
& \phantom{\int} +\sum_{\alpha,\beta=1}^{\mu+1} \nabla_\ell ( f_\alpha \mvardif{A}{M_p}) \g\indi{^\ell^k} \g\indi{^p^q}\nabla_q \Big(\mvardif{B}{M_k} (\mpd{f_\beta}{\tilde{z}_\alpha}  \mvardif{C}{\tilde{z}_\beta} + f_\beta \mvardif{^2 C}{\tilde{z}_\beta \delta \tilde{z}_\alpha}) - \mvardif{C}{M_k}(\mpd{f_\beta}{\tilde{z}_\alpha}  \mvardif{B}{\tilde{z}_\beta} + f_\beta \mvardif{^2 B}{\tilde{z}_\beta \delta \tilde{z}_\alpha})\Big) {}_{(3)}\\
& \phantom{\int} + \sum_{\alpha,\beta=1}^{\mu+1} \nabla_\ell ( \mvardif{A}{M_p} \g\indi{^p^q}\nabla_q(f_\alpha))\g\indi{^\ell^k} [\mvardif{B}{M_k} ( \mpd{f_\beta}{\tilde{z}_\alpha}\mvardif{C}{\tilde{z}_\beta} + f_\beta \mvardif{^2 C}{\tilde{z}_\beta \delta \tilde{z}_\alpha} ) - \mvardif{C}{M_k}(\mpd{f_\beta}{\tilde{z}_\alpha}\mvardif{B}{\tilde{z}_\beta} + f_\beta \mvardif{^2 B}{\tilde{z}_\beta \delta \tilde{z}_\alpha} )]{}_{(4)}\\
& \phantom{\int}-\sum_{\alpha,\beta=1}^{\mu+1}  \nabla_q( f_\alpha \mvardif{A}{\tilde{z}_\alpha})\g\indi{^q^p} [\mvardif{^2 B}{M_p \delta M_k}\g\indi{^k^\ell}\nabla_\ell ( f_\beta \mvardif{C}{\tilde{z}_\beta}) - \mvardif{^2 C}{M_p \delta M_k}\g\indi{^k^\ell}\nabla_\ell( f_\beta \mvardif{B}{\tilde{z}_\beta} )]{}_{(5)}\\
& \phantom{\int} -\sum_{\alpha,\beta=1}^{\mu+1}[\mvardif{B}{M_k}\g\indi{^k^\ell}\nabla_\ell (f_\beta \mvardif{^2C}{\tilde{z}_\beta \delta M_p} \g\indi{^p^q}\nabla_q(f_\alpha \mvardif{A}{\tilde{z}_\alpha} )) - \mvardif{C}{M_k }\g\indi{^k^\ell}\nabla_\ell (f_\beta \mvardif{^2 B}{\tilde{z}_\beta  \delta M_p} \g\indi{^p^q}\nabla_q( f_\alpha \mvardif{A}{\tilde{z}_\alpha}))]{}_{(6)}\,\dx\\
=& \int_\Omega \sum_{\alpha,\beta=1}^{\mu+1} \mvardif{A}{M_p} \g\indi{^p^q}\nabla_q \Big(f_\alpha [\mvardif{^2 B}{M_k \delta \tilde{z}_\alpha}\g\indi{^k^\ell}\nabla_\ell (f_\beta \mvardif{C}{\tilde{z}_\beta} ) - \mvardif{^2 C}{M_k \delta \tilde{z}_\alpha}\g\indi{^k^\ell}\nabla_\ell(f_\beta \mvardif{B}{\tilde{z}_\beta} )]\Big){}_{(1)}\\
& \phantom{\int} +\sum_{\alpha,\beta=1}^{\mu+1}  f_\alpha\mvardif{A}{M_p} \g\indi{^p^q}[\mvardif{B}{M_k} \nabla_q( \g\indi{^k^\ell}\nabla_\ell (\mpd{f_\beta}{\tilde{z}_\alpha}\mvardif{C}{\tilde{z}_\beta} + f_\beta \mvardif{^2 C}{\tilde{z}_\beta \delta \tilde{z}_\alpha})) - \mvardif{C}{M_k}\nabla_q( \g\indi{^k^\ell}\nabla_\ell(\mpd{f_\beta}{\tilde{z}_\alpha}\mvardif{B}{\tilde{z}_\beta} + f_\beta \mvardif{^2 B}{\tilde{z}_\beta \delta \tilde{z}_\alpha}))] {}_{(2.a)} \\
& \phantom{\int} +\sum_{\alpha,\beta=1}^{\mu+1} \mvardif{A}{M_p}\g\indi{^p^q} [\nabla_q( f_\alpha \mvardif{B}{M_k})\g\indi{^k^\ell}\nabla_\ell (\mpd{f_\beta}{\tilde{z}_\alpha}\mvardif{C}{\tilde{z}_\beta} + f_\beta \mvardif{^2 C}{\tilde{z}_\beta \delta \tilde{z}_\alpha}) - \nabla_q( f_\alpha\mvardif{C}{M_k})\g\indi{^k^\ell}\nabla_\ell(\mpd{f_\beta}{\tilde{z}_\alpha}\mvardif{B}{\tilde{z}_\beta} + f_\beta \mvardif{^2 B}{\tilde{z}_\beta \delta \tilde{z}_\alpha})] {}_{(2.b)} \\
& \phantom{\int} + \sum_{\alpha,\beta=1}^{\mu+1}\nabla_\ell ( f_\alpha \mvardif{A}{M_p})\g\indi{^\ell^k} \g\indi{^p^q} [\mvardif{B}{M_k}\nabla_q (\mpd{f_\beta}{\tilde{z}_\alpha}  \mvardif{C}{\tilde{z}_\beta} + f_\beta \mvardif{^2 C}{\tilde{z}_\beta \delta \tilde{z}_\alpha}) - \mvardif{C}{M_k}\nabla_q (\mpd{f_\beta}{\tilde{z}_\alpha}  \mvardif{B}{\tilde{z}_\beta} + f_\beta \mvardif{^2 B}{\tilde{z}_\beta \delta \tilde{z}_\alpha})] {}_{(3.a)}\\
& \phantom{\int} +  \sum_{\alpha,\beta=1}^{\mu+1}f_\alpha \nabla_\ell (\mvardif{A}{M_p})\g\indi{^\ell^k}\g\indi{^p^q} [\nabla_q(\mvardif{B}{M_k}) (\mpd{f_\beta}{\tilde{z}_\alpha}  \mvardif{C}{\tilde{z}_\beta} + f_\beta \mvardif{^2 C}{\tilde{z}_\beta \delta \tilde{z}_\alpha}) - \nabla_q(\mvardif{C}{M_k})(\mpd{f_\beta}{\tilde{z}_\alpha}  \mvardif{B}{\tilde{z}_\beta} + f_\beta \mvardif{^2 B}{\tilde{z}_\beta \delta \tilde{z}_\alpha})]  {}_{(3.b)}\\
& \phantom{\int} +\sum_{\alpha,\beta=1}^{\mu+1} \mvardif{A}{M_p}\nabla_\ell ( f_\alpha) \g\indi{^\ell^k} \g\indi{^p^q}[\nabla_q(\mvardif{B}{M_k}) (\mpd{f_\beta}{\tilde{z}_\alpha}  \mvardif{C}{\tilde{z}_\beta} + f_\beta \mvardif{^2 C}{\tilde{z}_\beta \delta \tilde{z}_\alpha}) -\nabla_q(\mvardif{C}{M_k})(\mpd{f_\beta}{\tilde{z}_\alpha}  \mvardif{B}{\tilde{z}_\beta} + f_\beta \mvardif{^2 B}{\tilde{z}_\beta \delta \tilde{z}_\alpha})] {}_{(3.c)}\\
& \phantom{\int} +\sum_{\alpha,\beta=1}^{\mu+1} \nabla_q(f_\alpha) \g\indi{^q^p} \nabla_\ell ( \mvardif{A}{M_p})\g\indi{^\ell^k} [\mvardif{B}{M_k} ( \mpd{f_\beta}{\tilde{z}_\alpha}\mvardif{C}{\tilde{z}_\beta} + f_\beta \mvardif{^2 C}{\tilde{z}_\beta \delta \tilde{z}_\alpha} ) - \mvardif{C}{M_k}(\mpd{f_\beta}{\tilde{z}_\alpha}\mvardif{B}{\tilde{z}_\beta} + f_\beta \mvardif{^2 B}{\tilde{z}_\beta \delta \tilde{z}_\alpha} )]{}_{(4.a)}\\
& \phantom{\int} + \sum_{\alpha,\beta=1}^{\mu+1}\nabla_\ell ( \g\indi{^q^p}\nabla_q f_\alpha ) \mvardif{A}{M_p}\g\indi{^\ell^k} [\mvardif{B}{M_k} ( \mpd{f_\beta}{\tilde{z}_\alpha}\mvardif{C}{\tilde{z}_\beta} + f_\beta \mvardif{^2 C}{\tilde{z}_\beta \delta \tilde{z}_\alpha} ) - \mvardif{C}{M_k}(\mpd{f_\beta}{\tilde{z}_\alpha}\mvardif{B}{\tilde{z}_\beta} + f_\beta \mvardif{^2 B}{\tilde{z}_\beta \delta \tilde{z}_\alpha} )] {}_{(4.b)}\\
& \phantom{\int}-\sum_{\alpha,\beta=1}^{\mu+1}  \nabla_q( f_\alpha \mvardif{A}{\tilde{z}_\alpha})\g\indi{^q^p} [\mvardif{^2 B}{M_p \delta M_k}\g\indi{^k^\ell}\nabla_\ell ( f_\beta \mvardif{C}{\tilde{z}_\beta}) - \mvardif{^2 C}{M_p \delta M_k}\g\indi{^k^\ell}\nabla_\ell( f_\beta \mvardif{B}{\tilde{z}_\beta} )]{}_{(5)}\\
& \phantom{\int} -\sum_{\alpha,\beta=1}^{\mu+1}[\mvardif{B}{M_k}\g\indi{^k^\ell}\nabla_\ell (f_\beta \mvardif{^2C}{\tilde{z}_\beta \delta M_p} \g\indi{^p^q}\nabla_q(f_\alpha \mvardif{A}{\tilde{z}_\alpha} )) - \mvardif{C}{M_k }\g\indi{^k^\ell}\nabla_\ell (f_\beta \mvardif{^2 B}{\tilde{z}_\beta  \delta M_p} \g\indi{^p^q}\nabla_q( f_\alpha \mvardif{A}{\tilde{z}_\alpha}))]{}_{(6)}\,\dx.
\end{align*}
Then under cyclic summation the terms $(1)$ together with $(6)$; $(2.a)$; $(2.b)$ together with $(3.a)$; $(3.b)$; $(3.c)$ together with $(4.a)$; $(4.b)$; and $(5)$ vanish.
For mixed terms with $\{\cdot,\cdot\}_3$ we observe, that 
\begin{align*}
&\{A,\{B,C\}_3\}_2+ \{A,\{B,C\}_2\}_3\\
=& \int_\Omega \sum_{\alpha=1}^{\mu+1} \tilde{z}_\alpha \mvardif{A}{M_p}\g\indi{^p^q}\nabla_q \Big(\mvardif{ }{\tilde{z}_\alpha} \int_\Omega \sum_{\beta=1}^{\mu+1} \mvardif{B}{M_k}\g\indi{^k^\ell}\nabla_{\ell} ( f_\beta \mvardif{C}{\tilde{z}_\beta} ) - \mvardif{C}{M_k}\g\indi{^k^\ell}\nabla_{\ell}( f_\beta \mvardif{B}{\tilde{z}_\beta}  ) \, \dx^\prime\Big) \\
 & \phantom{\int} \qquad \qquad - \sum_{\alpha=1}^{\mu+1} \tilde{z}_\alpha \nabla_q (\mvardif{A}{\tilde{z}_\alpha}) \g\indi{^q^p} \mvardif{ }{M_p} \int_\Omega \sum_{\beta=1}^{\mu+1}\mvardif{B}{M_k}\g\indi{^k^\ell}\nabla_{\ell} (f_\beta \mvardif{C}{  \tilde{z}_\beta} ) - \mvardif{C}{M_k}\g\indi{^k^\ell}\nabla_{\ell}( f_\beta \mvardif{B}{\tilde{z}_\beta} ) \, \dx^\prime\,\dx \\
&  +  \int_\Omega \sum_{\beta=1}^{\mu+1} \mvardif{A}{M_p}\g\indi{^p^q}\nabla_q \Big( f_\beta \mvardif{ }{\tilde{z}_\beta} \int_\Omega \sum_{\alpha=1}^{\mu+1} \tilde{z}_\alpha [\mvardif{B}{M_k}\g\indi{^k^\ell}\nabla_{\ell} \mvardif{C}{\tilde{z}_\alpha} - \mvardif{C}{M_k}\g\indi{^k^\ell}\nabla_{\ell} \mvardif{B}{\tilde{z}_\alpha}] \, \dx^\prime \Big)\\
 & \phantom{\int} \qquad \qquad - \sum_{\beta=1}^{\mu+1}\nabla_q( f_\beta \mvardif{A}{\tilde{z}_\beta} )\g\indi{^q^p} \mvardif{ }{M_p}\int_\Omega \sum_{\alpha=1}^{\mu+1} \tilde{z}_\alpha [\mvardif{B}{M_k}\g\indi{^k^\ell}\nabla_{\ell} \mvardif{C}{\tilde{z}_\alpha} - \mvardif{C}{M_k}\g\indi{^k^\ell}\nabla_{\ell} \mvardif{B}{\tilde{z}_\alpha}] \, \dx^\prime  \,\dx\\
=& \int_\Omega \sum_{\alpha,\beta=1}^{\mu+1}  \tilde{z}_\alpha \mvardif{A}{M_p}\g\indi{^p^q}\nabla_q \Big( \mvardif{^2 B}{M_k \delta \tilde{z}_\alpha}\g\indi{^k^\ell}\nabla_{\ell} (f_\beta \mvardif{C}{\tilde{z}_\beta}  ) - \mvardif{^2 C}{M_k \delta \tilde{z}_\alpha}\g\indi{^k^\ell}\nabla_{\ell}(f_\beta \mvardif{B}{\tilde{z}_\beta}  )\Big){}_{(1)}\\
& \phantom{\int} +\sum_{\alpha,\beta=1}^{\mu+1} \tilde{z}_\alpha \mvardif{A}{M_p}\g\indi{^p^q}\nabla_q \Big( \mvardif{B}{M_k}\g\indi{^k^\ell}\nabla_{\ell} (\mpd{ f_\beta}{\tilde{z}_\alpha}\mvardif{C}{\tilde{z}_\beta} +  f_\beta \mvardif{^2 C}{\tilde{z}_\beta \delta \tilde{z}_\alpha}) - \mvardif{C}{M_k}\g\indi{^k^\ell}\nabla_{\ell}(\mpd{ f_\beta}{\tilde{z}_\alpha}\mvardif{B}{\tilde{z}_\beta} +  f_\beta \mvardif{^2 B}{\tilde{z}_\beta \delta \tilde{z}_\alpha})\Big)  {}_{(2)}\\
& \phantom{\int} +\sum_{\alpha,\beta=1}^{\mu+1} \nabla_{\ell} (\tilde{z}_\alpha \mvardif{A}{M_p})\g\indi{^\ell^k}\g\indi{^p^q}\nabla_q\Big(\mvardif{B}{M_k}(\mpd{ f_\beta}{\tilde{z}_\alpha}\mvardif{C}{\tilde{z}_\beta} +  f_\beta \mvardif{^2 C}{\tilde{z}_\beta \delta \tilde{z}_\alpha}) - \mvardif{C}{M_k}(\mpd{ f_\beta}{\tilde{z}_\alpha}\mvardif{B}{\tilde{z}_\beta} +  f_\beta \mvardif{^2 B}{\tilde{z}_\beta \delta \tilde{z}_\alpha})\Big) {}_{(3)}\\
& \phantom{\int} - \sum_{\alpha,\beta=1}^{\mu+1} \tilde{z}_\alpha \nabla_q (\mvardif{A}{\tilde{z}_\alpha})\g\indi{^q^p}[ \mvardif{^2 B}{M_p \delta M_k}\g\indi{^k^\ell}\nabla_{\ell} (f_\beta \mvardif{C}{  \tilde{z}_\beta} ) - \mvardif{^2 C}{M_p \delta M_k}\g\indi{^k^\ell}\nabla_{\ell}( f_\beta \mvardif{B}{\tilde{z}_\beta} )] {}_{(4)}\\
& \phantom{\int}  -\sum_{\alpha,\beta=1}^{\mu+1}  [\mvardif{B}{M_k}\g\indi{^k^\ell}\nabla_{\ell} (f_\beta \mvardif{^2 C}{\tilde{z}_\beta \delta M_p}  \tilde{z}_\alpha \g\indi{^p^q}\nabla_q \mvardif{A}{\tilde{z}_\alpha}) - \mvardif{C}{M_k}\g\indi{^k^\ell}\nabla_{\ell}( f_\beta \mvardif{^2 B}{\tilde{z}_\beta \delta M_p} \tilde{z}_\alpha \g\indi{^p^q}\nabla_q \mvardif{A}{\tilde{z}_\alpha})]{}_{(5)} \\
& \phantom{\int} +\sum_{\beta=1}^{\mu+1} \mvardif{A}{M_p}\g\indi{^p^q}\nabla_q \Big(  f_\beta [\mvardif{B}{M_k}\g\indi{^k^\ell}\nabla_{\ell} \mvardif{C}{\tilde{z}_\beta} - \mvardif{C}{M_k}\g\indi{^k^\ell}\nabla_{\ell} \mvardif{B}{\tilde{z}_\beta}]\Big)  {}_{(6)}\\
& \phantom{\int} + \sum_{\alpha,\beta=1}^{\mu+1} \mvardif{A}{M_p} \g\indi{^p^q}\nabla_q \Big( f_\beta  \tilde{z}_\alpha [\mvardif{^2 B}{M_k \delta \tilde{z}_\beta}\g\indi{^k^\ell}\nabla_{\ell} \mvardif{C}{\tilde{z}_\alpha} - \mvardif{^2 C}{M_k \delta \tilde{z}_\beta}\g\indi{^k^\ell}\nabla_{\ell} \mvardif{B}{\tilde{z}_\alpha}]\Big)  {}_{(7)} \\
& \phantom{ \int } + \sum_{\alpha,\beta=1}^{\mu+1} \mvardif{A}{M_p}\g\indi{^p^q}\nabla_q \Big( f_\beta \tilde{z}_\alpha [\mvardif{B}{M_k}\g\indi{^k^\ell}\nabla_{\ell} \mvardif{^2 C}{\tilde{z}_\alpha \delta \tilde{z}_\beta} - \mvardif{C}{M_k}\g\indi{^k^\ell}\nabla_{\ell} \mvardif{^2 B}{\tilde{z}_\alpha \delta \tilde{z}_\beta}] \Big) {}_{(8)}\\
& \phantom{\int} +\sum_{\alpha,\beta=1}^{\mu+1} \nabla_{\ell}(  f_\beta \mvardif{A}{M_p}) \g\indi{^\ell^k} \g\indi{^p^q}\nabla_q\Big( \tilde{z}_\alpha[\mvardif{B}{M_k}\mvardif{^2 C}{\tilde{z}_\alpha \delta \tilde{z}_\beta} - \mvardif{C}{M_k} \mvardif{^2 B}{\tilde{z}_\alpha \delta \tilde{z}_\beta}]\Big) {}_{(9)}\\
& \phantom{\int} + \sum_{\alpha,\beta=1}^{\mu+1}\tilde{z}_\alpha\nabla_{\ell}( \nabla_q( f_\beta)\g\indi{^q^p} \mvardif{A}{M_p}) \g\indi{^\ell^k} [\mvardif{B}{M_k}\mvardif{^2 C}{\tilde{z}_\alpha \delta \tilde{z}_\beta} - \mvardif{C}{M_k} \mvardif{^2 B}{\tilde{z}_\alpha \delta \tilde{z}_\beta}]{}_{(10)}\\
& \phantom{\int} - \sum_{\alpha,\beta=1}^{\mu+1} \tilde{z}_\alpha \nabla_q(f_\beta \mvardif{A}{\tilde{z}_\beta}  )\g\indi{^q^p}[\mvardif{^2 B}{M_p \delta M_k}\g\indi{^k^\ell}\nabla_{\ell} \mvardif{C}{\tilde{z}_\alpha} - \mvardif{^2 C}{M_p \delta M_k}\g\indi{^k^\ell}\nabla_{\ell} \mvardif{B}{\tilde{z}_\alpha}]  {}_{(11)}\\
& \phantom{\int}  - \sum_{\alpha,\beta=1}^{\mu+1} \tilde{z}_\alpha [\mvardif{B}{M_k}\g\indi{^k^\ell}\nabla_{\ell} (\mvardif{^2 C}{\tilde{z}_\alpha \delta M_p} \g\indi{^p^q}\nabla_q(\mvardif{A}{\tilde{z}_\beta}  f_\beta)) - \mvardif{C}{M_k}\g\indi{^k^\ell}\nabla_{\ell} (\mvardif{^2 B}{\tilde{z}_\alpha \delta M_p}  \g\indi{^p^q}\nabla_q(f_\beta \mvardif{A}{\tilde{z}_\beta}  ))] {}_{(12)} \, \dx\\
=& \int_\Omega \sum_{\alpha,\beta=1}^{\mu+1}  \tilde{z}_\alpha \mvardif{A}{M_p}\g\indi{^p^q}\nabla_q \Big( \mvardif{^2 B}{M_k \delta \tilde{z}_\alpha}\g\indi{^k^\ell}\nabla_{\ell} (f_\beta \mvardif{C}{\tilde{z}_\beta}  ) - \mvardif{^2 C}{M_k \delta \tilde{z}_\alpha}\g\indi{^k^\ell}\nabla_{\ell}(f_\beta \mvardif{B}{\tilde{z}_\beta}  )\Big){}_{(1)}\\
& \phantom{\int} + \sum_{\alpha,\beta=1}^{\mu+1}   \tilde{z}_\alpha \mvardif{A}{M_p}\g\indi{^p^q} [\mvardif{B}{M_k}\nabla_q( \g\indi{^k^\ell}\nabla_{\ell} (\mpd{ f_\beta}{\tilde{z}_\alpha}\mvardif{C}{\tilde{z}_\beta} +  f_\beta \mvardif{^2 C}{\tilde{z}_\beta \delta \tilde{z}_\alpha})) - \mvardif{C}{M_k}\nabla_q (\g\indi{^k^\ell}\nabla_{\ell}(\mpd{ f_\beta}{\tilde{z}_\alpha}\mvardif{B}{\tilde{z}_\beta} +  f_\beta \mvardif{^2 B}{\tilde{z}_\beta \delta \tilde{z}_\alpha}))] {}_{(2.a)}\\
& \phantom{\int} + \sum_{\alpha,\beta=1}^{\mu+1}\tilde{z}_\alpha \mvardif{A}{M_p}  \g\indi{^p^q} [\nabla_q(\mvardif{B}{M_k})\g\indi{^k^\ell}\nabla_{\ell} (\mpd{ f_\beta}{\tilde{z}_\alpha}\mvardif{C}{\tilde{z}_\beta} +  f_\beta \mvardif{^2 C}{\tilde{z}_\beta \delta \tilde{z}_\alpha}) -  \nabla_q(\mvardif{C}{M_k})\g\indi{^k^\ell}\nabla_{\ell}(\mpd{ f_\beta}{\tilde{z}_\alpha}\mvardif{B}{\tilde{z}_\beta} +  f_\beta \mvardif{^2 B}{\tilde{z}_\beta \delta \tilde{z}_\alpha})]  {}_{(2.b)}\\
& \phantom{\int} + \sum_{\alpha,\beta=1}^{\mu+1}  \tilde{z}_\alpha\nabla_{\ell} ( \mvardif{A}{M_p}) \g\indi{^\ell^k} \g\indi{^p^q} [\mvardif{B}{M_k}\nabla_q(\mpd{ f_\beta}{\tilde{z}_\alpha}\mvardif{C}{\tilde{z}_\beta} +  f_\beta \mvardif{^2 C}{\tilde{z}_\beta \delta \tilde{z}_\alpha}) - \mvardif{C}{M_k}\nabla_q(\mpd{ f_\beta}{\tilde{z}_\alpha}\mvardif{B}{\tilde{z}_\beta} +  f_\beta \mvardif{^2 B}{\tilde{z}_\beta \delta \tilde{z}_\alpha})]{}_{(3.a)}\\
& \phantom{\int} + \sum_{\alpha,\beta=1}^{\mu+1} \tilde{z}_\alpha \nabla_{\ell} (\mvardif{A}{M_p})\g\indi{^\ell^k} \g\indi{^p^q}[\nabla_q(\mvardif{B}{M_k})(\mpd{ f_\beta}{\tilde{z}_\alpha}\mvardif{C}{\tilde{z}_\beta} +  f_\beta \mvardif{^2 C}{\tilde{z}_\beta \delta \tilde{z}_\alpha}) - \nabla_q(\mvardif{C}{M_k})(\mpd{ f_\beta}{\tilde{z}_\alpha}\mvardif{B}{\tilde{z}_\beta} +  f_\beta \mvardif{^2 B}{\tilde{z}_\beta \delta \tilde{z}_\alpha})] {}_{(3.b)}\\
& \phantom{\int} + \sum_{\alpha,\beta=1}^{\mu+1} \mpd{ f_\beta}{\tilde{z}_\alpha} \nabla_{\ell} (\tilde{z}_\alpha) \g\indi{^\ell^k} \mvardif{A}{M_p} \g\indi{^p^q} [\nabla_q(\mvardif{B}{M_k}) \mvardif{C}{\tilde{z}_\beta} - \nabla_q(\mvardif{C}{M_k})\mvardif{B}{\tilde{z}_\beta} ] {}_{(3.c.i)}\\
& \phantom{\int} + \sum_{\alpha,\beta=1}^{\mu+1}  \mpd{ f_\beta}{\tilde{z}_\alpha} \nabla_{\ell} (\tilde{z}_\alpha) \g\indi{^\ell^k}\mvardif{A}{M_p}\g\indi{^p^q}[\mvardif{B}{M_k}\nabla_q(\mvardif{C}{\tilde{z}_\beta}) - \mvardif{C}{M_k}\nabla_q(\mvardif{B}{\tilde{z}_\beta}) ] {}_{(3.c.ii)}\\
& \phantom{\int} + \sum_{\alpha,\beta=1}^{\mu+1} \mvardif{A}{M_p}\g\indi{^p^q}\nabla_q(\mpd{ f_\beta}{\tilde{z}_\alpha}) \nabla_{\ell} (\tilde{z}_\alpha)\g\indi{^\ell^k}[\mvardif{B}{M_k}\mvardif{C}{\tilde{z}_\beta} - \mvardif{C}{M_k} \mvardif{B}{\tilde{z}_\beta} ]   {}_{(3.c.iii)}\\
& \phantom{\int} + \sum_{\alpha,\beta=1}^{\mu+1} \mvardif{A}{M_p}\g\indi{^p^q}\nabla_{\ell} (\tilde{z}_\alpha) \g\indi{^\ell^k} [\nabla_q(f_\beta\mvardif{B}{M_k})  \mvardif{^2 C}{\tilde{z}_\beta \delta \tilde{z}_\alpha} - \nabla_q(f_\beta \mvardif{C}{M_k}) \mvardif{^2 B}{\tilde{z}_\beta \delta \tilde{z}_\alpha}] {}_{(3.d.i)}\\
& \phantom{\int} + \sum_{\alpha,\beta=1}^{\mu+1}  f_\beta\mvardif{A}{M_p}  \g\indi{^p^q}  \nabla_{\ell} (\tilde{z}_\alpha) \g\indi{^\ell^k} [\mvardif{B}{M_k}\nabla_q\mvardif{^2 C}{\tilde{z}_\beta \delta \tilde{z}_\alpha} - \mvardif{C}{M_k}\nabla_q\mvardif{^2 B}{\tilde{z}_\beta \delta \tilde{z}_\alpha}]  {}_{(3.d.ii)}\\
& \phantom{\int} - \sum_{\alpha,\beta=1}^{\mu+1} \tilde{z}_\alpha \nabla_q (\mvardif{A}{\tilde{z}_\alpha})\g\indi{^q^p}[ \mvardif{^2 B}{M_p \delta M_k}\g\indi{^k^\ell}\nabla_{\ell} (f_\beta \mvardif{C}{  \tilde{z}_\beta} ) - \mvardif{^2 C}{M_p \delta M_k}\g\indi{^k^\ell}\nabla_{\ell}( f_\beta \mvardif{B}{\tilde{z}_\beta} )] {}_{(4)}\\
& \phantom{\int}  -\sum_{\alpha,\beta=1}^{\mu+1}  [\mvardif{B}{M_k}\g\indi{^k^\ell}\nabla_{\ell} (f_\beta \mvardif{^2 C}{\tilde{z}_\beta \delta M_p}  \tilde{z}_\alpha \g\indi{^p^q}\nabla_q \mvardif{A}{\tilde{z}_\alpha}) - \mvardif{C}{M_k}\g\indi{^k^\ell}\nabla_{\ell}( f_\beta \mvardif{^2 B}{\tilde{z}_\beta \delta M_p} \tilde{z}_\alpha \g\indi{^p^q}\nabla_q \mvardif{A}{\tilde{z}_\alpha})]{}_{(5)} \\
& \phantom{\int} + \sum_{\beta=1}^{\mu+1}  \mvardif{A}{M_p}\g\indi{^p^q}\nabla_q( f_\beta)[\mvardif{B}{M_k}\g\indi{^k^\ell}\nabla_{\ell} \mvardif{C}{\tilde{z}_\beta} -  \mvardif{C}{M_k} \g\indi{^k^\ell}\nabla_{\ell} \mvardif{B}{\tilde{z}_\beta}]  {}_{(6.a)}\\
& \phantom{\int} + \sum_{\beta=1}^{\mu+1} f_\beta \mvardif{A}{M_p}\g\indi{^p^q} [\mvardif{B}{M_k}\nabla_q (\g\indi{^k^\ell}\nabla_{\ell} \mvardif{C}{\tilde{z}_\beta}) - \mvardif{C}{M_k}\nabla_q (\g\indi{^k^\ell}\nabla_{\ell} \mvardif{B}{\tilde{z}_\beta})]  {}_{(6.b)}\\
& \phantom{\int} + \sum_{\beta=1}^{\mu+1}   f_\beta \mvardif{A}{M_p}\g\indi{^p^q} [\nabla_q(\mvardif{B}{M_k})\g\indi{^k^\ell}\nabla_{\ell} \mvardif{C}{\tilde{z}_\beta} - \nabla_q( \mvardif{C}{M_k}) \g\indi{^k^\ell}\nabla_{\ell} \mvardif{B}{\tilde{z}_\beta}] {}_{(6.c)}\\
& \phantom{\int}   + \sum_{\alpha,\beta=1}^{\mu+1} \mvardif{A}{M_p}  \g\indi{^p^q}\nabla_q \Big( f_\beta  \tilde{z}_\alpha [\mvardif{^2 B}{M_k \delta \tilde{z}_\beta}\g\indi{^k^\ell}\nabla_{\ell} \mvardif{C}{\tilde{z}_\alpha} - \mvardif{^2 C}{M_k \delta \tilde{z}_\beta}\g\indi{^k^\ell}\nabla_{\ell} \mvardif{B}{\tilde{z}_\alpha}]\Big) {}_{(7)} \\
& \phantom{ \int } + \sum_{\alpha,\beta=1}^{\mu+1} f_\beta  \mvardif{A}{M_p}\g\indi{^p^q}\nabla_q (\tilde{z}_\alpha) [\mvardif{B}{M_k}\g\indi{^k^\ell}\nabla_{\ell} \mvardif{^2 C}{\tilde{z}_\alpha \delta \tilde{z}_\beta} - \mvardif{C}{M_k}\g\indi{^k^\ell}\nabla_{\ell} \mvardif{^2 B}{\tilde{z}_\alpha \delta \tilde{z}_\beta}]  {}_{(8.a)}\\
& \phantom{ \int } + \sum_{\alpha,\beta=1}^{\mu+1} f_\beta \tilde{z}_\alpha \mvardif{A}{M_p}\g\indi{^p^q}  [\mvardif{B}{M_k}\nabla_q (\g\indi{^k^\ell}\nabla_{\ell} \mvardif{^2 C}{\tilde{z}_\alpha \delta \tilde{z}_\beta} ) - \mvardif{C}{M_k}\nabla_q (\g\indi{^k^\ell}\nabla_{\ell} \mvardif{^2 B}{\tilde{z}_\alpha \delta \tilde{z}_\beta})]  {}_{(8.b)}\\
& \phantom{ \int } + \sum_{\alpha,\beta=1}^{\mu+1} \tilde{z}_\alpha\mvardif{A}{M_p}\g\indi{^p^q} [\nabla_q  (f_\beta \mvardif{B}{M_k})\g\indi{^k^\ell}\nabla_{\ell} \mvardif{^2 C}{\tilde{z}_\alpha \delta \tilde{z}_\beta} - \nabla_q  (f_\beta \mvardif{C}{M_k})\g\indi{^k^\ell}\nabla_{\ell} \mvardif{^2 B}{\tilde{z}_\alpha \delta \tilde{z}_\beta}]  {}_{(8.c)}\\
& \phantom{ \int } +\sum_{\alpha,\beta=1}^{\mu+1} \nabla_q( \tilde{z}_\alpha) \g\indi{^q^p} \nabla_{\ell}(  f_\beta \mvardif{A}{M_p}) \g\indi{^\ell^k}[\mvardif{B}{M_k}\mvardif{^2 C}{\tilde{z}_\alpha \delta \tilde{z}_\beta} - \mvardif{C}{M_k} \mvardif{^2 B}{\tilde{z}_\alpha \delta \tilde{z}_\beta}] {}_{(9.a)}\\
& \phantom{ \int } +\sum_{\alpha,\beta=1}^{\mu+1} \tilde{z}_\alpha\nabla_{\ell}( f_\beta) \g\indi{^\ell^k} \mvardif{A}{M_p} \g\indi{^p^q} [\nabla_q(\mvardif{B}{M_k})\mvardif{^2 C}{\tilde{z}_\alpha \delta \tilde{z}_\beta} - \nabla_q(\mvardif{C}{M_k}) \mvardif{^2 B}{\tilde{z}_\alpha \delta \tilde{z}_\beta}]  {}_{(9.b)}\\
& \phantom{ \int } +\sum_{\alpha,\beta=1}^{\mu+1} \tilde{z}_\alpha f_\beta \nabla_{\ell}( \mvardif{A}{M_p}) \g\indi{^\ell^k}\g\indi{^p^q} [\nabla_q(\mvardif{B}{M_k})\mvardif{^2 C}{\tilde{z}_\alpha \delta \tilde{z}_\beta} - \nabla_q(\mvardif{C}{M_k}) \mvardif{^2 B}{\tilde{z}_\alpha \delta \tilde{z}_\beta}] {}_{(9.c)}\\
& \phantom{ \int } + \sum_{\alpha,\beta=1}^{\mu+1} \tilde{z}_\alpha\nabla_{\ell}(  f_\beta \mvardif{A}{M_p})\g\indi{^\ell^k}\g\indi{^p^q}[\mvardif{B}{M_k}\nabla_q\mvardif{^2 C}{\tilde{z}_\alpha \delta \tilde{z}_\beta} - \mvardif{C}{M_k}\nabla_q \mvardif{^2 B}{\tilde{z}_\alpha \delta \tilde{z}_\beta}] {}_{(9.d)}\\
& \phantom{\int} + \sum_{\alpha,\beta=1}^{\mu+1}\tilde{z}_\alpha\nabla_q( f_\beta) \g\indi{^q^p}\nabla_{\ell}(\mvardif{A}{M_p})\g\indi{^\ell^k} [\mvardif{B}{M_k}\mvardif{^2 C}{\tilde{z}_\alpha \delta \tilde{z}_\beta} - \mvardif{C}{M_k} \mvardif{^2 B}{\tilde{z}_\alpha \delta \tilde{z}_\beta}] {}_{(10.a)}\\
& \phantom{\int} + \sum_{\alpha,\beta=1}^{\mu+1} \tilde{z}_\alpha\mvardif{A}{M_p}\nabla_{\ell}( \g\indi{^p^q}\nabla_q f_\beta) \g\indi{^\ell^k}[\mvardif{B}{M_k}\mvardif{^2 C}{\tilde{z}_\alpha \delta \tilde{z}_\beta} - \mvardif{C}{M_k} \mvardif{^2 B}{\tilde{z}_\alpha \delta \tilde{z}_\beta}]{}_{(10.b)}\\
& \phantom{\int} - \sum_{\alpha,\beta=1}^{\mu+1} \tilde{z}_\alpha \nabla_q(f_\beta \mvardif{A}{\tilde{z}_\beta}  )\g\indi{^q^p}[\mvardif{^2 B}{M_p \delta M_k}\g\indi{^k^\ell}\nabla_{\ell} \mvardif{C}{\tilde{z}_\alpha} - \mvardif{^2 C}{M_p \delta M_k}\g\indi{^k^\ell}\nabla_{\ell} \mvardif{B}{\tilde{z}_\alpha}]  {}_{(11)}\\
& \phantom{\int}  - \sum_{\alpha,\beta=1}^{\mu+1} \tilde{z}_\alpha [\mvardif{B}{M_k}\g\indi{^k^\ell}\nabla_{\ell} (\mvardif{^2 C}{\tilde{z}_\alpha \delta M_p} \g\indi{^p^q}\nabla_q(\mvardif{A}{\tilde{z}_\beta}  f_\beta)) - \mvardif{C}{M_k}\g\indi{^k^\ell}\nabla_{\ell} (\mvardif{^2 B}{\tilde{z}_\alpha \delta M_p}  \g\indi{^p^q}\nabla_q(f_\beta \mvardif{A}{\tilde{z}_\beta}  ))] {}_{(12)} \, \dx.
\end{align*}
Then under cyclic summation all terms vanish except of $(3.c.i)$ and $(6.c)$. This can be seen if one considers $ \sum_{\alpha=1}^{\mu+1}\tpd{f_\beta}{\tilde{z}_\alpha} \nabla_\ell \tilde{z}_\alpha  = \nabla_\ell f_\beta$ and the combination $(1)$ together with $(12)$; $(2.a)$; $(2.b)$ together with $(3.a)$; $(3.b)$;  $(3.c.ii)$ together with $(6.a)$; 
$(3.c.iii)$; $(3.d.i)$ together with $(9.a)$; $(3.d.ii)$ together with $(8.a)$; $(4)$ together with $(11)$; $(5)$ together with $(7)$; $(6.b)$; $(8.b)$; $(8.c)$ together with $(9.d)$; $(9.b)$ together with $(10.a)$; $(9.c)$; and $(10.b)$. If one does the same expansion for $\{A,\{B,C\}_3\}_1+ \{A,\{B,C\}_1\}_3$, one additional term
\begin{align*}
& \int_\Omega -\sum_{\alpha,\beta=1}^{\mu+1} \nabla_q(f_\beta \mvardif{A}{\tilde{z}_\beta}  )\g\indi{^q^p}[\mvardif{B}{M_k}\g\indi{^k^\ell}\nabla_{\ell} \mvardif{C}{M_p} - \mvardif{C}{M_k}\g\indi{^k^\ell}\nabla_{\ell} \mvardif{B}{M_p}] {}_{(13)} \,\dx\\
=& \int_\Omega - \sum_{\beta=1}^{\mu+1} \mvardif{A}{\tilde{z}_\beta}\nabla_q(f_\beta)\g\indi{^q^p} [\mvardif{B}{M_k}\g\indi{^k^\ell}\nabla_{\ell} \mvardif{C}{M_p} - \mvardif{C}{M_k}\g\indi{^k^\ell}\nabla_{\ell} \mvardif{B}{M_p}]  {}_{(13.a)}\\
& \phantom{\int} - \sum_{\beta=1}^{\mu+1} f_\beta  \nabla_q(\mvardif{A}{\tilde{z}_\beta}  )\g\indi{^q^p} [\mvardif{B}{M_k}\g\indi{^k^\ell}\nabla_{\ell} \mvardif{C}{M_p} - \mvardif{C}{M_k}\g\indi{^k^\ell}\nabla_{\ell} \mvardif{B}{M_p}]{}_{(13.b)}\,\dx
\end{align*}
appears, while the terms $(3.c.\ast)$ and $(6.a)$ do not appear, since $f_\alpha$ and $\tilde{z}_\alpha$ are independent of $\mathbf{M}$. Note, that term $(13)$ was not a part of $\{A,\{B,C\}_3\}_2+ \{A,\{B,C\}_2\}_3$, since $\tpd{M_k}{\tilde{z}_\alpha}=0$. 
However, under the cyclic summation as for $\{\cdot,\{\cdot,\cdot\}_3\}_2+ \{\cdot,\{\cdot,\cdot\}_2\}_3$ all terms of $\{\cdot,\{\cdot,\cdot\}_3\}_1+ \{\cdot,\{\cdot,\cdot\}_1\}_3$ vanish unless $(13)$.  Finally, one notice that $(13.a)$ together with $(3.c.i)$ as well as $(6.c)$ together with $(13.b)$ annul each other in a cyclic sum. Therefore, the Jacobi identity follows with the previous calculations and
\begin{align*}
\{A,\{B,C\}\}= & \{A,\{B,C\}_{1}\}_{1} + [ \{A,\{B,C\}_{2}\}_{2} + \{A,\{B,C\}_{2}\}_{1} + \{A,\{B,C\}_{1}\}_{2}]\\
 &\qquad + \{A,\{B,C\}_{3}\}_{3} - \sum_{i=1}^{2} [\{A,\{B,C\}_{3}\}_{i} + \{A,\{B,C\}_{i}\}_{3}]. \qedhere 
\end{align*}
\end{proof}
	\end{appendices}			

\end{document}